\documentclass[12pt,a4paper]{article}
\usepackage{jheppub}
\usepackage{rotating}
\usepackage{amsmath,amssymb,amsfonts,pstricks,mathtools,epsfig}
\usepackage{float}
\usepackage{subfig}
\usepackage{scalefnt}
\usepackage{wrapfig}
\usepackage{fancybox}
\newcommand{\bea}{\begin{eqnarray}\displaystyle}
\newcommand{\eea}{\end{eqnarray}}

\newlength{\arrow}
\newcommand{\cA}{\mathcal{A}}
\newcommand{\Li}{\mathrm{Li}}
\newcommand{\cM}{\mathcal{M}}
\newcommand{\cO}{\mathcal{O}}
\newcommand{\C}{\mathbb{C}}
\newcommand{\PP}{\mathbb{P}}
\newcommand{\I}{\mathrm{i}}
\newcommand{\de}{\mathrm{d}}
\renewcommand{\sf}{\mathrm{sf}}
\newcommand{\inst}{\mathrm{inst}}
\newcommand{\IP}[1]{\langle{#1}\rangle}   
\newcommand{\ti}[1]{\textit{#1}}
\newcommand{\eps}{\epsilon}

{\setlength{\fboxsep}{15pt}
\setlength{\mylength}{\linewidth}%
\addtolength{\mylength}{-2\fboxsep}%
\addtolength{\mylength}{-2\fboxrule}%
\Sbox
\minipage{\mylength}%
\setlength{\abovedisplayskip}{0pt}%
\setlength{\belowdisplayskip}{0pt}%
\equation}%
{\endequation\endminipage\endSbox
\[\fbox{\TheSbox}\]}

\newcommand\cW{{\mathcal W}}
\renewcommand\sf{{\mathrm{sf}}}
\newcommand\N{{\mathcal N}}
\newcommand{\insfigscaled}[3]{\begin{figure}[htbp] \centering \includegraphics[scale=#2]{#1-crop.pdf} \caption{#3} \label{fig:#1} \end{figure}}
\newcommand\abs[1]{\lvert #1 \rvert}
\newcommand\R{{\mathbb R}}

\newcommand\re{{\mathrm {Re}}}
\newcommand\im{{\mathrm {Im}}}
\newcommand\half{\frac12}
\begin{document}
\title{Twistorial Topological Strings and a\\ $tt^*$ Geometry for ${\cal N}=2$ Theories in $4d$}

\author[\dag]{Sergio Cecotti,}
\author[\S]{Andrew Neitzke,}
\author[\ast]{Cumrun Vafa}
\affiliation[\dag]{International School of Advanced Studies (SISSA),
via Bonomea 265, 34136 Trieste}
\affiliation[\S]{The University of Texas at Austin, Math Department, Austin, TX 78712-1202, USA }
\affiliation[\ast]{Jefferson Physical Laboratory, Harvard University, Cambridge, MA 02138, USA}

\abstract{We define twistorial topological strings by considering $tt^*$ geometry of the 4d ${\cal N}=2$ supersymmetric theories
on the Nekrasov-Shatashvili ${1\over 2}\Omega$ background, which leads to quantization of
the associated hyperK\"ahler geometries.  We show that in one limit it reduces to the refined
topological string amplitude.  In another limit it is a solution to a quantum Riemann-Hilbert
problem involving quantum Kontsevich-Soibelman operators.
In a further limit it encodes
the hyperK\"ahler integrable systems studied by GMN. 
In the context of AGT conjecture, this perspective leads to a twistorial extension of Toda.  The 2d index of the ${1\over 2}\Omega$
theory leads to the recently introduced index for ${\cal N}=2$ theories in 4d.
 The twistorial topological string can alternatively be viewed, using the work of
Nekrasov-Witten, as studying the vacuum geometry of 4d $ {\cal N}=2$ supersymmetric theories on $T^2\times I$ where
$I$ is an interval with specific boundary conditions at the two ends. }

\maketitle

\section{Introduction}

Supersymmetric theories have a rich vacuum structure.  On the other hand studying degenerate states as a function of parameter
space in a quantum mechanical system  is well known to lead to Berry's connection on the parameter space.  Combining these
two ideas,  it is natural
to ask what is the geometry of the vacua for supersymmetric quantum theories.  It is most natural to study
this when we consider the space to be a compact flat geometry such as tori.
This question has been answered for theories with 4 supercharges in $d=2,3,4$ dimensions \cite{Cecotti:1991me,Cecotti:2013mba} leading
to a highly nontrivial geometry known as $tt^*$.  For more supersymmetry the vacuum geometry in a sense becomes too
rigid and more universal and thus less interesting.   It is natural to ask if there is any
way which we can get a non-trivial vacuum geometry out of theories with say $8$ supercharges, and in particular
for $4d$ theories with ${\cal N}=2$ supersymmetry (for other attempts in this direction see \cite{papa}).

Motivated by the similarity between $tt^*$ geometry for theories with 4 supercharges and open
topological string amplitudes, in \cite{Vafa:2014lca} a twistorial extension of topological string was proposed.
The main aim of this paper is to make this more precise and compute the corresponding amplitudes
in some simple cases.  Translating the proposal in \cite{Vafa:2014lca}, we come up with a natural
definition of twistorial topological string, in terms of the corresponding target space physics.
For topological B-model the target physics involves type IIB theories on local Calabi-Yau threefolds
and  for A-model it involves M-theory compactifications on local Calabi-Yau threefolds times a circle.
In both cases we end up with a theory in 4 dimensions with ${\cal N}=2$ supersymmetry.
 The basic
idea is to consider the ${1\over 2}\Omega$ background \cite{Nekrasov:2009rc} with some parameter $\epsilon_1$.  In
M-theory picture this involves rotating the 3-4 plane by $\epsilon_1$  as we go around the 5-th circle
(and doing a compensating rotation in the non-compact Calabi-Yau 3-fold to preserve supersymmetry).  In the B-model it is more implicit
but can be viewed as mirror to the above operation.  As argued in \cite{Nekrasov:2009rc} in such a case we end up
with a theory in 2d which has ${\cal N}=2$ supersymmetry with infinitely many discrete vacua where
the Coulomb branch parameters are quantized $\vec a =\vec k \epsilon_1$ with a mass gap.
This allows us to study the associated $tt^*$ geometry, by putting the theory on a circle of length $R={1/{\tilde \epsilon_2}}$.
The theory will have natural D-branes labeled by vacua $\vec k$, and a phase $\zeta$ depicting the choice of which combination of two supercharges we preserve on the D-brane.  $tt^*$ geometry \cite{Cecotti:1991me} can be used \cite{Hori:2000ck} to compute 
the wave function of such D-branes when we take the overlap
of these states with vacua of the theory.  The phase $\zeta$ can be extended to the full complex plane excluding
$0$ and $\infty$ and will play
the role of twistor parameter for us.  The D-brane amplitudes define the twistorial topological string amplitudes.
One can show that in the limit $R\rightarrow 0$ and $\zeta\rightarrow 0$ keeping $\zeta/R \equiv \epsilon_2$ finite,
we get a discretization of refined topological strings at Coulomb branch parameters given by $\vec a= \vec k \epsilon_1$,
which is sufficient to give an unambiguous perturbative expansion in $\epsilon_i$.  In this limit, the amplitudes reduce
to that of refined topological strings, or equivalently to the full $\Omega$ background with parameters $\epsilon_1,\epsilon_2$.

One can also interpret this structure in terms of the geometry of ${\cal N}=2$ supersymmetric vacua in $4d$  along the lines
of \cite{Nekrasov:2010ka}.  This leads to a direct interpretation of twistorial topological strings in terms of a $tt^*$ geometry
for the ${\cal N}=2$ theories in $4d$.
Consider the $4d$ theory on $T^2\times I$ where $I$ is an interval of length $L$ and $T^2$ with radii $1/\epsilon_1, {1/ {\tilde \epsilon_2}}$ (and
the tilt of the $T^2$ given by an additional angle $\theta$ leading to the complex moduli of torus $\tau=\tfrac{\theta}{2\pi}+i{\epsilon_1\over {\tilde \epsilon_2}}).$   On one end of the the interval $I$ we put a D-brane which is related to Dirichlet condition along one of the cycles
of the $T^2$ for electric gauge components (and its supersymmetric completion).  On the other end we have a ${1\over 2}{\Omega}$
deformation which can be viewed as a D-brane (brane of ``opers'') of a 3d theory obtained from compactification of the 4d theory on the same circle.  In other words, from the perspective of the resulting 3d theory we have a space given by $S^1\times I$ where
the supersymmetry is reduced to 4 supercharges by the D-branes on both ends.  This results in vacua labeled by $\vec k$, which we
can study in the usual $tt^*$ setup, treating $S^1$ as the circle in 2d.  The D-brane wave function of this geometry,
in the limit the length $L$ of the interval  goes to infinity, leads to the twistorial topological string.

In the limit $\epsilon_1\rightarrow 0$, we find evidence that the theory reduces to the hyperK\"ahler geometry studied by Gaiotto-Moore-Neitzke in \cite{Gaiotto:2008cd}.
More precisely, we obtain a quantum version of this geometry by keeping $\theta$ finite, what
we call the $\theta$-limit, and obtain
a quantum Riemann-Hilbert problem for the line operators.  The twistorial partition function
is a wave function associated to this quantum Riemann-Hilbert problem.  There is a further limit, a `classical
limit' where $\theta \rightarrow  0$, where we make contact with the standard version of the story 
of \cite{Gaiotto:2008cd}.
In this limit we expect that the twistorial topological string partition function gets related
to the objects introduced in \cite{Neitzke:2011za} as part of the construction of a hyperholomorphic line 
bundle over the hyperK\"ahler moduli space which is the target of the 3d sigma model.  
We show that this is indeed the case for some simple examples.  We can also consider, in the $\theta$ limit, to make the 2d time to be correlated with the 
phase of the supersymmetry we preserve on $S^1$ as in \cite{Cecotti:2010qn}.  In this context we make
contact with the work \cite{Cecotti:2010fi}, where the trace of the monodromy of the $tt^*$ geometry in this limit
was the object of study.

We can also study other twistorial invariant (i.e.\! wall-crossing invariant) objects in the $tt^*$ setup.  In particular we study
the metric on the ground state vacua (leading to Berry's connection).  Among the vacua, there is a distinguished one, corresponding
to the insertion of the identity operator in topologically twisted theory.  Studying its norm $\langle {\overline 0}|0\rangle$ leads to a partition function which
depends only on masses of the $4d$ theory as well as the $(\epsilon_1,{\tilde \epsilon_2},\theta)$.  It is the twistorial
extension of combining topological string amplitudes with anti-topological string amplitudes.  In the usual $\Omega$ background,
a similar object has been related to partition function of the $4d$ theory on $S^4$ \cite{Pestun:2007rz}, and in the context of M5 branes on Riemann
surfaces the resulting amplitudes have been related to Toda theories  \cite{AGT,wyllard}.  In these cases we find a twistorial
extension of the resulting theories.
Another object one studies in the 2d setup is the CFIV index \cite{Cecotti:1992qh}.  We provide evidence  that in the
limit $\epsilon_1\rightarrow 0$, $\theta \rightarrow 0$  this index becomes equivalent to the recently studied
AMNP index \cite{Alexandrov:2014wca} of the associated 4d theories.  In addition, studying the R-flow of the 2d
theory \cite{Cecotti:2010qn,Cecotti:2011iy}  leads to the 4d quantum KS monodromy studied in \cite{Cecotti:2010fi}.

In a sense twistorial topological string can be viewed as quantizing the hyperK\"ahler geometry associated
to circle compactifications of $4d$ ${\cal N}=2$ theories, where one of the parameters ($\epsilon_1$) quantizes the Coulomb branch base, and another parameter ($\theta$) quantizes the Jacobian fiber of the hyperK\"ahler space.

For a different approach to a ``twistorial'' extension of the topological string see \cite{Alexandrov:2010ca}.

\bigskip

The organization of this paper is as follows:
In section 2 we review the definition of twistorial open topological string \cite{Vafa:2014lca}.  In section 3 we define
twistorial closed topological strings.  We do this in two ways:  One is to use large $N$ dualities of topological strings, which
we review and use as a spring
board for a twistorial definition of closed topological string.  We also give alternative, more general definition
of twistorial topological string without employing large $N$ dualities, using $tt^*$ of ${1\over 2}\Omega$ background.
We also reinterpret this in terms of studying $tt^*$ geometry by placing branes on the boundaries of the space.
Furthermore we discuss the various interesting limits one can take, including in particular  the $\theta$-limit.
We study the $\theta$-limit in more detail, and relate it to a quantum Riemann-Hilbert problem, in section 4.
In section 5 we study twistorial extension of matrix models by studying (2,2) supersymmetric LG matrix models and the associated
D-brane wave functions, and solve explicitly the twistorial extension of the Gaussian matrix model and a number of related
examples (which have abelian $tt^*$ geometries).  This example leads, by large $N$ duality, to the twistorial extension of the conifold (a.k.a.\! $N=2$ SQED) which is discussed in section 6.  In section 7 we evaluate the three point function for the twistorial Liouville theory (both the twistorial conformal block as well as the twistorial 3--point function).
We also show how the AMNP index is related to the CFIV index.
In section 8 we study the classical limit (C-limit) of the twistorial topological string and make contact with the
hyperholomorphic line bundle on moduli space of ${\cal N}=2$ studied in \cite{Neitzke:2011za},  as well as the twistorial line operators studied in \cite{Gaiotto:2008cd}.  In section 9 we close by presenting some concluding remarks.
Some technical details and extensions of the ideas discussed are relegated to appendices.

\section{Open twistorial topological string}

In this section we first quickly review $tt^*$ geometry and then show how it is connected to the open
topological string.

\subsection{A lightning review of $tt^*$} \label{s:ttstar-review}

Consider a quantum field theory in 2 dimensions with $(2,2)$ SUSY, which is \ti{massive}, i.e.\! there is a 
discrete set of $m$ vacua, each with a mass gap.  Putting such a theory on a spatial circle of length $R$, we obtain
a Hilbert space with an $m$-dimensional ground state subspace.  Varying parameters of the theory (deforming by chiral 
operators), we get a Berry connection on the bundle of Hilbert spaces over parameter space, 
which restricts to a unitary connection $D$ on 
the $m$-dimensional ground state bundle.  Moreover, we have the $tt^*$ equations 
\cite{Cecotti:1991me,Cecotti:1992rm}:  if we define the ``improved'' connection
$$\nabla_\zeta = D + R {C\over \zeta}, \qquad {\overline \nabla}_{\zeta} = {\overline D} + R \zeta\, {\overline C}$$
where $C$ denotes the action of the chiral operators on the vacua, and $\zeta \in \C^\times$ is arbitrary, then
$$[\nabla_\zeta, \nabla_\zeta] = [\nabla_\zeta, {\overline \nabla}_\zeta] = [{\overline \nabla}_\zeta, \overline{\nabla}_\zeta]=0.$$
We refer to $\nabla_\zeta$, $\overline{\nabla}_\zeta$ as the $tt^*$ Lax connection.

Our major objects of study will be flat\footnote{ From the higher--dimensional hyperK\"ahler perspective of \cite{Cecotti:2013mba} $\psi$ is a (non--flat) section of the vacuum hyperholomorphic bundle which is holomorphic in complex structure $\zeta$. } sections 
$\psi$ of the $tt^*$ Lax connection, obeying
$$\nabla_\zeta \psi={\overline \nabla}_\zeta \psi = 0.$$
There is a distinguished set of such sections $D_b$, $b = 1, \dots, N$,
obtained, as we explain below, from \ti{boundary states} corresponding to a distinguished 
set of D-branes.
These D-branes break half the supersymmetry; which half of the supersymmetry they preserve is 
characterized by an angle $\phi$, which is related to the $\zeta$ appearing
above by $\zeta = \exp(\I \phi)$.
Thus, for the flat sections arising from D-branes 
the parameter $\zeta$  
is restricted to have $|\zeta|=1$.

Vacua of the theory also play a distinguished role.   They are in 1-1 correspondence
with the chiral ring elements.  For the chiral
ring element $a$, the vacuum state $\langle a|$ is obtained by performing the path integral
over a ``cigar'' geometry, with a topological twist near the tip, and the chiral operator 
$a$ inserted at the tip \cite{Cecotti:1991me}.  This gives a holomorphic section of the vacuum
bundle.  One can also choose a unitary section of the vacua, by suitably normalizing them.
Thus, to associate wave functions to D-branes $D_b$ we consider
$$\psi^a_b = \langle a | D_b \rangle.$$
In particular, letting $a$ be the identity operator gives a distinguished element
$$\psi^0_b = \langle 0 | D_b \rangle.$$  
Sometimes we will drop the superscript for this distinguished wave function
and denote it simply as $\psi_b$.  Note that this will depend on the choice of basis
for the vacuum.   The holomorphic versus the unitary basis differ by the normalization
factor $1/\sqrt{\langle 0 |{\overline 0}\rangle}$.  Both bases will be useful for us.  We will be implicit about which
choice of basis we make for the vacuum, until the examples sections.

Now suppose the theory is an $\N=2$ Landau-Ginzburg model, with chiral multiplet fields $X_i$ and 
a holomorphic superpotential $W(X_i)$.
In this case the distinguished D-branes can be described explicitly; they impose Dirichlet type 
boundary conditions on the $X_i$, restricting them to
Lagrangian cycles $D_b$.  Each cycle $D_b$ 
is a ``Lefschetz thimble'' beginning from a critical point of $W$,
along which $\re (W / \eps_2) \to \infty$ \cite{Hori:2000ck}.
The chiral ring elements can also be described explicitly: indeed the chiral ring
is the Jacobian ring ${\cal R} = \mathbb{C}[X_i] / \langle \partial_j W \rangle$,
so each chiral ring element $a$ corresponds to some holomorphic function $f^a(X)$.

The explicit computation of the $\psi^a_b$
is in general very difficult, and no closed form for them is known,
except in a handful of cases. However, there is a limit in which 
they simplify: fix some $\eps_2 \in \C$ and take
\begin{equation}\label{asymm-limit}
\zeta = R\,\epsilon_2, \qquad R \rightarrow 0.
\end{equation}
We call \eqref{asymm-limit} the \emph{asymmetric limit}.
For a Landau-Ginzburg theory, we then have an explicit formula:
$$\lim_{asym} \psi^a_b = \int_{D_b} \de X \ f^a(X)\ \exp(-W(X)/\epsilon_2)$$
and in particular
\begin{equation} \label{asymm-flat}
\lim_{asym} \psi^0_b = \int_{D_b} \de X \ \exp(-W(X)/\epsilon_2).
\end{equation}
However, we emphasize that there is something unphysical about this limit:  we have continued $\zeta$ away from the locus 
$\abs{\zeta} = 1$, so that the corresponding state $\psi_b$ no longer has a direct interpretation as a D-brane
in the original theory.  This is like taking a non-unitary deformation of the theory,
in which we set $\overline W = 0$ and replace $W \rightarrow W / \epsilon_2$.

\subsection{Connection with open topological strings}\label{s:connectionopentopstrings}

The kind of $\N=(2,2)$ theories we have just discussed can naturally arise from string theory, as follows.
Fix a non-compact Calabi-Yau threefold $CY$, with a non-compact holomorphic curve $Y \subset CY$.
We consider the Type IIB superstring on
$CY \times \R^4$, with D3-branes on a subspace $Y \times \R^2$.  
The theory admits $\Omega$-deformation \cite{Nekrasov:2002qd}, with parameters
\begin{itemize}
\item $\eps_1$ for a rotation in the $x^3-x^4$ plane (transverse to the brane),
\item $\eps_2$ for a rotation in the $x^1-x^2$ plane (along the brane).
\end{itemize}
If we hold $\eps_2 = 0$ (the Nekrasov-Shatashvili limit \cite{Nekrasov:2009rc} or
$\half\Omega$ background), this system has 2-dimensional Poincar\'e invariance
and $\N=(2,2)$ supersymmetry.\footnote{ Although we are focusing on the B-model to be concrete,
all of this discussion has a parallel version in the A-model; the corresponding physical picture would 
involve M-theory on an $\R^4$ bundle over $CY \times S^1$, 
where as we go around $S^1$ we rotate $\R^4$ by angles $\epsilon_1$, $\epsilon_2$.}

Indeed, if we consider a single D3-brane, 
the theory can be described as
a Landau-Ginzburg model, where the superpotential $W = W(X, \eps_1, \vec t\,)$ is a ``holomorphic 
Chern-Simons''-type functional, a function of fields $X$ representing
deformations of the holomorphic curve 
$Y$ \cite{Aganagic:2000gs}, depending on background parameters $\vec t$ controlling
the complex structure moduli of $CY$.
Thus, the theory with the full $\Omega$-background turned on can be viewed as a deformation 
of this 2-dimensional Landau-Ginzburg model; this viewpoint will be useful momentarily.

The physical setup just discussed has an analogue in the topological string:
we consider the B model on $CY$ with a brane on 
$Y \subset CY$.  It has been found in this 
case \cite{Dijkgraaf:2002fc,Dijkgraaf:2002vw,Aganagic:2011mi} 
that the refined open 
topological string partition function is
$$ Z_{open} = \int \de X\ \exp\!\left[-{1\over \epsilon_2} W(X, \epsilon_1, \vec t\;)\right]. $$
Now let us consider a slightly fancier situation, where we have $N$ branes rather than one,
and a particular choice for $CY$, as follows.
Consider a hypersurface in $\C^4$ of the form
$$y^2 = W'(x)^2 + uv$$
where $W(x)$ is a polynomial of degree $n+1$, and $W'(x)$ has $n$ simple zeroes.
Each of these zeroes gives a conifold singularity; blowing each of them up to an exceptional cycle $Y_i \simeq \PP^1$ gives a
smooth Calabi-Yau threefold, which we take to be our $CY$.
Now we can wrap $k_i$ D3-branes around the cycles $Y_i \times \R^2$, as we considered above.  Let $N = \sum k_i$.  

The corresponding open topological string amplitude is known to be
\cite{Dijkgraaf:2002fc,Dijkgraaf:2002vw,Aganagic:2011mi}
\begin{equation} \label{Zopen}
Z_{open}(\vec k, \eps_1, \eps_2) = \int_{D_{\vec k}} \de x^j \;\Delta (x)^{\epsilon_1\over \epsilon_2}\ \exp\!\Big(-\frac{1}{\eps_2} \sum_j W(x^j) \Big).
\end{equation}
Here $\Delta(x)$ is the squared Vandermonde,
$$\Delta(x)=\prod_{j_1 \neq j_2}(x^{j_1} - x^{j_2}),$$
and the integration cycle $D_{\vec k}$ is defined by integrating $k_i$
of the $x^j$ along the steepest-descent contour emanating from
the $i$-th critical point of $W$ (along this contour $\re (W / \eps_2) \to +\infty$ while $\im (W / \eps_2)$ remains fixed,
so that the integral is convergent).

Now here is the key point:  \eqref{Zopen} 
can be identified with the asymmetric limit of a $tt^*$ flat section in the physical $\N=(2,2)$ theory!
Indeed, in this case the physical theory is a gauged Landau-Ginzburg model, where the field $\Phi$ is an $N \times N$ matrix, 
we have a gauge group $U(N)$, and the superpotential is $\mathrm{Tr}\,W(\Phi)$.
Upon integrating out the gauge dynamics the Landau-Ginzburg model is replaced by an effective
version, where the fields are just the eigenvalues $x^j$ of $\Phi$,
with the superpotential
\begin{equation} \label{Eefft}
\mathcal{W}^{eff}(x) = \sum_j W(x^j) + {\epsilon_1} \sum_{j_1 \neq j_2} \log (x^{j_1} - x^{j_2}).
\end{equation}
We now revisit the formula \eqref{asymm-flat} for the 
asymmetric limit of the 
$tt^*$ flat sections corresponding to D-branes of this 2-dimensional model.
The integration cycle $D_{\vec k}$
is the Lefschetz thimble attached to the critical point of $W^{eff}$ 
labeled by $\vec k$.\footnote{ More precisely, this is the description for $\eps_1 = 0$; for $\epsilon_1 \not=0$ the
critical points are deformed, as we discuss later in this paper, but their 
labeling does not change.}
Thus, \eqref{asymm-flat} is \ti{identical} 
to \eqref{Zopen}:
$$Z_{open}(\vec k, \eps_1, \eps_2) = \lim_{asym} \psi^0_{\vec k}(\eps_1, \eps_2).$$

It was this observation that
motivated the definition of the ``twistorial open topological string'' in \cite{Vafa:2014lca}.  Namely,
on the $tt^*$ geometry side we can move away from the asymmetric limit, and this
means that we have a deformation on the topological string side as well:  we \ti{define} \cite{Vafa:2014lca}
$$\psi_{open}^{twist}(\vec k, \epsilon_1, {\tilde \epsilon_2}, \zeta) = \psi^0_{\vec k}(\epsilon_1, {\tilde \epsilon_2}, \zeta), $$
where we introduced the notation
$$\tilde \epsilon_2=R^{-1}$$
with $R$ the length of the circle which appears in the $tt^*$ story.\footnote{ Note that rescaling the length by a factor
of $R$ changes $W\rightarrow RW$.} 

More generally, for any choice of $CY$, we expect that
the ordinary open topological string can be recovered as the asymmetric limit of the $tt^*$ flat section 
corresponding to the D-brane, and thus for any $CY$
\ti{we define the twistorial open topological string partition function 
to be $\psi^0_b$, i.e.\! the overlap between the $tt^*$ flat section corresponding to a boundary condition $b$ 
and the topological ground state.}

Now, note that in the above example the superpotential $\mathcal{W}^{eff}$ is actually
multivalued due to the logarithm.
This introduces a wrinkle in the $tt^*$ geometry story:  we need to consider
a cover of the field space, on which $\mathcal{W}^{eff}$ is single-valued.
In particular, each vacuum $\vec k$ gets replaced by an integer's worth of vacua on this cover.
It is convenient to work in a different basis for the vacua, 
introducing a phase $\theta$ Fourier dual to this integer.
Thus the coupling constants of the twistorial topological string include, in addition to the
parameters $\epsilon_1$ and $\tilde \epsilon_2$, the new circle-valued parameter 
$\theta$.
Similar additional parameters were encountered in \cite{Cecotti:2013mba}, where it was 
shown that solutions of the resulting extended version of $tt^*$ can be 
understood as hyperholomorphic connections
over the extended parameter space; in the examples we consider below, we will find the same
structure.

Additional angular variables will emerge, which
also allows us to make contact with the setup of \cite{Gaiotto:2008cd,Gaiotto:2010be,Gaiotto:2011tf},
as we now explain.
 Consider the local Calabi-Yau geometry
$$uv=f(x,y).$$
This can be interpreted as in geometric engineering context  as a 4d theory with SW curve $f(x,y)=0$.  Furthermore we
can consider B-branes given by $u=0, x=x_0$ parameterized by a point $x_0$ on the SW curve.  In the uncompactified worldsheet 2d theory
of this brane we get a 2d theory which (in the $\epsilon_1=0$ case) has a superpotential $W(x_0)$ with $dW=y(x_0) dx_0$ where $ydx $ is
the SW differential \cite{Aganagic:2000gs}.  From the 4d viewpoint, this can be interpreted as a surface operator \cite{Gukov:2006jk} whose moduli is parameterized by $x_0$.
The $tt^*$ geometry for this theory would be, by the definition above, the open twistorial topological string.  Note however,
in this setup we have extra parameters in the target space geometry having to do with the choice of the electric  and magnetic
Wilson lines around the circumference of $S^1$ in the $tt^*$ geometry.  This is consistent with the fact that the 2d LG theory has
a multi-valued superpotential and extra parameters can also be introduced for it.  In fact this case has been studied from
the perspective of target $4d$ theory in \cite{Gaiotto:2011tf} (see also \cite{Cecotti:2013mba}).  In particular it is shown there that as we take $x_0$ around a cycle $\gamma$
of the SW curve the D-brane wave function picks up monodromy $X_\gamma$, where $X_\gamma$ can be interpreted as the line
operators of \cite{Gaiotto:2008cd,Gaiotto:2010be}.  In this way we make a connection between twistorial aspects of hyperK\"ahler geometry
associated with $4d$, ${\cal N}=2$ theories compactified on a circle, with open twistorial topological string.  More precisely,
as we will discuss later in the paper, this connection arises in the limit $\epsilon_1,\theta \rightarrow 0$.

\section{Closed twistorial topological strings}\label{s:twiststringgs}

In the last section we have defined the open twistorial topological string in terms of 
the $tt^*$ geometry associated to the corresponding physical D-brane.  
Now we would like to give a compatible twistorial extension of
the closed sector of the topological string.  
There are two ways this can be done in principle.  In \S\ref{s:largeNduality} below we use
the large $N$ duality of the topological string 
\cite{Gopakumar:1998ki,Dijkgraaf:2009pc,Aganagic:2011mi} to give one such definition. 
Then in \S\ref{s:target} we give another definition
purely from the target space point of view, and argue that the two approaches
are equivalent.
Finally in \S\ref{s:nw} we reformulate
this definition in terms of 
the results of \cite{Nekrasov:2009rc}, which provides a direct $4d$ $tt^*$ interpretation.

For concreteness, we continue to consider only the topological B model throughout this section,
though similar considerations apply to the A model.

\subsection{Large $N$ duality and closed twistorial topological strings}\label{s:largeNduality}

For a closed topological string setup which has an open topological string 
dual, we can simply define the closed twistorial topological string partition function
to be the same as the open twistorial topological string partition function:
$$\psi^{twist}(\vec t,\epsilon_1,{\tilde \epsilon_2},\theta, \zeta)=\psi_{open}^{twist}(\vec k,\epsilon_1,{\tilde \epsilon_2},\theta, \zeta).$$
Here the moduli on the two sides are related by
\begin{equation}\label{largeNrel}
\vec t = \epsilon_1 \vec k,
\end{equation}
where the components $t_i$ of $\vec t$ are the closed string moduli, and the components $k_i$ of $\vec k$ are the numbers of branes wrapped around cycles $Y_i$ 
on the open string side.
The closed 
twistorial topological string is {\it not} symmetric under the exchange $\epsilon_1\leftrightarrow {\tilde \epsilon_2}$,
unlike the usual closed topological string; this is 
why we have been using the notation $\tilde \epsilon_2$.

Note that according to this definition the closed twistorial topological string does not make sense
for arbitrary values of $t_i$, since the $k_i$ in \eqref{largeNrel} have to be integers
(although in some examples below we will see
that the partition function admits a natural continuation away from the integral locus.)  
However, in a perturbative expansion around $\epsilon_1 = 0$, we would not see this quantization; 
then we expect to get functions defined on the full Coulomb branch (arbitrary $\vec t\,$).  

Let us consider again the main example discussed in \S\ref{s:connectionopentopstrings}. The holographic dual 
of the open topological string theory we considered there is the 
closed topological string for the local Calabi-Yau hypersurface
$$y^2=W'(x)^2+P(x)+uv,$$
where $P(x)$ is a polynomial of degree $n-1$, 
whose coefficients are fixed in terms of $\vec k$ 
as in \cite{Dijkgraaf:2002fc}.
Thus, we may define the closed twistorial topological string 
for this local Calabi-Yau as
$$\psi^{twist}(\vec t = \vec k \eps_1, \epsilon_1,{\tilde \epsilon_2},\theta, \zeta)=\psi_{open}^{twist}(\vec k,\epsilon_1,{\tilde \epsilon_2},\theta, \zeta).$$

To recover the usual closed topological string from our twistorial extension, we repeat
what we did for the open sector: namely, we take $\zeta \rightarrow 0$, $\theta\rightarrow 0$, and $\epsilon_2\rightarrow \infty$ 
while holding ${\tilde \epsilon_2} \zeta=\epsilon_2$ finite.  As we argued in \S\ref{s:connectionopentopstrings}, in this limit
the open twistorial topological string reduces to the ordinary open topological string; via the 
topological large $N$ duality, 
we then recover the ordinary closed topological string partition function:
$$Z^{top}(\vec t,\epsilon_1,\epsilon_2)=\lim_{{\tilde \epsilon_2}\rightarrow \infty,\ (\zeta,\theta) \rightarrow 0, \atop\zeta {\tilde \epsilon_2}=\epsilon_2}
\psi^{twist}(\vec t,\epsilon_1,{\tilde \epsilon_2},\theta, \zeta).$$
Note that in this limit we still have the quantization constraint that each $t_i$ is an integer multiple of $\eps_1$, which
suffices to define $Z^{top}$ in a perturbative expansion in $\epsilon_1$.

\subsection{Target space interpretation of closed twistorial topological string} \label{s:target}

In this section we give an alternative definition of the 
closed twistorial topological string, which is more general
and does not use a large $N$ duality.  Its only disadvantage is
that it is not easy to compute it when an explicit dual open string description is lacking.

Suppose given an $\N=2$ theory in 4 dimensions.  
We may place this theory in the ${1\over 2} \Omega$
background with parameter $\epsilon_1$, corresponding to a rotation in the $x^3-x^4$ plane.
We thus get a theory in 2 dimensions with ${\cal N}=(2,2)$ supersymmetry.  
Now we consider $tt^*$ geometry for this 2d theory. This involves studying the Hilbert
space on a circle $S^1_R$, and we take the circle radius
to be $$R = 1 / \tilde \epsilon_2.$$
Moreover, in this theory there is a global symmetry corresponding to the rotation in the $x^3-x^4$ plane,
and we can also turn on a Wilson line around $S^1_R$ for this symmetry, with holonomy $e^{\I \theta}$
(thus as we go around $S^1_R$ we are rotating the $x^3-x^4$ plane by an angle $\theta$ while compensating
it with the a $U(1)\subset SU(2)$ R-symmetry to maintain supersymmetry).

It is known from \cite{Nekrasov:2009rc} that in this 2-dimensional theory 
we have a discrete set of vacua labeled by integer
vectors $\vec k = (k_1, \dots, k_r)$, with $r$ the complex 
dimension of the Coulomb branch of the original $\N=2$
theory.
Thus, there is a natural basis of supersymmetric D-branes
for this 2-dimensional theory, labeled
by the discrete parameter $\vec{k}$ together with a continuous parameter $\zeta$ determining
which half of \textsc{susy} the D-brane preserves.
In parallel to what we did above, we will consider the
overlap between such a D-brane state and the topological 
vacuum; we propose to define the closed 
twistorial topological string partition function $\psi^{twist}$
to be this overlap.
To do so, we need to relate the parameter $\vec{k}$
to the Coulomb branch 
moduli $\vec t$ on which $\psi^{twist}$ is supposed 
to depend. 
 When our $\N=2$ theory is a Lagrangian gauge 
theory, and when we work in the classical approximation, 
the results of \cite{Nekrasov:2009rc} would lead 
to the identification $\vec t = \vec k \eps_1$.  More generally
this $\vec t$ should be viewed as $\eps_1$ deformation of the Coulomb branch parameter, in terms of which the
Nekrasov partition function is expressed.
Thus we define
$$\psi^{twist}(\vec t = \vec k \eps_1, \epsilon_1,{\tilde \epsilon_2}, \theta, \zeta) = \langle 0| \vec k, \theta, \zeta \rangle_{\epsilon_1,\; R={1/ {\tilde \epsilon_2}}}.$$

One tricky point requires discussion.  Our description of the vacua is not symmetric under electric-magnetic duality: 
in writing $\vec t = \vec k \eps_1$, we are saying the vacua correspond to points where the Coulomb branch scalars $t_i$ \ti{in a particular electric-magnetic 
duality frame} are quantized. In fact, in our description we used a basis which is electric-magnetic dual to that used 
in \cite{Nekrasov:2009rc}.
In the 2-dimensional theory, this asymmetry 
between electric and magnetic can be understood as coming from the boundary conditions imposed at spatial infinity; alternatively this boundary condition can also be understood as coming from a boundary condition at infinity in the original 4-dimensional theory;
this mechanism was discussed in \cite{Nekrasov:2010ka}.

\subsection{Equivalence via physical large $N$ duality} \label{compare-largeN}

We have now offered two definitions of the closed twistorial topological string. 
Each involves $tt^*$ geometry for some 2-dimensional field theory; in one
case the theory was living on the noncompact part of the worldvolume of a 
D3-brane of the Type IIB superstring in $\half \Omega$-background; in the other case
the theory came from taking a 4-dimensional $\N=2$ theory and turning on a $\half \Omega$-background.
We will now argue that, in cases where
they are both applicable, the two definitions are actually equivalent, because the two 
2-dimensional theories in question are equivalent.

The basic idea of the equivalence is a physical version of the open/closed duality 
of topological strings.  Such ideas have been used before.
In particular, in \cite{VN} the topological open/closed 
duality was embedded in the Type IIA superstring, which was later applied to derive
nonperturbative results for ${\cal N}=1$ supersymmetric theories in four dimensions 
\cite{CIV,DVn}.  
The Type IIB version of this involves D5-branes wrapping $\C\mathbb{P}^1$ in $CY$,
and filling the 4-dimensional
spacetime, so that there is 4-dimensional Poincar\'e invariance
on both sides of the duality.  
In this paper, on the other hand, we have been discussing D3-branes,
which fill only 2 of the 4 noncompact dimensions.
However, once we have turned on the $\half \Omega$-background we have only 2-dimensional 
Poincar\'e invariance, whether or not we have the D3-branes present; thus at least on symmetry
grounds there is no reason why there cannot be an open/closed duality in this setting as well.
Moreover, such a duality has been considered before in the A model, in that case 
involving 5-dimensional and 3-dimensional theories \cite{aga,aga1};
see also \cite{Taki:2010bj}.

The main new point we make here is that the vacuum structures in the two theories match.
Indeed, in the open/closed duality story, we meet the quantization law $\vec{t} = \vec{k} \eps_1$, where $\vec{k}$ keeps track of the number
of D-branes. On the other hand, when we put an $\N=2$ theory in $\half \Omega$-background,
as we have reviewed above, the vacua are also labeled 
by integers $\vec{k}$.
This is a good consistency check,
and part of the motivation for writing $\vec t = \vec k \eps_1$
in that context as well.

Strictly speaking, there is a slight mismatch here:  
to obtain all the vacua of the $\N=2$ theory in $\half \Omega$-background, 
we should allow the components of $\vec{k}$ to be arbitrary integers;
on the other hand the numbers of D3-branes would naively be restricted to be 
\ti{positive} integers.  It would be interesting to clarify this point.
A possible resolution could involve 
replacing the matrix model with gauge group $U(N)$ by a supermatrix model 
with gauge group $U(N+M|M)$, for 
$M\rightarrow \infty$.
As discussed in \cite{Dijkgraaf:2003xk}, this would allow eigenvalues to occur with 
negative multiplicity.

\subsection{Extra phases and hyperK\"ahler geometry} \label{s:m-review}

So far we have defined $\psi^{twist}$ to be a function
of Coulomb branch moduli $\vec t$ (which are quantized for finite $\eps_1$, but become continuous in the limit 
$\eps_1 \to 0$).
In examples which we consider below, we will find evidence that $\psi^{twist}$ as $\epsilon_1\rightarrow 0$ should in fact be viewed not as 
a function on the Coulomb branch, but rather as the 
restriction of something involving a larger space $\cM$.
The relevant $\cM$ is the ``Seiberg-Witten integrable 
system'' which has been considered before in many works, 
e.g.\!\! \cite{Martinec:1995by,Donagi:1997sr,Donagi:1996cf}, and 
played a key role in the study of BPS states 
of $\N=2$ theories in \cite{Gaiotto:2008cd}. 
Let us quickly recall the basics.

$\cM$ is the total space of a complex integrable system 
arising as a torus fibration over the Coulomb branch.
To describe the torus fibers concretely, choose 
an electric-magnetic splitting; then the fibers
can be coordinatized by angles $\theta_{e^i}$, $\theta_{m^i}$,
where $e$, $m$ refer to ``electric'' and ``magnetic'',
and $i = 1, \dots, r$.
From the point of view of the $\N=2$ theory of the last
section, $\cM$ arises as the moduli space of the theory we obtain by compactification to 3 dimensions on $S^1$.  Here the electric coordinates 
$\theta_{e^i}$ are the holonomies of the abelian
gauge fields around $S^1$, while $\theta_{m^i}$ arise
from dualization of the 3-dimensional gauge field.
This compactified theory has $\N=4$ supersymmetry in 3 dimensions, from which it follows that $\cM$ carries
a natural hyperK\"ahler metric, as discussed in \cite{Seiberg:1996nz}.

Concretely, what we are proposing is that 
$\psi^{twist}$ is best viewed as depending on the angular parameters
$\vec \theta_e$ as well as $\vec t$.
Of course, in our discussion so far we have not seen these extra 
parameters appear explicitly.  We will argue in Section 
\ref{ss:thetaquant} below that in fact they are fixed to 
\begin{equation} \label{thetae-fixed}
\vec \theta_e = \vec k \theta
\end{equation}
which explains why we have not seen them so far.

Nevertheless it is sometimes useful to keep these extra 
parameters in mind, as we will see below.  In fact, as $\epsilon_1\rightarrow 0$
one can see due to periodicity of $\theta^i$, that $t^i$ and $\theta^i$ become effectively independent
variables.
In particular, in the $C$-limit 
which we define in \S\ref{s:c-limit-def} below, 
we will identify the twistorial topological 
string partition function (rescaled) 
with the quantity $\Psi$ considered in \cite{Neitzke:2011za}, 
which did depend on the extra parameters 
$\vec\theta_e$; to make this comparison, 
we will need to use \eqref{thetae-fixed}.

\subsection{Relation to Nekrasov-Witten} \label{s:nw}

In the target space description of the closed twistorial topological string, given in \S\ref{s:target}, 
we considered an $\N=2$ theory placed in 
$\half \Omega$-background, with parameter $\eps_1$ corresponding
to rotation in the $x^3-x^4$ plane. There is an alternative perspective on the $\half \Omega$-background, due to Nekrasov-Witten \cite{Nekrasov:2010ka}, which gives some further insight into this setup.

Nekrasov-Witten argue that the theory
with $\half \Omega$-background is equivalent 
to a theory without $\Omega$-background.  In the new theory the metric in the $x^3-x^4$ directions is modified
to a cigar, whose asymptotic radius is $1 / \eps_1$.
More precisely, this equivalence is supposed to hold everywhere 
except for the tip of the cigar; at the tip,
the equivalence breaks down, so from the point of
view of the new theory there is some nontrivial insertion there.
Now suppose that we compactify this new theory 
from 4 to 3 dimensions along the circle direction of this
cigar. Away from the tip, then, we just have the compactification of the original $\N=2$ theory to 
three dimensions on $S^1$. In the IR, this compactification
gives rise to a 3d sigma model into the hyperK\"ahler space $\cM$ reviewed in \S\ref{s:m-review}.

The cigar becomes a half-line in the 
spacetime of the compactified theory, 
parameterized by $r = \sqrt{(x^3)^2 + (x^4)^2}$.
To describe the situation more completely, we should now consider the boundary conditions on this half-line.
At $r = 0$, the compactification produces
a boundary condition of the 3d sigma model,
corresponding to the local physics at the 
tip of the cigar. Nekrasov-Witten argue that this boundary condition restricts the sigma model field to a certain subspace
$\cO \subset \cM$.  $\cO$ is a complex submanifold with respect to one of the complex structures of $\cM$, and also Lagrangian 
with respect to the corresponding holomorphic symplectic form.\footnote{ The complex structure in question on $\cM$ lies on the equator of the twistor sphere of $\cM$; precisely \ti{which} point of the equator we get is determined by the phase of the $\half \Omega$-deformation parameter $\eps_1$. Here we have taken $\eps_1$ to be real; having made this choice, we get a definite point of the equator, sometimes referred to as ``complex structure $K$''. The fact that $\cO$ is complex Lagrangian in this structure can then be summarized by 
saying that $\cO$ is an $(A,A,B)$ brane on $\cM$.}
We do not know an explicit description of $\cO$ in general;
however, if
our $\N=2$ theory happens to be a theory of class $\mathcal{S}$,
then $\cM$ is a moduli space of solutions of Hitchin equations, and Nekrasov-Witten propose that in this case the $\cO$ is the space of \ti{opers}.
At $r = \infty$
(or $r = L$ after regulating) we should also fix a boundary
condition. This boundary condition is not dictated by the local
physics of the original 4d $\N=2$ theory; rather it corresponds 
to a choice of boundary condition in 4d, the same choice 
which we discussed at the end of \S\ref{s:target}, which picks
out a particular electric/magnetic duality frame.
From the 2d point of view it corresponds to another 
brane $\cO'$ in the hyperK\"ahler space $\cM$.

If we now compactify on the $r$ interval to the $x^1 - x^2$ plane,
the vacua of the resulting two-dimensional theory come from 
configurations where the sigma model field is constant 
on the interval.  Such configurations correspond to 
points of intersection between $\cO$ and $\cO'$.
On the other hand, the theory we obtain by this compactification
is just the original $\N=2$ theory in $\half \Omega$-background;
thus the vacua of this theory correspond to these intersection points.
This is one of the key observations of \cite{Nekrasov:2010ka}.

\begin{figure}[t!]
  \begin{center}
    \includegraphics[width=10cm]{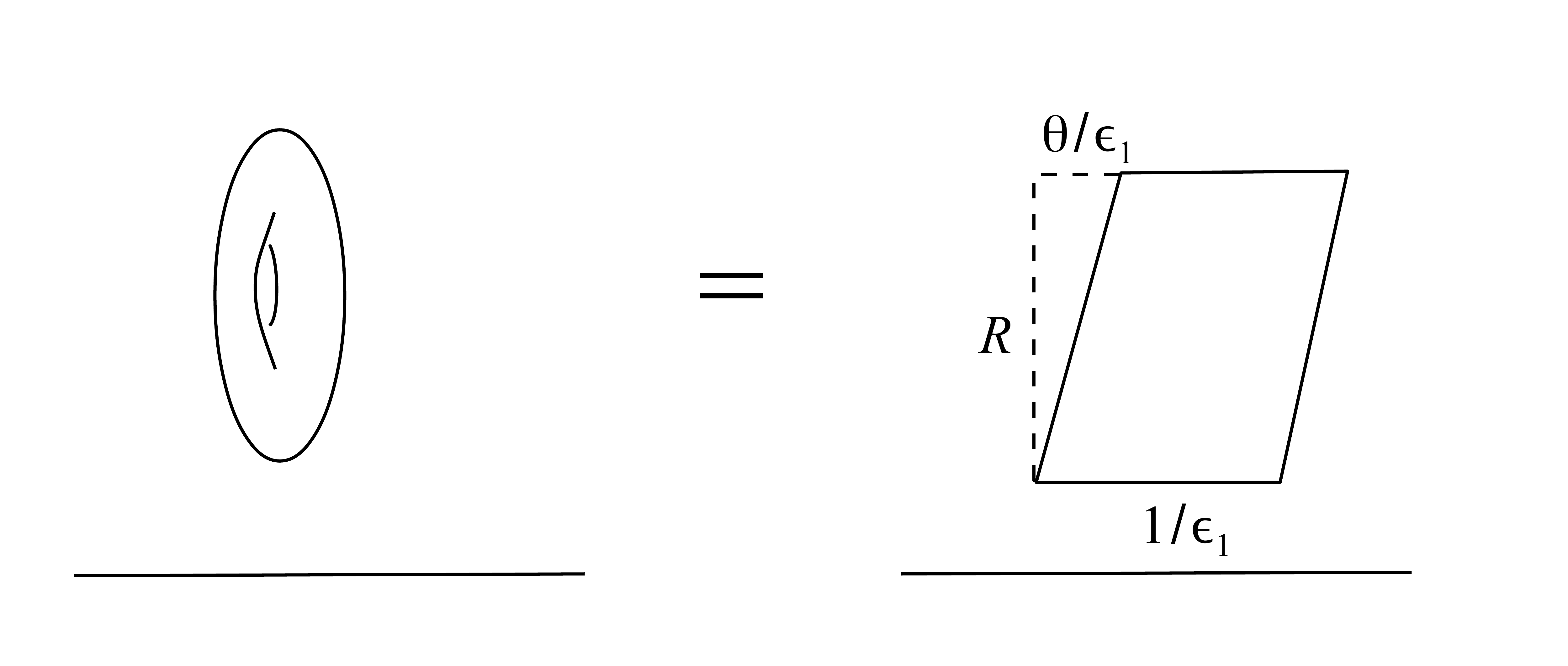}
\caption{This figure shows the geometry of the 3d space in the Nekrasov-Witten picture.
The geometry is that of $T^2$ fibered over a line ${\bf R}$ where on one end of the line we have the
${1\over 2} \Omega$ background and on the other the electric (or magnetic) D-brane boundary conditions.}\label{torus}
  \end{center}
\end{figure}

To apply this point of view to the twistorial topological string,
we do $tt^*$ geometry for the two-dimensional theory we obtain by 
this reduction, as we did in the previous sections; 
in other words, we put the two-dimensional theory on a spatial circle,
of radius $R = 1 / \tilde \epsilon_2$ (say in the $x^2$ direction), and turn on a Wilson line $e^{\I \theta}$ 
around that circle for the $U(1)$ symmetry coming from rotations in the $x^3 - x^4$ plane (combined
with R-symmetry to maintain supersymmetry).
From the viewpoint of the original 4d $\N=2$ theory, 
we are considering a geometry which in the bulk looks like a 2-torus fibration 
over the $r-x^1$ plane --- indeed we compactified on two circles, 
with radii $1/\epsilon_1$ and $1/{\tilde \epsilon_2}$.
If $\theta \neq 0$ then this torus is not rectangular;
its complex structure parameter is given by
\begin{equation}\label{whattau}
\tau = \I{\epsilon_1\over {\tilde \epsilon_2}}+\frac{\theta}{2\pi}.
\end{equation}
See Fig.\ref{torus}. 

We have boundary conditions at $r=0$ and $r=L$ as described above.
Thus the $tt^*$ geometry we consider, from this point of view,
is describing the ground state geometry of the 4d $\N=2$ theory compactified
on the torus with these particular boundary conditions, in the limit 
$L \to \infty$.  (Keeping $L$ finite would give a natural extension of the twistorial topological string,
which we will not consider in this paper, but would be interesting to study.)

Finally, our definition of $\psi^{twist}$ is as in the previous sections.
We consider the massive vacua, corresponding 
to the intersection points $\vec k$ between $\cO$ and $\cO'$ above.
Each such vacuum corresponds to a D-brane $D_{\vec k}(\zeta)$ 
of the two-dimensional theory in the $x^1-x^2$ directions,
and we compute the overlap
$$\psi^{twist} = \langle 0 \vert D_{\vec k}(\zeta) \rangle.$$
Let us try to interpret this from our present point of view.
The state $\langle 0 \vert$ corresponds to the topological 
path integral on a cigar in the $x^1 - x^2$ directions.
In terms of the torus compactification to the $r - x^1$ plane, 
this means we are inserting a boundary condition at 
$x^1 = 0$, corresponding to the shrinking of the $x^2$ circle.
We also have a second boundary condition at $x^1 \to \infty$, 
coming from the $D$-brane $D_{\vec k}(\zeta)$,
and the boundary conditions at $r = 0$ and $r = L$ as described above.
Thus in the end $\psi^{twist}$ is given by a path integral in the 2-dimensional 
theory over a strip with these 4 boundary conditions, in the limit $L \to \infty$.

\subsection{Quantization of the hyperK\"ahler integrable system}\label{ss:thetaquant}

As we have been discusing, the twistorial topological string can be viewed as 
the $tt^*$ geometry associated with the ${1\over 2} \Omega $ background, with parameter $\epsilon_1$.  
As such, the Coulomb branch parameters are discretized.  Here we argue that
this discretization actually extends to a discretization of the full hyperK\"ahler space
$\cM$. 

We take
the radius of the 2d circle where we are considering the $tt^*$ geometry to be $R$ and consider in addition
a twist in the 3-4 plane by $\theta$ as we go around the circle. We take the time direction to correspond to the 2d geometry
along the cylinder and we base our Hilbert space on the 2d circle.
 For a moment let us consider the limit where $\epsilon_1=0$.
In this case we simply have the compactification of the 4d ${\cal N}=2$ theory on a circle of radius $R$, where as we go
around the circle we rotate the other two spatial coordinates by $\theta$.  On this geometry one can consider line defects $X_\gamma(\zeta)$,
studied in \cite{Gaiotto:2010be}, which correspond to supersymmetric Wilson-'t Hooft lines in the IR, 
where $\gamma$ denotes an element of the charge lattice and $\zeta$ 
(taken to be a phase) controls which half of the supersymmetry the line operator preserves.

As was argued in \cite{Gaiotto:2010be,Cecotti:2009uf,Cecotti:2010fi}, in this background the line defects 
become non-commutative:
$$X_\gamma(\zeta)X_{\gamma'}(\zeta)=q^{\langle \gamma, \gamma'\rangle} X_{\gamma'}(\zeta)X_{\gamma}(\zeta)$$
where
$$q={\rm exp}(i\theta),$$
and $\langle \gamma, \gamma'\rangle$ denotes the symplectic inner product on the charge lattice.

Now consider turning on the ${1\over 2}\Omega$ background, letting $\epsilon_1 \neq 0$.  In
the later sections of this paper, we will find evidence
through examples that in this context the above commutation relations should 
be modified to
\begin{equation}\label{quanti}
X_\gamma(\zeta)X_{\gamma'}(\zeta)=q(\zeta)^{\langle \gamma, \gamma'\rangle} X_{\gamma'}(\zeta)X_{\gamma}(\zeta)
\end{equation}
where
\begin{equation}\label{qdef}
q(\zeta)={\rm exp}\!\!\left[-{2\pi R\epsilon_1\over \zeta} +i\theta-2\pi R{\overline \epsilon_1} \zeta\right]={\rm exp}\!\!\left[-{2\pi\epsilon_1\over {\tilde \epsilon_2}\zeta} +i\theta-{2\pi{\overline \epsilon_1} \zeta\over {\tilde \epsilon_2}}\right]
\end{equation}
In the ``semi-flat'' limit $R \gg 1$, this commutation relation can be interpreted as follows.  In this limit
we have 
$$X_\gamma(\zeta)= \exp\!\!\left[ -{2\pi Ra_\gamma \over \zeta}+i \theta_\gamma + 2\pi R {{\overline a}_\gamma \zeta}\right]$$
and so the relation \eqref{quanti} would follow from
\begin{equation}\label{commrel1}
\begin{aligned}
[a_\gamma,a_{\gamma'}]&=-\zeta \langle \gamma, \gamma' \rangle {\epsilon_1\over 2\pi R} =-\frac{1}{2\pi}\,\zeta \langle \gamma,\gamma'\rangle\; \epsilon_1 {\tilde \epsilon_2},\\ [{\overline a}_\gamma,{\overline a}_{\gamma'}]&=-{1\over \zeta} \langle \gamma, \gamma' \rangle {{\overline \epsilon_1}\over 2\pi R} =-\frac{1}{2\pi}{1\over \zeta} \langle \gamma,\gamma'\rangle\; {\overline \epsilon_1} {\tilde \epsilon_2},
\end{aligned}
\end{equation}
\begin{equation}\label{commrel2}
[\theta_\gamma,\theta_{\gamma '}]=-i\,\langle \gamma, \gamma'\rangle\, \theta.
\end{equation}

Moreover, in this limit we will 
find that the twistorial topological string partition function 
behaves like a wave function. Namely, relative to the electric/magnetic splitting picked out by our boundary condition, we have (see eqn.\eqref{Ztwist-NStype} below)
$$\psi^{twist} \sim {\rm exp}[ F(a_{e_i}) /\zeta^2 \epsilon_1{\tilde \epsilon_2}+{\overline F}({\overline a_{e_i}}) \zeta^2 /{\overline \epsilon_1}{\tilde \epsilon_2}],
$$
where we write $\{e_i\}$ for a basis of the electric charges, 
and $F(a_{e_i})$ is the prepotential of our ${\cal N}=2$ theory.
As we change the electric/magnetic basis, this formula changes, because the prepotential 
$F$ changes by a Legendre 
transform. On the other hand the above commutation relations for the $a_\gamma$ 
would imply that a wave function of the $a_{e_i}$ would transform by Fourier transform.
At least to leading order in the quantization parameter, this matches.
This is compatible with the interpretation of the holomorphic
anomaly \cite{Bershadsky:1993cx} as making the topological string 
an element of a Hilbert space acted on by operators with the 
above commutation relation \cite{Witten:1993ed}.

As shown in \cite{Gaiotto:2008cd}, the line operators $X_\gamma$ can be viewed as providing local Darboux coordinates for the holomorphic symplectic structures on the 
space $\cM$ which we reviewed in \S\ref{s:m-review}.  
In this context, what \eqref{quanti} means is that $\cM$ is being quantized in 
the twistorial topological string.  Roughly speaking,
$\epsilon_1$ is a quantization parameter for the base of the hyperK\"ahler geometry (Coulomb branch) and $\theta$ is a quantization parameter for the torus fiber.

As we have already noted, in the $\half \Omega$-background
we get a discretization of the Coulomb branch: $a_{e_i}=k_i \epsilon_1$.  We now argue
that in the $\half \Omega$-background we also naturally get a discretization of the 
$\theta_{e_i}$, so that we have
a `twistorial triplet' discretization\footnote{ More precisely, the discretization 
is shifted by a half-integer
as is seen in the context of large $N$ dualities.  This is also related to the ``quadratic refinement'' of 
\cite{Gaiotto:2008cd}.}:
$$\big(a_{e_i}, \theta_{e_i}, \bar a_{e_i}\big)=k_i\big(\epsilon_1,\theta, \bar\epsilon_1\big).$$
To argue for the discretization 
of the electric angles $\theta_{e_i}$, we first recall how in the  $R\rightarrow \infty$ limit
the arguments of Nekrasov-Shatashvili give the discretization of $a_{e_i}$: we have an effective superpotential 
which in terms of magnetic variables looks like
$$W={1\over \epsilon_1} W(\Sigma_{m_i},\epsilon_1)-k_i \Sigma_{m_i},$$
so that at the critical points
$$a_{i_e}=\Sigma_{i_e}=\partial_{\Sigma_{m_i}} W(\Sigma_{m_i},\epsilon_1)=k_i \epsilon_1.$$
On the other hand, viewed as a superfield $\Sigma_{m_i}$ has a top component $F_{m_i}$ which is the magnetic flux
in the 2d plane, which integrated on the cigar geometry
of $tt^*$ leads to the magnetic holonomy $ \theta_{m_i}$ around $S^1$.  Therefore the above $W$ implies that the wave function $\psi^{twist}$
has a $\theta_{m_i}$ dependence given by
$${\rm exp}(i k_i \theta_{m_i})$$
Given the fact that $[\theta_{e_i},\theta_{m_i}]=-i\theta$, this wave function gives a quantization $\theta_{e_i}=k_i \theta$, as we wished to
show.

\subsection{Interesting limits}

\begin{figure}[t!]
  \begin{center}
    \includegraphics[width=10cm]{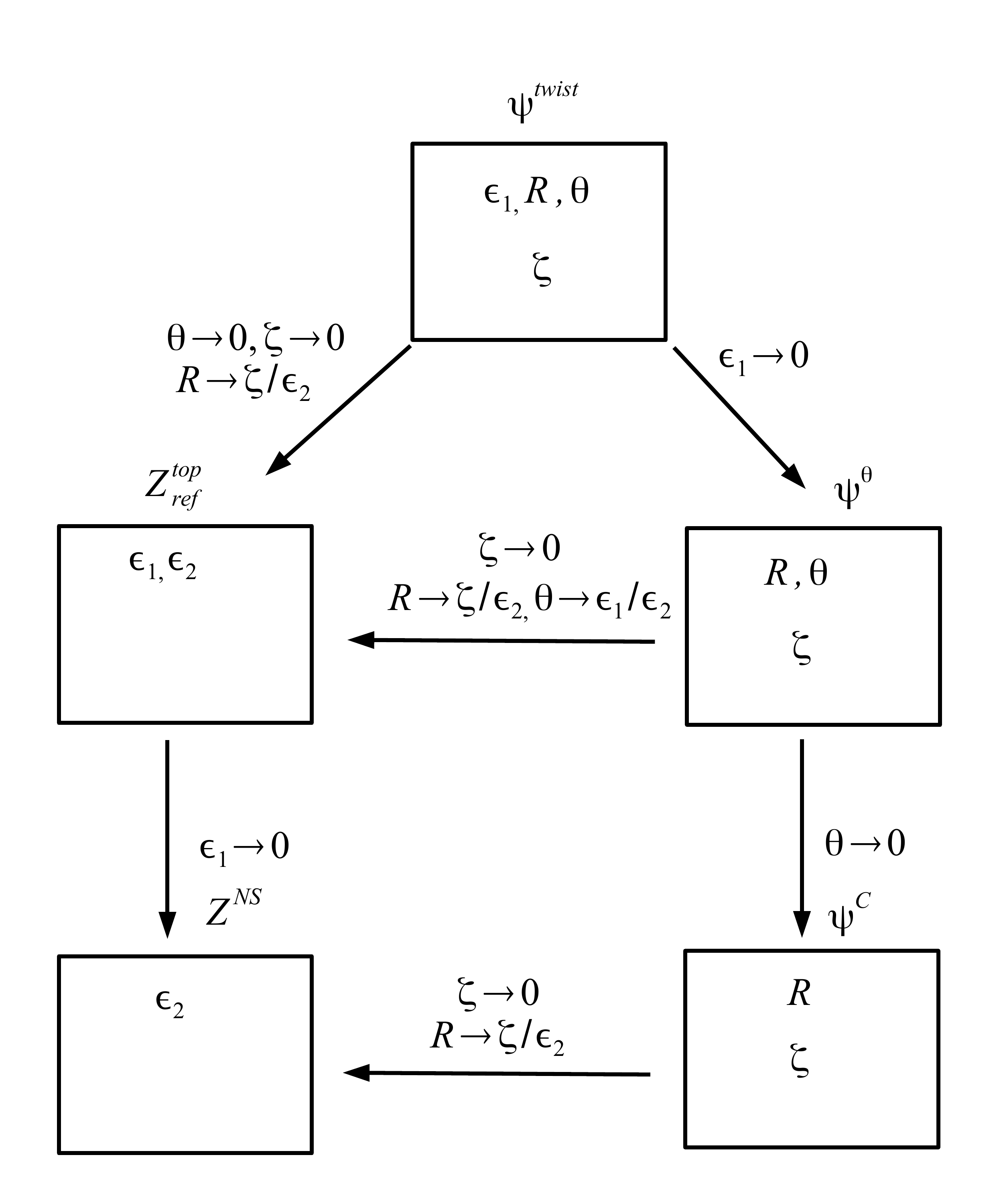}
\caption{The diagram of various limits.  The quantities in the boxes denote the parameters
the object depends on.  $\psi^{twist}$ is the twistorial partition function which depends
on all three coupling constants $(\eps_1,R,\theta)$, where $R={1\over \tilde \eps_2}$, and the twistor parameter $\zeta$.
The left column denotes the limits leading to refined and NS limits of topological strings.
The right column denotes the $\theta$-limit, and the $C$-limit.  Various reductions are
shown by arrows and the limits we need to take are indicated next to the arrows.}\label{brane-figure}
  \end{center}
\end{figure}

So far we have formulated what we mean generally by the closed twistorial topological string partition function: it is a D-brane amplitude in the $tt^*$ geometry associated to a 4-dimensional field theory in $\half \Omega$-background.
In general, though, this D-brane amplitude would be very complicated to compute.  In order
to get some handle on it, in this section we point out some simplifying 
limits that we shall consider later in this paper.  The twistorial 
topological amplitude depends on the parameters
$(\epsilon_1,{\tilde \epsilon}_2, \theta, \zeta)$.  The interesting limits to consider will involve
taking various of these parameters to $0$ or $\infty$, while holding other parameters
 or combinations of parameters fixed.
Fig. \ref{brane-figure} shows the various limit we take starting with the twistorial topological
string amplitude $\psi^{twist}$.  The limits on the left correspond to refined topological string partition function
and its NS limit and the right column corresponds to the various twistorial limits.  We discuss these limits next.
 
\subsubsection{The asymmetric limit} \label{s:lim-asymm}

The asymmetric limit is the limit in which we take $\zeta \rightarrow 0, {\tilde \epsilon_2}\rightarrow \infty$ keeping
$\epsilon_2 = {\tilde \epsilon_2} \zeta$ finite, and also set $\theta = 0$.
In this limit, as we have described above, we expect to obtain the closed refined topological
string amplitudes:
$$\lim_{\zeta \to 0} \psi^{twist}(\epsilon_1, {\tilde \epsilon_2}=\epsilon_2 / \zeta, \theta =0, \zeta) = Z^{top}(\epsilon_1,\epsilon_2).$$
Here, more precisely, to make the right side well defined 
we should specify in which polarization we write $Z^{top}$: we mean the real polarization determined by 
an electric-magnetic splitting, the same splitting 
which we have discussed above.
Also, as already noted, $\psi^{twist}$ is strictly speaking defined on a discrete subset of the Coulomb branch, 
$\vec a = \eps_1 \vec k$ for an integer vector $\vec k$. Still, in the perturbative expansion 
in $\eps_1$ around $0$, the $a_i$ would appear continuous. This matches the usual situation for
$Z^{top}$.

This limit is a \ti{simplifying} limit for $\psi^{twist}$ in several senses.
First, since $Z^{top}$ is holomorphic in the $a_i$, in this limit we expect
$\psi^{twist}$ to become holomorphic.
We also expect the emergence of a $\mathbb{Z}_2$ symmetry $\epsilon_1\leftrightarrow \epsilon_2$, 
which is also not there in the full $\psi^{twist}$
(our definition makes clear that 
$\epsilon_1,{\tilde \epsilon_2}$ are not on the same footing.)

We could of course take a further limit, sending either 
$\epsilon_1\rightarrow 0$ or $\epsilon_2\rightarrow 0$,
leading to the NS limit of the topological string:
$$Z^{top}(\epsilon_1\rightarrow 0,\epsilon_2)\rightarrow {\rm exp}\!\big[ W^{NS}(a^i,\epsilon_2)/\epsilon_1\big]$$
$$Z^{top}(\epsilon_1,\epsilon_2\rightarrow 0)\rightarrow {\rm exp}\!\big[ W^{NS}(a^i,\epsilon_1)/\epsilon_2\big]$$
where $W^{NS}$ is the Nekrasov-Shatashvili superpotential for the theory in
${1\over 2}\Omega$-background.

\subsubsection{An NS-like limit}

A second limit we could consider is $\tilde \epsilon_2 \rightarrow 0$.
This corresponds to taking the radius of the circle on which we are compactifying our 2d theory 
to be $R \to \infty$.
In this limit the vacua of this theory become decoupled, so that the $tt^*$ geometry is 
dominated by classical contributions.  In this limit, we expect the
brane wave function attached to a given vacuum (twistorial topological string amplitude) 
to be dominated by the value of the Nekrasov-Shatashvili 
effective superpotential $W^{NS}$ at the corresponding critical point.
More precisely, we expect (in the unitary gauge) to get contributions both from $W^{NS}$ and $\overline{W}^{NS}$,
\begin{equation} \label{Ztwist-NStype}
\psi^{twist}(\epsilon_1,{\tilde \epsilon_2}, \theta, \zeta)  \xrightarrow{\tilde\epsilon_2\rightarrow 0} {\rm exp}\!\left[{W^{NS}(a,\epsilon_1)\over \zeta {\tilde \epsilon_2}}+\zeta {{\overline W}^{NS}({\overline a},{\overline \epsilon_1})\over {\tilde \epsilon_2}} \right]
\end{equation}

\subsubsection{The $\theta$-limit}\label{sss:thetalimt}

Next we consider the limit $\epsilon_1\rightarrow 0$, while keeping all the other parameters fixed;
we call this the $\theta$-limit.

This is a somewhat subtle limit:  it corresponds to making 
the coupling $q(\zeta)$, given in eqn.\eqref{qdef} above, 
effectively $\zeta$-independent except
in small neighborhoods of $\zeta = 0$ and $\zeta = \infty$.

As we will discuss in the explicit examples below, in this limit we 
find singular behavior of the form
\begin{equation}\label{Sergio}
\psi^{twist} \xrightarrow{\epsilon_1\rightarrow 0}\psi^\theta ({\tilde \epsilon_2}, \theta ,\zeta) = {\rm exp}\!\!\left[{W^{NS}(a,\theta\tilde\epsilon_2 \zeta)\over \zeta {\tilde \epsilon_2}}+\zeta {{\overline W}^{NS}({\overline a},\theta\tilde\epsilon_2 / \zeta)\over {\tilde \epsilon_2}} + \cdots \right]
\end{equation}
where $...$ represents terms which are nonsingular in the limit $\eps_1 \to 0$
and moreover vanish as $\tilde \eps_2 \to 0$.
It is crucial here that we have {\it not} taken the limit $\theta \rightarrow 0$.  

Note that in this limit we are turning off the ${1\over 2}\Omega$-background.  Thus our setup is approaching
the original 4d ${\cal N}=2$ theory on a circle of radius $R={1/ {\tilde \epsilon_2}}$, and the vacua
at $\vec a = \vec k \eps_1$ become continuous.
The values of $\vec a$ which we can reach still lie on a real subspace of the Coulomb branch, 
determined by the phase of $\eps_1$; we expect that $\psi^\theta$ admits a further analytic continuation from this real locus to the full space.
Moreover, 
recall that the components $\theta_{e_i}$ of $\vec \theta_e$ are all periodic variables, fixed in terms of 
$a_i$ by the relation $\vec \theta_{e} = \vec k \theta$, $\vec a = \vec k \eps_1$.  
Remarkably, in the limit $\eps_1 \to 0$, the locus of 
points $(\vec a, \vec \theta_e)$ which we can access fills up the whole parameter-space, 
i.e.\! $\vec \theta_e$ becomes continuous and arbitrary in this limit. (Said more precisely, 
for any desired target point $(\vec a, \vec \theta_e)$, there is a way of tuning the
vectors $\vec k$ as $\eps_1 \to 0$ in such a way that in the limit we hit the target point.)

Thus altogether we expect that the $\theta$-limit of the partition function will be a function of the form
$$\psi^{\theta}(\vec a, \vec \theta_e, R, \theta, \zeta)$$
where we have replaced $\tilde \epsilon_2$ by 
$R=1/{\tilde \epsilon_2}$, to emphasize its role as the radius of the circle on 
which we compactify the 4d theory.

In the next section, motivated by examples, we propose that $\psi^\theta$ should 
be considered as the solution to a ``quantum'' Riemann-Hilbert
problem, with the quantum parameter
$$q=\exp (i \theta).$$
Moreover, in the `classical limit' $\theta \rightarrow 0$, we argue that this quantum Riemann-Hilbert problem becomes equivalent to the Riemann-Hilbert problem studied in \cite{Gaiotto:2008cd} incorporating the Kontsevich-Soibelman
  wall crossing \cite{Kontsevich:2008fj} in finding the expectation values of line operators
of the 2d theory, wrapped on the circle.  The $\theta \not=0$ limit extends this to the refined
wall crossing \cite{Cecotti:2009uf,Dimofte:2009bv, 
Dimofte:2009tm,Cecotti:2010fi}.  In these cases the line operators become {\it actual operators} acting
on a Hilbert space satisfying commutation relations
$$X_{\gamma}X_{\gamma'}=q^{\langle \gamma,\gamma'\rangle}  X_{\gamma'}X_{\gamma}$$
where $\gamma,\gamma'$ are charges in the central charge lattice and $\langle , \rangle$ denotes the corresponding symplectic product.
In this context $\psi^\theta$ should be viewed as a wave function in this Hilbert space.
Moreover as we change the phase of $\zeta$ and cross the phases of BPS central charges, we have the action of 
quantum dilogarithm operators on $\psi^\theta$.  
In the same sense the line operators
get conjugated by quantum dilogarithm operators.  In this context the monodromy of the quantum dilogarithm operators
representing wall crossing that was studied in \cite{Cecotti:2010fi} represent the 2d $tt^*$ monodromy.

\subsubsection{The $\theta$-limit $\rightarrow C$-limit} \label{s:c-limit-def}

A further ``classical'' limit, which we call the $C$-limit, is 
obtained by starting with the $\theta$-limit and then 
taking $\theta \to 0$.

In this limit, we will see in explicit examples below that $\psi^{\theta}\sim {\rm exp} {[f/\theta]}$.  Thus we define the $C$-limit amplitude by
$$\psi^C (\vec a, \vec \theta, R, \zeta)=\lim_{\theta\rightarrow 0}\big[ \psi^\theta(\vec a, \vec \theta, \theta, R, \zeta)\big]^{\theta/2\pi}.$$
As we will see in \S\ref{s:c-limit} below,
$\psi^C$ will turn out to be identified with a
key geometric quantity which entered into the work \cite{Neitzke:2011za}.  The aim of that work was to construct a certain hyperholomorphic connection over
the hyperK\"ahler moduli space $\cM$ reviewed in \S\ref{s:m-review}. 
Thus, in the $C$-limit we are recovering 
information about the classical hyperK\"ahler geometry of $\cM$.

\subsubsection{The $\theta$-limit $\rightarrow Top$-limit} \label{s:thetatotop}

In \S\ref{s:lim-asymm} we have described one way of recovering the usual refined topological string partition function from $\psi^{twist}$, by setting $\theta = 0$ and taking the asymmetric limit $\zeta \to 0$ with $R / \zeta = 1/\eps_2$ fixed.

Here is another way.  We can begin with the $\theta$-limit $\eps_1 = 0$, then take the asymmetric limit $\zeta \rightarrow 0$ and $R\rightarrow 0$ keeping $R/\zeta =1/\epsilon_2$ finite, and then \ti{rename} $\theta \rightarrow \eps_1 / \eps_2$, thus reintroducing the parameter $\eps_1$.
We also set the angles $\vec \theta_e=0$.

In this limit we expect to get back the topological string again (because the dependence
on topological string coupling constants is in the form $q(\zeta)$):
$$\psi^\theta \rightarrow Z^{top}\!\left(a^i,\epsilon_1=\theta \eps_2, {1\over \eps_2}={R\over \zeta}\right).$$

\subsubsection{The ($C$-limit or $Z^{top}$) $\rightarrow$ $NS$- limit} \label{s:corztons}

Finally, we can go to the NS limit of the refined topological string in either of two ways.  
Either we can start from the full twistorial topological string, take its $C$-limit, and then take the asymmetric limit
(where we take $\zeta \to 0, R \to 0$ holding $R / \zeta = 1/\eps_2$ fixed), or we can simply 
begin with $Z^{top}$, then take the $\eps_1 \to 0$ limit of $(Z^{top})^{\epsilon_1}$.
In either way, we recover the usual NS limit of the refined topological string.

\section{The $\theta$--limit and the quantum Riemann--Hilbert problem}\label{ss:thetalim}

As we have noted, the twistorial topological string gets simplified in the $\theta$-limit where $\epsilon_1\rightarrow 0$.  In addition,
starting from this limit we can get, by further reductions, the classical wall-crossing as well as
aspects of the hyperK\"ahler geometry studied in \cite{Gaiotto:2008cd}.
Moreover in a different limit we can obtain the refined topological string amplitudes.  So the $\theta$-limit is quite rich.
In this section we propose a computational scheme for the 
$\theta$--limit of the twistorial topological string 
based on  a plausible physical picture of the twistorial brane amplitudes using BPS structure of the $4d$, $ {\cal N}=2$ theory.  Our conclusions will be checked in the next two sections by comparing with the exact twistorial amplitudes of the Abelian geometries in the same limit.
Further evidence is provided by the fact that the proposed formalism reproduces the TBA equations of \cite{Gaiotto:2008cd} in the appropriate limit.

\subsection{General picture from dual matrix LG models}\label{ss:genpicctu}

To orient our ideas, we start with some heuristic considerations on the twistorial topological string as defined by the large--$N$ limit of the $tt^*$ brane amplitudes of the matrix LG models in eqn.\eqref{Eefft}. (These models will be analyzed more precisely in section \ref{s:lgmatrixmodels}). For definiteness we focus on the cubic LG model
\begin{equation}\label{oritmod}
\mathcal{W}=\sum_{i=1}^N (X_i^3/3-tX_i)+\epsilon_1\!\!\!\!\!\sum_{1\leq i<j\leq N}\!\! \log(X_i-X_j)^2,
\end{equation}
 where we identify field configurations up to permutations of the $X_i$. At $\epsilon_1= 0$ the model reduces to the
symmetric tensor product of $N$ copies of the 
mass--perturbed $A_2$ minimal model, each copy having two susy vacua at $X_i=\pm\sqrt{t}$. The point vacua of \eqref{oritmod} at $\epsilon_1=0$ then are labeled by $(N_+,N_-)$ and the corresponding D-branes by $|N_+,N_-\rangle$ where
$N_+$ (resp.\! $N_-\equiv N-N_+$) is the number of eigenvalues $X_i\big|_\text{vac.}$ equal to
$+\sqrt{t}$ (resp.\! $-\sqrt{t}$). In total we have $N+1$ vacua. The one--field LG model $W(X)=X^3/3-tX$
has a $tt^*$ Lax connection $\mathcal{A}^\zeta$ which takes values in $SL(2,\mathbb{C})$ and the brane amplitude
is a flat section of the vector bundle corresponding to
the fundamental representation $\mathbf{2}$. At $\epsilon_1=0$ the brane amplitudes of  the matrix LG model \eqref{oritmod} are flat sections of the
$SL(2,\mathbb{C})$ connection $\mathcal{A}^\zeta$ in the spin $N/2$ representation. The Stokes matrices of the associated Riemann--Hilbert problem are
elements of  $SL(2,\mathbb{C})$, and (at 
$\epsilon_1=0$) the two Stokes matrices $S_\pm$ of the matrix model \eqref{oritmod}, acting on the
$N+1$ dimensional space of D-branes, are just these group elements written as matrices in the $\boldsymbol{N+1}$ representation, i.e.\!
\begin{equation}\label{rrrrqqq}
S_\pm=\exp(J_\pm )\Big|_{\boldsymbol{N+1}\ \text{irrepr.}}.
\end{equation}

When $\epsilon_1\neq 0$ the vacuum structure becomes subtler: although the closed holomorphic one--form $d\mathcal{W}$ is still well defined, it is no longer exact due to the non--trivial fundamental group $\pi_1$ of the field space $(\mathbb{C}^N\setminus\text{diagonal})/\mathfrak{S}_N$, isomorphic to the braid group $\mathcal{B}_N$. As a result each of the $N+1$ vacua gets promoted to a $\theta$--vacuum  and in particular the vacuum amplitudes
for the D-branes gets realized as $\langle \theta |N_+,N_-\rangle$. $q=e^{i\theta}$ acts as a character of $\mathcal{B}_N$: 
switching on a non--zero $\theta$ is equivalent to the insertion in the amplitude of the corresponding chiral primary
 \begin{equation}\label{ohhh34}
 \mathcal{O}(\theta)=C(\theta)\, \prod_{i<j}(X_i-X_j)^{\theta/\pi},
 \end{equation}
 where $C(\theta)$ is some normalization factor.
From eqn.\eqref{ohhh34} it is obvious that braiding the two consecutive fields $X_i$ and $X_{i+1}$ introduces an extra factor of $q$, and that the power of $q$ counts the number of such elementary braiding operations.

 The three parameters $(\epsilon_1,\theta,\bar\epsilon_1)$
 become three coordinates in the periodic $tt^*$ geometry of the LG model (see \cite{Cecotti:2013mba} and next section) which get unified in a single twistorial object
 \begin{equation}
 q(\zeta)=\exp\!\Big(\!-2\pi \epsilon_1/({\tilde \epsilon_2}\zeta)+i\theta-2\pi \bar\epsilon_1\,\zeta /{\tilde \epsilon_2}\Big).
 \end{equation}
 
 The $\theta$--limit is taking $\epsilon_1\to0$ while keeping $\theta$ fixed, so that
 $q(\zeta)$ becomes a constant independent of $\zeta$, while the amplitudes are still `quantum' in the sense that $q\neq 1$.
 One also sends $N$ to infinity, keeping the Coulomb branch parameters $a_\pm\equiv N_\pm\epsilon_1$ finite. 
 In a sense we are making the Coulomb branch parameters commutative in this limit (i.e.\! classical) but keeping the
 fiber parameters non-commutative (quantum).
In this qualitative discussion, we take $N$ large but finite, and $|\epsilon_1|\lll |\sqrt{t}|$, while
 $\theta$ and $N\epsilon_1$ are taken to be of order one. $\sqrt{t}$ is assumed to be somehow larger than $N\epsilon_1$.
 In this regime the low energy configurations consist of $N_+$ fields fluctuating around the $+\sqrt{t}$ classical vacuum and
 $N_-=N-N_+$ fields fluctuating around the $-\sqrt{t}$ one.
 As $\epsilon_1\to 0$, the only communication between
 these two sectors is through the BPS solitons connecting the separated classical vacua; these
 solitons have masses $\geq 4|\sqrt{t}|$ and their effects are exponentially suppressed for large $\sqrt{t}$. For $\epsilon_1\lll 1$ these solitons are
small deformations of the BPS solitons of the one--field model $W(X)=X^3/3-tX$. Note, in particular, that processes in which several eigenvalues $X_i$ change sign, $\pm\sqrt{t}\longleftrightarrow \mp\sqrt{t}$, are  suppressed by large powers of the exponentially small number $e^{-4|\sqrt{t}|/\tilde\epsilon_2}$; hence these solitonic transitions may change the large integers $N_\pm$ only
by $O(1)$ corrections, that is, they may change the Coulomb branch parameters $a_\pm$ only at the $O(\epsilon_1)$ level. In the $\theta$--limit, $\epsilon_1\to 0$, the values of $a_\pm$ get completely frozen. This is why the $\theta$--limit was introduced in the first place:
one wants to simplify the problem by reducing to a \emph{classical} (i.e.\! non--fluctuating) Coulomb branch, while keeping the angles to be quantum (the angles are the coordinates on the fiber of the hyperK\"ahler geometry, which is endowed with a natural symplectic structure, and hence a canonical quantization). Moreover, the BPS phase of the $N_\pm$--changing solitons is equal to
$\pm\arg\sqrt{t}$ up to $O(\epsilon_1/\sqrt{t})$;
then in the $\theta$--limit their BPS rays $\pm\ell_m$ have  \emph{sharp} positions\footnote{ The main difficulty in formulating the RH problem without taking the $\theta$--limit, is that one has to work with Stokes rays whose position is subject to quantum fluctuations. Then, even if the general RH problem exists, it is not too convenient for concrete computations.}.  This can be understood in another way:  Taking
$N_i\rightarrow \infty$ while keeping $N_i\epsilon_1$ fixed, is the same as the large $N$ limit of matrix models.  We know
that in that limit a spectral curve emerges which for this class of models was studied in \cite{Dijkgraaf:2002fc}.   The corresponding
spectral curve has a 1-form $ydx$, where $y$ can be viewed as the derivative of the {\it effective} superpotential
$$f(x,y)=y^2-W'(x)^2+...=0$$
and the effective BPS central charges of the 2d theory get related to $\oint ydx$ around the cycles of $f(x,y)=0$ curve.
These can in turn be interpreted as the central charges of the 4d theory on the SW curve $f(x,y)=0$.  In this context the phase of the
twistorial parameter $\zeta$ controls the jumps associated with either the 4d or the 2d BPS states, depending on one's perspective.
The main point is that in the $\theta$-limit these jumps have become sharp.
In the cubic super potential example above we get the SW curve
$$f(x,y)=y^2-(x^2-t)^2+ax+b,$$
where $a,b$ are determined in terms of $N_+$ and $N_-$.
This shows that the jumps of the D-brane wave function in the $\theta$-limit is sharp.  However, it
remains to compute it.  Here we motivate what this is, based on the following observations.
The jumps should be a universal property of the geometry, and given the symplectic symmetry
of the problem it should be the same for the jumps associated with the A-cycles, or the B-cycles
of the theory.
In the limit that $t\ggg0$, the system reduces to two decoupled A-cycles where the associated $t_i=\epsilon_1 N_+$, $\epsilon_1N_-$.
As we shall show in the next two sections, a crucial property of the $tt^*$ solutions for the decoupled $A$--cycles is
that, in the $\theta$--limit, the Stokes jumps
of their brane amplitudes
at the BPS rays in the $\zeta$--plane (the positive and negative imaginary axis for $\epsilon_1$ real) are  given by multiplication by the quantum dilogs\footnote{ For convergence reasons, one assumes $\theta$ to have a small positive imaginary part. We set $2\pi/\tilde \epsilon_2=1$ to simplify the notation. We write $f$ and $e$ for the flavor and electric charges, respectively.}
\begin{gather}\label{rrrqasv}
\prod_{k=0}^\infty\big(1-e^{-a_\pm/\zeta+i\theta_\pm -\bar a_\pm\zeta}e^{i(k+1/2)\theta}\big)^{-1} \equiv\boldsymbol{\Psi}(X_\pm(\zeta);q)\\
\intertext{where $X_\pm(\zeta)$ are the GMN line operators associated to the charges $f\pm e$ of the BPS states of the effective theories $\mathcal{W}^\mathrm{eff}_\pm$}
X_\pm(\zeta)\equiv e^{-a_\pm/\zeta+i\theta_\pm -\bar a_\pm\zeta},\qquad q\equiv e^{i\theta}\\
\text{with\ \ $a_\pm= a_f\pm a_e$,\ \ $\theta_\pm=\theta_f\pm\theta_e$,}
\end{gather} 
  and the expression \eqref{rrrqasv} becomes exact as
 $N_\pm\to\infty$ with $a_\pm\equiv N_\pm\epsilon_1$ fixed. 
 We note that eqn.\eqref{rrrqasv} is the formula
 expected in 4d from the refined version
 \cite{Cecotti:2009uf,Dimofte:2009bv, 
Dimofte:2009tm}
of the
 Kontsevich--Soibelman (KS) wall crossing formula \cite{Kontsevich:2008fj}.
Indeed, the quantum Stokes jump at the ray $\ell_\gamma$ of a BPS hypermultiplet of charge $\gamma$ is given by the (adjoint) action of the quantum dilog \cite{Cecotti:2010fi,Cecotti:2009uf,Dimofte:2009bv, 
Dimofte:2009tm}
 \begin{equation}\label{opksjum}
 \boldsymbol{\Psi}(\mathsf{X}_\gamma;q)
 \end{equation}
 where $\mathsf{X}_\gamma$ is the quantum torus algebra element associated to the charge $\gamma\in\Gamma$, whose expectation value is identified with the Darboux coordinate 
$X_\gamma(\zeta)$ of \cite{Gaiotto:2008cd}.
 Eqn.\eqref{rrrqasv} thus states that, in the $\theta$--limit,  
 the brane amplitude $\Phi(\zeta)$ of \eqref {oritmod} has jumps at the rays $\ell_\gamma$ associated to charges $\gamma$ of the form $\pm e\pm f$ which
 have the expected quantum KS form \eqref{opksjum}.  
 
So far we have explained how the electric line operators appear in the $\theta$-limit.
The 4d magnetically charged solitons correspond in the 2d model \eqref{oritmod} to BPS solitons connecting vacua with \emph{different} values of $N_\pm$. As already discussed, the leading such solitons connect vacua with $N_+\rightarrow N_+\pm 1$, and their BPS rays $\pm\ell_m$ are sharp in the $\theta$--limit.
Physically, the magnetic line function  $X_m(\zeta)$ is identified with the expectation value of the operator $\mathsf{X}_m$ which implements the transition of a single eigenvalue field $X_i$ from $-\sqrt{t}$ to $+\sqrt{t}$. 
In terms of
$a_\pm$ and $\theta_\pm$, $\mathsf{X}_m$ implements the shifts
\begin{equation}
a_\pm \to a_\pm\pm \epsilon_1,\qquad \theta_\pm \to \theta_\pm\pm \theta.
\end{equation}
Then, acting on a brane wave--function written as a function of $(a_\pm, \theta_\pm, \bar a_\pm)$,
\begin{equation}
\begin{split}
\mathsf{X}_m&=\exp\!\Big(\epsilon_1(\partial_{a_+}-\partial_{a_-})+\bar\epsilon_1(\partial_{\bar a_+}-\partial_{\bar a_-})+\theta(\partial_{\theta_+}-\partial_{\theta_-})\Big)\\
&\xrightarrow{\ \text{$\theta$--limit}\ } \exp\!\Big(\theta(\partial_{\theta_+}-\partial_{\theta_-})\Big).
\end{split}
\end{equation}
Although this identification is a bit heuristic,
it may be given a more precise meaning by looking at the exact solution of the low--energy effective models.  
Defining $\mathsf{X}_\pm$ as the operator which acts
on the above wave--functions as multiplication by $X_\pm(\zeta)$, we get
the commutation relations
\begin{equation}\label{comrel}
\mathsf{X}_m\,\mathsf{X}_\pm =q^{\pm 1}\, \mathsf{X}_\pm\,\mathsf{X}_m,
\end{equation}
which yield the correct quantum torus algebra for the
4d model corresponding to the large $N$ limit of 
\eqref{oritmod} which is the $A_3$ Argyres--Douglas (AD) model whose BPS quiver has the form \cite{Cecotti:2010fi,Cecotti:2011rv}
\begin{equation}
\bullet \rightarrow \bullet\rightarrow \bullet
\end{equation}
Besides, the 2d analysis leads (in the regime considered) to a 4d BPS spectrum which correctly
matches the one in the minimal BPS chamber
\cite{Cecotti:2010fi} of the $A_3$ AD theory, which consist of three hypermultiplets of charges $e+f$, $e-f$, and $m$.
 
Then, to reproduce the exact structure expected from the refined version of the 4d KS wall crossing formula, it remains to show that the $\theta$--limit Stokes jumps are given by the action of the operator \eqref{opksjum}
also at the magnetic BPS rays $\pm \ell_m$. Since $\mathsf{X}_m$ is the operator which makes a single eigenvalue to jump from $-\sqrt{t}$ to $+\sqrt{t}$,
which corresponds to the element $J_+\in\mathfrak{gl}(2,\mathbb{C})$, lifting to the covering LG model before modding out $\mathfrak{S}_N$, a naive application of formula 
\eqref{rrrrqqq} would produce, at $\theta=\epsilon_1=0$, the magnetic--ray Stokes matrix
\begin{equation}
S_m=(1-\mathsf{X}_m^{(1)})^{-1}(1-\mathsf{X}_m^{(2)})^{-1}(1-\mathsf{X}_m^{(3)})^{-1}\cdots (1-\mathsf{X}_m^{(N)})^{-1},\label{kkkkma}
\end{equation}
where $\mathsf{X}_m^{(j)}$ acts on the 
$j$--th factor LG model.
Switching on a non--zero $\theta$ and modding out $\mathfrak{S}_N$, identifies the several operators $\mathsf{X}_m^{(j)}$, and in addition introduces
in the expression \eqref{kkkkma} powers of $q$
which  keep track of the braiding numbers of the eigenvalues 
$\{X_i(t)\}$ for each BPS soliton connecting the two vacua. 
This suggests replacing $\mathsf{X}_m^{(j)}\to \mathsf{X}_m\,q^{j-1}$ in the previous formula, with the result 
\begin{equation}
\begin{split}
S_m\big|_{q\neq 1}&\approx 
(1-\mathsf{X}_m)^{-1}(1-\mathsf{X}_m q)^{-1}(1-\mathsf{X}_m q^2)^{-1}\cdots (1-\mathsf{X}_m q^{N-1})^{-1}\\
&\xrightarrow{\ \ \text{large $N$}\ \ }\ \boldsymbol{\Psi}(\mathsf{X}_m;q),
\end{split}
\end{equation}
Given the symmetry between electric and magnetic, and in view of the result
\eqref{rrrqasv} for the electric/flavor jumps, this formula is very natural.\footnote{ Other arguments lead to the same 
conclusion. Suppose that the jump at the magnetic phase is given by some unknown function $f(q,\mathsf{X}_m)$. The phase--ordered product of the Stokes 
operators in the $\zeta$--plane is equal to the 2d
quantum monodromy $H$ \cite{Cecotti:1992rm} of the $(2,2)$ matrix LG model. Then the 2d monodromy would be
\begin{equation*}
H=f(q^{-1},\mathsf{X}_m^{-1})^{-1}\,\boldsymbol{\Psi}(\mathsf{X}_+^{-1};q)\,\boldsymbol{\Psi}(\mathsf{X}_-;q)f(q,\mathsf{X}_m)\,\boldsymbol{\Psi}(
\mathsf{X}_+;q)\,\boldsymbol{\Psi}(\mathsf{X}_-^{-1};q)
\end{equation*}
In a unitary theory the spectrum of $H$ should belong to the unit circle \cite{Cecotti:1992rm}.
In facts, if the model flows in the UV to a good SCFT,
the 2d monodromy has finite order $r$, $H^r=1$,
and the order of its adjoint action $\mathrm{Ad}(H)$ is a divisor of $r$. Assuming the matrix LG model has a good UV limit,
we may compute the order of the adjoint monodromy
in the UV limit, $N\epsilon_1\to 0$, where the effect of the Vandermonde coupling is totally negligible. We are reduced to the monodromy of the $A_2$ minimal model,
and hence the order of  $\mathrm{Ad}(H)$ is $3$. 
Using the commutation relations \eqref{comrel},
the equation $\mathrm{Ad}(H)^3=1$ is written as a functional equation for the unknown function $f(q,X)$. Comparing with the 4d quantum monodromy of the $A_3$ Argyres--Douglas model \cite{Cecotti:2010fi}, we see that the functional equation for $f(q;X)$ is equivalent
 to the usual pentagonal identity for the quantum dilog $\boldsymbol{\Psi}(q;X)$. }

\subsection{A Quantum Riemann-Hilbert problem}

The brane amplitudes of an ordinary $(2,2)$ model, having a finite number $m$ of vacua, are the solutions to a Riemann--Hilbert (RH) problem for matrices $\Psi(\zeta)$ of size $m$ which operate on the vacuum vector space $\mathbb{C}^m$ \cite{Dubrovin:1992yd,Cecotti:1992rm}. For finite $N$, a matrix LG model of the form \eqref{Eefft}
has $\theta$--vacua \cite{Cecotti:2013mba}, and the space of vacua takes the form $\mathbb{C}^m\otimes L^2(S^1)$, where $m=O(N^{n-1})$. The brane amplitudes are now the solutions of an infinite--size
RH problem for operators acting on the Hilbert space
 $\mathbb{C}^m\otimes L^2(S^1)$. Except for the special case $n=1$, as $N\to \infty$, also $m\to \infty$ and the space $\mathbb{C}^m$
 gets replaced by a Hilbert space direct factor $\mathcal{H}$, so that the
 full twistorial topological string amplitudes may be
 thought of as the solutions to a Riemann--Hilbert problem for quantum operators acting on the Hilbert space $\mathcal{H}\otimes L^2(S^1)$.
 The resulting quantum RH problem is however extremely hard
 to formulate in concrete terms, let alone to solve. One looks for a limit in which the quantum RH problem admits a formulation which is difficult but still reasonable. Formally, the $\theta$--limit is such a limit, although the limit itself is delicate in the sense that its definition requires appropriate regularizations and/or analytic continuations.
 
 The idea is as follows.  We first formulate a quantum-Riemann Hilbert problem which characterizes the operators
 $X_\gamma (\zeta)$ and then use that to define a state in the Hilbert space, which we identify as the $\theta$-limit
 of the twistorial topological string amplitude $\psi^\theta$.
 
 \subsubsection{The operators $\hat X_\gamma(\zeta)$}
At the full quantum level, the holomorphic Darboux coordinates $X_\gamma(\zeta)$ (with $\gamma\in\Gamma$), get replaced by quantum operators $\hat  X_\gamma(\zeta)$. Most of the time, we will suppress from the notation the dependence of these operators on the other variables (Coulomb branch parameters and couplings) and write only the dependence on the twistor variable $\zeta$ which is taken to be valued in $\mathbb{C}^\times$, that is, $\zeta$ takes values in the twistor sphere minus the North and South poles.

It is convenient to think of the phase of $\zeta$ as time. So we also write $\hat X_\gamma(\rho\, e^{it})$ with $\rho\in \mathbb{R}_+$. In this vein, an useful analogy is to think of the operators $\log\hat X_\gamma(\rho e^{it})$ as two dimensional quantum \emph{chiral} fields where $t$ plays the role of time and $\log\rho$ of the space coordinate. In fact, the operators $\hat X(\zeta)$ are required to satisfy the \emph{equal time} commutation relations 
\begin{equation}\label{eqtime}
\hat X_\gamma(\rho\, e^{it})\: \hat X_{\gamma^\prime}(\rho^\prime e^{it}) = q^{\langle \gamma,\gamma^\prime\rangle}\: \hat X_{\gamma^\prime}(\rho^\prime e^{it}) \: \hat X_{\gamma}(\rho\, e^{it}) 
\end{equation}
where $q=\exp(2\pi i\tau)$.  Thus $\log\hat X_\gamma(\rho\, e^{it})$ satisfy `canonical' equal time commutation relations. We shall refer to equation \eqref{eqtime} as the \emph{equal time quantum torus algebra}.

In correspondence with this analogy, we introduce the following time--ordering operation $T$
\begin{equation}
T\,\hat X_\gamma(\rho\, e^{it})\: \hat X_{\gamma^\prime}(\rho^\prime e^{it^\prime})= \begin{cases}
\hat X_\gamma(\rho\, e^{it})\: \hat X_{\gamma^\prime}(\rho^\prime e^{it^\prime}) & t > t^\prime \\
q^{\langle \gamma,\gamma^\prime\rangle}\:\hat X_\gamma(\rho\, e^{it^\prime})\: \hat X_{\gamma^\prime}(\rho^\prime e^{it}) 
& t^\prime> t.\end{cases}
\end{equation}

\medskip

The quantum operators $\hat X_\gamma(\zeta)$ are required to satisfy the same piece--wise holomorphic conditions as their classical counterparts $ X_\gamma(\zeta)$ except for two points:
\begin{enumerate}
\item As $R\rightarrow\infty$ they are asymptotic to semi--flat \emph{operators}
\begin{equation}
\hat X_\gamma^\mathrm{sf}(\zeta):= \exp\!\left(\frac{Z_\gamma}{\zeta}+i\theta_\gamma +\zeta\, \bar Z_\gamma\right)
\end{equation}
where the $\theta_\gamma$ satisfy the equal time CCR
\begin{equation}
\big[\theta_\gamma,\theta_{\gamma^\prime}\big]=-2\pi i \tau\, \langle\gamma,\gamma^\prime\rangle,
\end{equation}
Here $2\pi \tau =\theta $, and the $Z_\gamma$'s, being central, are $c$--number functions of the various parameters in the theory.
\item At times equal to the phase of some BPS particle, the $\hat X_\gamma(\zeta)$'s jump according to the \emph{quantum} WCF 
given by conjugation with suitable quantum dilogarithms rather than the classical one.  \end{enumerate}

\subsubsection{The quantum TBA integral equation}

The above conditions fix the $\hat X_\gamma(\zeta)$ to be the solution to  a $q$--TBA integral equation which is a $q$--deformation of the one written in \cite{Gaiotto:2008cd}.
Explicitly
\begin{equation}\label{qTBA}
\hat  X_\gamma(\zeta)= T\Bigg\{\!\exp\!\left[\sum_{\gamma^\prime}\frac{\Omega(\gamma^\prime)}{4\pi i} \int_{\ell_{\gamma^\prime}}\frac{d\zeta^\prime}{\zeta^\prime}\, \frac{\zeta+\zeta^\prime}{\zeta-\zeta^\prime}\, \log G_{\langle \gamma,\gamma^\prime\rangle}\big(q^{s_{\gamma^\prime}}\hat X_{\gamma^\prime}(\zeta^\prime); q\big)\right] \hat X_\gamma^\mathrm{sf}(\zeta)\Bigg\}
\end{equation}
where $\Omega(\gamma^\prime)$ and $\ell_\gamma$ are as in \cite{Gaiotto:2008cd}, $s_{\gamma^\prime}$ are the spins of the BPS particles, and $G_m(X;q)$ are Fock--Goncharov functions ($q$--deformed versions of $(1+X)^m$) which are defined by their basic property
\cite{Fock031,FockX2,FockX3}
\begin{equation}
\boldsymbol{\Psi}(X_{\gamma^\prime};q)^{-1}\:X_\gamma\: \boldsymbol{\Psi}(X_{\gamma^\prime};q)=G_{\langle \gamma,\gamma^\prime\rangle}\big(X_{\gamma^\prime};q\big)\, X_\gamma. 
\end{equation}
for $X_\gamma X_{\gamma^\prime}= q^{\langle\gamma,\gamma^\prime\rangle}\, X_{\gamma^\prime}X_\gamma$. In particular,
\begin{equation}\label{Gsym}
G_{-m}(X;q)=G_m(X; q^{-1})^{-1}.
\end{equation}

\smallskip

Eqn.\eqref{qTBA} is deduced and makes sense under the assumption that the \emph{equal phase} BPS states are mutually local.

\medskip

As $q\rightarrow 1$ \footnote{ Or, rather, as $q^{1/2}\rightarrow -1$, taking into account the \textit{quadratic refinement}.}, the above equations reduce to the classical TBA equations of \cite{Gaiotto:2008cd}. Formally, we may expand them in powers of $\tau \equiv (\log q)/2\pi i$. The zeroth order is classical TBA corresponding (from that viewpoint) to the energy of the ground state. Then we get an infinite sequence of integral equations by equating the order $\tau^n$ of the two sides of eqn.\eqref{qTBA}. Each equation contain the solutions to the previous integral equations. 
We discuss solutions of this TBA system for the Argyres-Douglas case in appendix B. 

One borrows from 
ref.\cite{Cecotti:2009uf} the identification of the phase $\arg\zeta$ of the twistor parameter $\zeta\in \mathbb{C}^*$ with the periodic time of an auxiliary quantum system whose operator algebra is the quantum torus 
\begin{equation}\label{qqtorsus}
\mathsf{X}_{\gamma}\,\mathsf{X}_{\gamma^\prime}=q^{\langle\gamma,\gamma^\prime\rangle}\, \mathsf{X}_{\gamma^\prime}\,\mathsf{X}_\gamma,\qquad \gamma,\gamma^\prime\in\Gamma, 
\end{equation}
defined by the Dirac electromagnetic pairing of charges $\langle\cdot,\cdot\rangle\colon \Gamma\times\Gamma\to\mathbb{Z}$. In particular,
one interprets the ordering in BPS phase of the Kontsevich--Soibelman product of 
symplectomorphisms \cite{Kontsevich:2008fj} as 
the usual time--ordering of (time--dependent) evolution operators in quantum physics \cite{Cecotti:2009uf}. We have argued in the previous subsection that the $\theta$--limit produces effective operators $\mathsf{X}_\gamma$ which generate the algebra \eqref{qqtorsus} of the auxiliary QM system. 

In order to complete the problem we need to find $\psi^theta$, i.e.\! the state in the Hilbert space which
corresponds in the electric basis, $\psi^\theta$:
$$\psi^\theta (a_e,\theta_e, R,\theta, \zeta)=\langle \theta_e |\psi (\zeta)\rangle$$
The state $|\psi (\zeta)\rangle$ is characterized by the jumps as we cross the phases of $\zeta $ for which there
is a BPS state.  This implies that $|\psi (\zeta )\rangle$ satisfies the following Riemann-Hilbert equation:
$$|\psi (\zeta)\rangle ={1\over 2\pi i}\sum_{(\gamma,s)\in BPS} \int_{l_\gamma} {d \zeta'\over \zeta'-\zeta}\Big( [\boldsymbol{\Psi}(q^s{\hat X}_\gamma(\zeta'),q)]^{(-1)^{2s }} -1\Big) |\psi(\zeta')\rangle$$
Moreover, the boundary condition we have for $|\psi(\zeta)\rangle$ is given by \eqref{Sergio}:
$$\langle \theta^i |\psi(\zeta)\rangle \xrightarrow{\zeta \rightarrow 0,\;\infty}  {\rm exp}\!\left[{F(a)\over \theta \zeta^2 {\tilde \epsilon_2}^2}+{{\overline F}({\overline a}) \zeta^2\over { \theta {\tilde \epsilon_2}^2}}\right]$$

This fixes the state $|\psi(\zeta)\rangle$, completing the formulation of our quantum Riemann-Hilbert problem.

\section{LG Matrix Models and Twistorial Matrix Models}\label{s:lgmatrixmodels}

We have seen how in the $\theta$-limit a quantum Riemann-Hilbert problem can be used
to formally solve for the partition function of the twistorial topological string, assuming one knows the spectrum
of the BPS state (including their spin data).  To solve for the full twistorial topological string partition function
without taking any limits is much harder.  In the case when we have a dual description of the 2d model,
as in the LG matrix model, we may be in a better shape.  This is why,
in this section we study in some detail the $tt^*$ geometry of the 2d $(2,2)$ matrix Landau--Ginzburg models of the class
discussed in section \ref{s:connectionopentopstrings}; a more general class is considered in appendix \ref{appendixmultimatrix}. After some generality  (\S.\,\ref{ss:themodels}),
in \S\S.\ref{chiringva},\,\ref{HSandvanvleck} we describe their chiral rings $\mathcal{R}$ in terms of the
associated Schroedinger equation \cite{Aganagic:2011mi,Eynard:2007kz}. In \S.\ref{s:gaussianmodel}
we solve exactly the $tt^*$ geometry for the basic example, the Gaussian model, and describe in detail the properties of its various $tt^*$ quantities.
In \S.\ref{ss:abelian} we introduce a more general class of models whose $tt^*$ geometry may be explicitly computed, and describe the corresponding $tt^*$ geometries in detail. In the last two subsections we present two additional explicit examples of exactly solved $tt^*$ geometries, namely the generalized and double LG Penner models.

\subsection{The models}\label{ss:themodels}

We consider the LG models with superpotential $\mathcal{W}$ of the form
\begin{equation}\label{whichmodels}
\mathcal{W}(e_1,e_2,\dots, e_N)=\sum_{i=1}^N W(X_i)+\beta \sum_{1\leq i<j\leq N} \log(X_i-X_j)^2,
\end{equation}
where $W(z)$ is a polynomial of degree $(n+1)$
\begin{equation}\label{Wpoly}
 W(z)=\frac{z^{n+1}}{n+1}+\sum_{k=2}^n t_k\, z^{n+1-k}.
\end{equation}
Here $\beta=\epsilon_1$ (in case $W$ is homogeneous, as in the Gaussian matrix model, it is
convenient to absorb a factor of $1/{\tilde \epsilon_2}$ into the fields and in this case
we can view $\beta=\epsilon_1/{\tilde \epsilon_2}$, up to a constant shift of $W$).
We stress that in eqn.\eqref{whichmodels} the independent chiral fields are not the matrix eigenvalues $X_i$ but rather their elementary symmetric 
functions $e_k$
\begin{equation}\label{esymfn}
 e_k=\sum_{1\leq j_1<j_2<\cdots <j_k\leq N}X_{j_1}X_{j_2}\cdots X_{j_k}.
\end{equation}
The change of fundamental degrees of freedom from $X_i$ to $e_k$ automatically projects
the model into its $\mathfrak{S}_N$--invariant sector (and introduces a Jacobian factor in the topological measure 
\cite{Cecotti:1992rm,Cecotti:1991me}).

In view of the application to other physical problems \cite{Dijkgraaf:2009pc,Chair:2014wpa}, as well as to connect with existing mathematical 
literature, we find convenient to enlarge the class of models to LG theories with superpotentials of the form \eqref{whichmodels}
with $W(z)$ a possibly multi--valued function such that its differential $dW= W^\prime(z)\,dz$ is a rational\footnote{ The class of models may be further generalized by
replacing the Riemann sphere $\mathbb{P}^1$ by a higher genus Riemann surface.} one--form on $\mathbb{P}^1$
normalized so that $z=\infty$ is a pole of maximal order. The number $n$ of susy vacua of the one--field (\textit{i.e.}\! $N=1$) model
is the number of zeros of the rational one--form $dW$, equal to its total pole order minus $2$. The Witten index of the $N$--field model is
then expected to be
\begin{equation}\label{numbervacua}
m={N+n-1\choose N}\approx \frac{N^{n-1}}{(n-1)!}\quad \text{for large }N.
\end{equation}
We shall make more precise statements on the number of susy vacua momentarily.

The generic degree $n+2$ rational differential has only simple poles
at the $n+2$ distinct points $\{w_1,w_2,\dots, w_{n+1},\infty\}$
\begin{equation}\label{genericWm}
 W^\prime(z)\,dz=\sum_{\ell=1}^{n+1} \frac{\mu_\ell}{z-w_\ell}\, dz,
\end{equation}
where we may take $w_1=0$ and $w_2=1$ by a field redefinition.
In ref.\cite{Dijkgraaf:2009pc} it was shown that the matrix LG model \eqref{whichmodels} with one--field superpotential 
\eqref{genericWm} corresponds to the $(n+2)$--point function of the Liouville theory on $\mathbb{P}^1$.
The models whose one--form 
have higher order poles may be obtained from \eqref{genericWm} by taking limits in which 
many ordinary singularities coalesce into higher
order ones. In particular the polynomial superpotential \eqref{Wpoly} is obtained by making the $n+2$ ordinary singularities
to coalesce into a single order $n+2$ pole at infinity. By considering these various
confluent limits, we get from \eqref{genericWm} $p(n+2)$ distinct models all with a number of vacua equal to
\eqref{numbervacua} (here $p(k)$ is the number of partitions of the integer $k$). This observation allows
 to study all rational models with a given $n$ in a unified way. 

The one--field superpotential $W(z)$ is, in general, a multi--valued function of $z$ which is well--defined only up to
the periods of the one--form $W^\prime\,dz$, that is, for the generic case \eqref{genericWm} up to
\begin{equation}
 \Delta W= 2\pi i\sum_{\ell=1}^{n+1} n_\ell\,\mu_\ell,\qquad n_\ell\in\mathbb{Z}. 
\end{equation}
Comparing with the general analysis in ref.\cite{Cecotti:2013mba}, we conclude that in such a rational model with $N>1$ chiral fields we have to introduce
$p+1$ vacuum angles $\theta_s$, where $p$ is the number of independent residues of the one--form $dW$ (i.e.\! the number of its
simple poles in $\mathbb{C}\equiv\mathbb{P}^1\setminus\{\infty\}$). 
\smallskip

A configuration of the chiral fields $e_k$ is most conveniently encoded in a degree $N$ monic polynomial in an indeterminate $z$ as
\begin{equation}
 P(z)=\sum_{k=0}^N(-1)^k\,e_k\,z^{N-k}=\prod_{i=1}^N (z-X_i)\equiv \det(z-X),
\end{equation}
where $e_0=1$.

\subsection{Chiral ring and vacuum configurations}\label{chiringva}

To describe the $tt^*$ geometry of the $(2,2)$ models \eqref{whichmodels}, we first have to find their chiral ring $\mathcal{R}$ 
\cite{Cecotti:1991me}. For generic $W(z)$ all classical vacua are non--degenerate (that is, massive); in this case, as complex algebras,
$\mathcal{R}\simeq \mathbb{C}^m$, $m$ being the number of supersymmetric vacua. The isomorphism is given by sending the class of a
general chiral superfield, 
represented by a holomorphic function $h(e_1,\dots, e_N)$, into the $m$--tuple of its values at the classical vacuum configurations.
In the case of the models \eqref{whichmodels} there is a special class of chiral operators, the
\emph{single--trace} operators, of the form
\begin{equation}
 \widehat{h}(e_1,\dots, e_N):=\sum_{k=1}^N h(X_i),
\end{equation}
where $h(z)$ is a holomorphic (polynomial) function. It is easy to show that all elements of $\mathcal{R}$ have a
 single--trace representative. 
Then the isomorphism $\mathcal{R}\simeq \mathbb{C}^d$ reduces to
\begin{equation}
 \mathcal{R}\ni \widehat{h}(e_1,\dots, e_N)\longmapsto \left(\oint_C h(z)\,\frac{P^\prime_{(1)}(z)}{P_{(1)}(z)}\,dz,\;
\cdots,\; \oint_C h(z)\,\frac{P^\prime_{(m)}(z)}{P_{(m)}(z)}\,dz\right)\in\mathbb{C}^m,
\end{equation}
where $P_{(a)}(z)$ is the polynomial specifying the $a$--th vacuum configuration, and $C$ is a large circle.
Finding $\mathcal{R}$ is then equivalent to computing the $m$ polynomials $P_{(a)}(z)$ describing the classical susy vacua.  

The classical vacua are the solutions $\{e_k\}$ to the system of equations
\begin{equation}\label{eee4cx}
\frac{\partial \mathcal{W}(e_j)}{\partial e_k}=0, \quad k=1,2,\dots, N,
\end{equation}
which are obviously equivalent to
\begin{equation}\label{ppqaz}
W^\prime(X_i)+\frac{2\beta}{\prod_{j\neq i}(X_i-X_j)}\,\sum_{j\neq i}\prod_{k\neq i,j}(X_i-X_k)=0,
\end{equation}
from which it is obvious that $\prod_{i<j}(X_i-X_j)^2\neq 0$, that is, on the vacuum configurations the $X_i$'s 
are all distinct and the discriminant of the associated polynomials $P(z)$ is non zero. From the definition
\begin{equation}
P(z)=\prod_i(z-X_i)\equiv z^n+\sum_{k=1}^n (-1)^k\, e_k\, z^{n-k},
\end{equation}
one gets
\begin{gather}
P^\prime(X_i)=\prod_{j\neq i}(X_i-X_j)\\
P^{\prime\prime}(X_i)=2\sum_{j\neq i}\prod_{k\neq i,j}(X_i-X_k).
\end{gather}
Using these identities, we may rewrite the equations \eqref{ppqaz} 
in terms of the polynomial $P(z)$ describing the vacuum configuration $\{X_i\}^{\mathfrak{S}_N}$ as
\begin{equation}
W^\prime(X_i)\,P^\prime(X_i)+\beta\, P^{\prime\prime}(X_i)=0.
\end{equation}
For the generic case, eqn.\eqref{genericWm}, this equation says that the degree $N+n-1$ polynomial
\begin{equation}
\Big( W^\prime(z)\,P^\prime(z)+\beta\,P^{\prime\prime}(z)\Big)\prod_{\ell=1}^{n+1}(z-w_\ell)=0
\end{equation}
has the $N$ distinct roots $X_i$, and hence it should be a multiple
of $P(z)$, the quotient being some polynomial
$f_{n-1}^{(a)}(z)$ of degree $n-1$
\begin{equation}\label{pppmn}
\beta\,P^{\prime\prime}(z)+W^\prime(z)\, P^\prime(z)=\frac{f^{(a)}_{n-1}(z)}{\prod_\ell(z-w_\ell)}\,P(z).
\end{equation}
All monic degree $N$ polynomials $P(z)$ which solve this linear second--order equation, 
for \emph{some} choice of the polynomial $f^{(a)}_{n-1}$, correspond to a classical vacuum.
The polynomial $f^{(a)}_{n-1}(z)$ depends on the classical vacuum configuration, and 
the index $a$ takes the values $a=1,2,\dots, m$ where $m$ is the number of classical vacua which, in the present
case, coincides with the Witten index $m$, eqn.\eqref{numbervacua}.
There is a one--to--one correspondence between susy vacua and distinct polynomials $f^{(a)}_{n-1}(z)$.
  Writing
\begin{equation}
\psi(z)=e^{W(z)/2\beta}\,P(z),
\end{equation}
we recast eqn.\eqref{pppmn} in the Schroedinger form (identifying $\hbar\equiv\beta$)
\begin{equation}\label{schr}
\left(-\beta^2\, \frac{\partial^2}{\partial z^2}+\frac{1}{4}(W^\prime(z))^2+
\frac{\beta}{2}W^{\prime\prime}(z)+\beta\,\frac{f^{(a)}_{n-1}(z)}{\prod_\ell(z-w_\ell)}\right)\psi(z)=0,
\end{equation}
which coincides with the Schroedinger equation discussed in a related context in refs.\cite{Aganagic:2011mi,Eynard:2007kz}.
For instance, in the Gaussian case, $W(z)=-z^2/2$, eqn.\eqref{schr} reduces to the 
Schroedinger equation for the harmonic oscillator, with energy eigenvalue $-(f_{0}+\frac{1}{2})$ 
and coordinate $x=z/\sqrt{2\beta}$. In this case, for each $N$ there is a \emph{unique} susy vacuum given by
\begin{equation}\label{herherm}
P(z)= (\beta/2)^{N/2}\, H_N(z/\sqrt{2\beta})
\end{equation}
where $H_N(x)$ is the $N$--th Hermite polynomial.

\subsection{Heine--Stieltjes and van Vleck polynomials}\label{HSandvanvleck}

To determine the chiral ring $\mathcal{R}$ we are
reduced to the following problem: \textit{Given the two polynomials 
\begin{equation}
 A(z)=\prod_{\ell=1}^{n+1}(z-w_\ell),\qquad B(z)= \frac{1}{2\beta}\,A(z)\, W^\prime(z),
\end{equation}
respectively of degree $n+1$ and $n$,  determine all degree $n-1$ polynomials $f_{n-1}(z)$ such that the differential equation
\begin{equation}\label{genLame}
 A(z)\, \frac{d^2 P}{dz^2}+2 B(z)\, \frac{d P}{dz}-f_{n-1}(z)\,P=0
\end{equation}
has a solution $P(z)$ which is a polynomial of degree $N$.} This is precisely the classical
Heine--Stieltjes problem, see \textit{e.g.}\! \S.6.8 of the book by Szeg\"o \cite{szego}. The degree $N$ polynomials
$P(z)$ describing a vacuum configuration are known as \emph{Heine--Stieltjes polynomials,} while
the degree $(n-1)$ polynomials $-f_{n-1}(z)$ as \emph{van Vleck polynomials}.
In 1878 Heine stated \cite{heine} that there are at most
\begin{equation}\label{vacuanumm}
m={N+n-1\choose N},
\end{equation}
polynomials $f_{n-1}(z)$, of degree $n-1$, counted with appropriate multiplicity, such that the generalized Lam\'e ODE
\eqref{genLame} has a polynomial solution $P(z)$ of degree exactly $N$. In fact, he proved that for
 generic\footnote{ The precise meaning of
`generic' is that the two polynomials $A(z)$ and $B(z)$ should be algebraically independent.} $A(z)$, $B(z)$ the number of solutions
is exactly $m$. This result is consistent with the Witten index computation in \S.\ref{ss:themodels}, since for particular values 
of the couplings
a few susy vacua may escape to infinity. Since 1878 many authors gave necessary conditions for the number of solutions to be
precisely $m$. Finally in 2008 Shapiro proved\footnote{ Shapiro theorem refers to the \emph{non--degenerate} case, that is,
at infinity the differential $dW$ has at most a single pole. However, if $dW$ has a higher order pole at $\infty$, \emph{a fortiori} susy vacua cannot 
escape since the scalar potential is bounded away from zero in a neighborhood of $\infty$.  } \cite{shapiro1,shapiro2} that there exists an $N_0$ such that for all $N>N_0$ 
we have exactly $m$ solutions (counted
with multiplicity).

Physically we interpret the result \eqref{vacuanumm} as the statement that each vacuum corresponds to one of the possible ways of distributing the $N$ eigenvalues of the matrix $X$ between the $n$ critical points of the one--field superpotential $W(z)$. Giving a precise meaning to this statement has been an active field of research in mathematics for more than a century, see \textit{e.g.}\! \cite{heine,HS2,HS3,HS4,HS5,HS5.5,HS6,HS7,HS8,HS9,HS10,HS11,HS12,HS13,HS14,HS15,HS16,HS17,HS18,HS19,shapiro1,shapiro2,shapiro3,shapiro4}.

A lot of properties of the Heine--Stieltjes and Van Vleck polynomials are discussed in the mathematical literature \cite{heine,HS2,HS3,HS4,HS5,HS5.5,HS6,HS7,HS8,HS9,HS10,HS11,HS12,HS13,HS14,HS15,HS16,HS17,HS18,HS19,shapiro1,shapiro2,shapiro3,shapiro4}.  
The best known cases are $n=1,2$. 
For $n=1$ the ODE \eqref{genLame} becomes the hypergeometric equation whose polynomial solutions are the
the Jacobi polynomials. Colliding two (resp.\! three) singularities we get the confluent hypergeometric equation 
(resp.\! the parabolic--cylinder equation) whose polynomial solutions are the Laguerre (resp.\! Hermite) polynomials.
The next case, $n=2$, leads to Heun polynomials \cite{heun1,heun2} and their various multi--confluent limits.

\subsection{The Gaussian matrix LG model}\label{s:gaussianmodel}

The simplest and most basic twistorial matrix theory is the
Gaussian model, that is, the $(2,2)$ Landau--Ginzburg model with superpotential
\begin{equation}
\mathcal{W}(e_k)=-\frac{1}{2}\sum_{i=1}^N X_i^2
+\beta\!\!\!\!\!\sum_{1\leq i<j\leq N}\log(X_i-X_j)^2,
\end{equation}
where the independent chiral superfields $e_k$ ($k=1,\dots, N$) are the elementary symmetric functions of the matrix eigenvalue superfields $X_i$, eqn.\eqref{esymfn}. The large $N$ duality maps this model to the B-model closed topological
string for the conifold, or equivalently to the $4d$ SQED.

\subsubsection{$tt^*$ geometry}

From eqn.\eqref{herherm} the Gaussian model has a single vacuum $\{e_k|_\text{vac}\}$ such that
\begin{equation}
\sum_{k=0}^N (-1)^k\, z^{N-k}\,e_k\big|_\text{vac}= (\beta/2)^{N/2}\,H_N(z/\sqrt{2\beta}),
\end{equation}
where $H_N(w)=(2w)^N+\cdots$ is the $N$--th Hermite polynomial. In particular,
\begin{equation}
e_1\big|_\text{vac}=0,\qquad e_2\big|_\text{vacuum}=-\beta {N\choose 2},
\end{equation}
\begin{equation}\label{szego}
\begin{split}
\prod_{1\leq i<j\leq N}(X_i-X_j)^2\Big|_\text{vacuum}&= \mathrm{Discr}\Big((\beta/2)^{N/2}\,H_N(z/\sqrt{2\beta})\Big)=\\
&=\beta^{N(N-1)/2} \prod_{k=1}^N k^k\equiv\beta^{N(N-1)/2}\,H(N),
\end{split}\end{equation}
where we used the Szeg\"o formula \cite{szego} for the discriminant of the Hermite polynomials, and $H(z)$ is the \emph{hyperfactorial} function, related to the Barnes $G$--function $G(z)$ as
\begin{equation}
H(z)=\exp\!\big[ z\,\log\Gamma(z+1)\big]\big/G(z+1),
\end{equation}
so that
\begin{equation}
\log H(z)=-\frac{1}{2}z\, \log2\pi +\frac{1}{2}(z+1)z+\int_0^z \log\Gamma(t+1)\,dt.
\end{equation}

The element of the chiral ring
\begin{equation}
\partial_\beta\mathcal{W}= \sum_{i<j}\log(X_i-X_j)^2
\end{equation}
takes on the vacuum the values 
\begin{equation}
\frac{N(N-1)}{2} \log\beta+\log H(N)+2\pi i k,\qquad k\in\mathbb{Z},
\end{equation}
where the term $2\pi i k$ takes into account the multiple determinations of the logarithm. We extend  the theory to a cover of field space in which the superpotential is univalued. The integer $k$ then labels the distinct susy vacua of the extended theory which cover the unique vacuum of the original model.  Following \cite{Cecotti:2010fi,Cecotti:2013mba}, we introduce 
the $\theta$--vacua, the angle $\theta$ being the Fourier dual to the integer $k$. Setting $x=\theta/2\pi$,
we represent the chiral operator $\partial_\beta\mathcal{W}$ acting on the $\theta$--vacua as the differential operator \cite{Cecotti:1991me,Cecotti:2010fi,Cecotti:2013mba}
\begin{equation}\label{cnbeta}
C_\beta= \frac{N(N-1)}{2} \log\beta+\log H(N)+\frac{\partial}{\partial x}\equiv \frac{\partial}{\partial x}+C(N,\beta).
\end{equation}
The $tt^*$ equations for the metric $G_N(\beta,x)$ then read
\cite{Cecotti:1991me,Cecotti:2010fi,Cecotti:2013mba}
\begin{equation}\label{333cxz}
\frac{\partial^2}{\partial\bar\beta\partial\beta}\log G_N+\frac{\partial^2}{\partial x^2}\log G_N=0,
\end{equation}
so that the function $h_N(\beta,x)\equiv \log G_N(\beta,x)$ is harmonic in $\mathbb{R}^3$ (with coordinates $(2\,\mathrm{Re}\,\beta,2\,\mathrm{Im}\,\beta,x)$), and periodic of period $1$ in $x$. $h_N(\beta,x)$ depends on
$\beta$ only trough $|\beta|$, and vanishes exponentially as $|\beta|\to\infty$ by the $tt^*$ 
 IR asymptotics \cite{Cecotti:1991me,Cecotti:1992rm,Cecotti:1992qh}.
 Moreover, the reality structure of $tt^*$ requires $h_N(\beta,x)$ to be an odd function\footnote{\label{ffoot} In the topological \emph{un--normalized}  $\theta$--basis one has 
\begin{equation*}\log G_N(x)+\log G_N(-x)=
 \log\!\Big| \text{Hessian of $\mathcal{W}\big|_\text{vacuum}$}\Big|^2\end{equation*}
 The statement in the text refers to the metric written in the topologically normalized $\theta$--vacua.} of $x$.
Hence the $tt^*$ metric may be written as a series of Bessel functions
\begin{equation}
h_N(\beta,x)\equiv \log G_N(\beta,x)=\sum_{m\geq 1} A_m(N)\,\sin(2\pi m x)\, K_0(4\pi m |\beta|),
\end{equation}
for some coefficients $A_m(N)$ to be determined using the appropriate boundary condition to be discussed momentarily. The harmonic function
\begin{equation}
V_N=\frac{1}{2}\,\frac{\partial h_N}{\partial x}
\end{equation} 
is the solution to the classical electrostatic problem with a charge distribution
\begin{equation}
\varrho_N(y_i,x)=\pi\,\delta(y_1)\,\delta(y_2) \sum_{m\geq 1} m\, A_m(N)\, \cos(2\pi m x).
\end{equation}
For instance, from the well--known identity \cite{besselsum1,besselsum2}
\begin{equation}\label{wellknownide}
\begin{split}
&\log\frac{z^2+y^2}{4}-4\sum_{k=1}^\infty\cos(2\pi k x)\;K_0(2\pi k\sqrt{z^2+y^2})=\\
&=-\frac{1}{\sqrt{z^2+y^2+x^2}}-\sum_{k\in\mathbb{Z}\atop k\neq0} \left(\frac{1}{\sqrt{z^2+y^2+(x-k)^2}}-\frac{1}{|k|}\right)-2\gamma,
\end{split}
\end{equation}
we see that
\begin{equation}
A_m=\frac{2}{\pi}\,\frac{1}{m}
\end{equation}
corresponds to a linear periodic  
array of charge one monopoles superimposed to a linear screening  constant charge distribution. Indeed,
\begin{equation}\label{n2distr}
2\sum_{m\geq 1} \cos(2\pi m x)= \delta_\mathbb{Z}(x)-1,\qquad
\text{where}\quad\delta_\mathbb{Z}(x)\equiv\sum_{k\in \mathbb{Z}}\delta(x-k).
\end{equation}
Comparing with \cite{Cecotti:2010fi,Cecotti:2013mba}, we see that (minus) the charge distribution \eqref{n2distr} gives the  Gaussian model with $N=2$.
We define the \emph{magnetic charge function} to be
\begin{equation}\label{magneticchargefunction}
F_N(z)=\frac{\pi}{2}\sum_{m\geq 1} A_m(N)\, e^{-2\pi m z},
\end{equation}
which is related to the linear charge distribution by
\begin{equation}\label{pppqqqX}
\rho_N(x)=\frac{1}{2\pi i}\partial_x \Big(F_N(-ix)-F_N(ix)\Big).
\end{equation}

\subsubsection{$tt^*$ Lax equations}

The brane amplitudes are flat sections of the
 the $tt^*$ Lax equations\footnote{ We have redefined $\zeta\to i/\zeta$ with respect to the usual 2d conventions in order to adhere to the standard 4d conventions.}
\begin{equation}
\Big(\overline{D}_\beta+i\zeta\, \overline{C}_\beta\Big)\Psi=\Big(D_\beta-\frac{i}{\zeta}C_\beta\Big)\Psi=0.
\end{equation}
The geometrical meaning of these equations
 is that the brane amplitudes $\Psi(\zeta)$ are $y$--independent \emph{holomorphic sections in complex structure $\zeta$} of a hyperholomorphic bundle $\mathcal{V}$ over a hyperK\"ahler manifold $\mathcal{H}$ of coordinates $(\beta, x+iy)$, translation in $y$ being symmetries of $\mathcal{H}$ and $\mathcal{V}$ \cite{Cecotti:2013mba}.

In the Gaussian case the $tt^*$ equations for the brane amplitude
$$\Psi_N(x)=\langle \theta=2\pi x\; |\;D\rangle$$ take the explicit form (cfr.\! eqn.\eqref{cnbeta})
\begin{gather}
\Big(\partial_{\bar\beta}-i\zeta\, \partial_x\Big)\log\Psi_N=-i\zeta\, \partial_x h_N-i\zeta\, \overline{C(N,\beta)}\\
\Big(\partial_{\beta}-i\zeta^{-1}\, \partial_x\big)\log\Psi_N=\partial_\beta h_N+i\zeta^{-1}\, C(N,\beta),\label{wwwv1}
\end{gather}
whose compatibility condition is eqn.\eqref{333cxz}.
We write
\begin{equation}\label{firstbox}
\log\Psi_N=\frac{i}{\zeta}\,U_N+\Phi_N-i\,\zeta\,\overline{U}_N,
\end{equation}
where
\begin{equation}\label{firstbox2}
\begin{split}
U_N&=\frac{N(N-1)}{2}\,\beta\,\log\beta+\beta\left(\log H(N)-\frac{N(N-1)}{2}\right)\equiv\\
&\equiv \sum_{k=1}^N \Big((k\beta)\big(\log(k\beta)-1\big)-\beta\big(\log\beta-1\big)\Big),
\end{split}
\end{equation}
and $\Phi_N$ is the solution to
\begin{align}\label{www1}
\Big(\partial_{\overline{\beta}}-i\zeta\, \partial_x\Big)\Phi_N&=-i\zeta \partial_x h_N\\
\Big(\partial_{\beta}-i\zeta^{-1}\, \partial_x\big)\Phi_N&=\partial_\beta h_N,\label{www2}
\end{align}
satisfying the appropriate boundary conditions.
If $\Phi_N$, $\Phi_N^\prime$ are two periodic solutions of \eqref{www1}\eqref{www2},  $\Delta\Phi_N\equiv\Phi_N-\Phi_N^\prime$ satisfies the homogeneous equations
\begin{equation}
\Big(\partial_{\bar\beta}-i\zeta\, \partial_x\Big)\Delta\Phi_N=\Big(\partial_{\beta}-i\zeta^{-1}\, \partial_x\Big)\Delta\Phi_N=0,\end{equation}
whose general solution is 
\begin{equation}\label{nonuni1}
\Delta\Phi_N=\log f\big(\beta/\zeta-ix+\bar\beta\zeta\big),
\end{equation}
with $f(z)$ an arbitrary analytic function of $z$ such that $f(z+i)=f(z)$.
To get the general solution of eqns.\eqref{www1}\eqref{www2},
it remains to find a particular solution.
Following \cite{Cecotti:2013mba}, we use the integral representation
\begin{equation}
K_0(4\pi m |\beta|)=\frac{1}{2}\int_\ell \exp\!\Big(-2\pi m\, \beta/s-2\pi m\,\bar\beta s\Big)\frac{ds}{s},
\end{equation}
(here $\ell\subset\C$ is a ray chosen so that the integral converges)
to rewrite the harmonic function $h_N$ in the form
\begin{equation}\label{xxxH}
h_N=\frac{1}{4i} \int_\ell \frac{ds}{s}\sum_{m\geq1}A_m(N) \Big(e^{-2\pi m(\beta/s-i x+\bar\beta s)}-e^{-2\pi m(\beta/s+i x+\bar\beta s)}\Big).
\end{equation}
Then we look for a particular solution to eqns.\eqref{www1}\eqref{www2} of the form
\begin{equation}
\Phi_N=\int_\ell ds\; \sum_{m\geq 1} \left(\phi_N(s,\zeta;m)\,e^{-2\pi m(\beta/s-i x+\bar\beta s)}+\widetilde{\phi}_N(s,\zeta;m)
\,e^{-2\pi m(\beta/s+i x+\bar\beta s)}\right).
\end{equation} 
Plugging this expression in \eqref{www1}\eqref{www2} one finds
\begin{align}
\phi_N(s,\zeta;m)&= -\frac{A_m(N)}{4i}\,\frac{1}{s}\,\frac{\zeta}{s-\zeta}\\
\widetilde{\phi}_N(s,\zeta;m)&=\phi_N(-s,\zeta;m),
\end{align}
so that,
\begin{equation}\label{3456}\begin{split}
\Phi_N(\beta,x, \zeta)
=
&-\frac{1}{2\pi i}\int_\ell \frac{ds}{s}\, \frac{\zeta}{s-\zeta}\,F_N(\beta/s-i x+\bar\beta\, s)-\\
&\quad-\frac{1}{2\pi i}\int_\ell \frac{ds}{s}\, \frac{\zeta}{s+\zeta}\,F_N(\beta/s+i x+\bar\beta\, s),
\end{split}\end{equation}
where $F_N(z)$ is the magnetic charge function \eqref{magneticchargefunction}.
The amplitude $\Psi_N(\beta,x,\zeta)$, eqn.\eqref{firstbox}, as a function of $\zeta$ is subjected to the Stokes phenomenon, 
which just means that the function $\Phi_N(\zeta)$ has a discontinuity
across  the two rays $\pm \ell\subset \mathbb{C}$
given by the residues at the poles of the integrals in eqn.\eqref{3456}
\begin{equation}\label{stokesgauss}
\mathrm{disc}\,\Phi_N\Big|_{\pm\zeta\in\ell}=-F_N(\pm \beta/\zeta\mp ix\pm \bar\beta\zeta).
\end{equation}
In particular,  $\Psi_N(\beta,x,\zeta)$ has a non trivial monodromy
as $\zeta\to e^{2\pi i}\zeta$
\begin{equation}
\Psi_N(\beta,x, e^{2\pi i}\zeta)=\exp\!\Big(F_N(\beta/\zeta-ix+\bar\beta)-F_N(-\beta/\zeta+ix-\bar\beta \zeta)\Big)\,\Psi_N(\beta,x,\zeta). 
\end{equation} 

\subsubsection{The asymmetric limit as a boundary condition}

Following \cite{Cecotti:2013mba}, we wish to identify
the above solution $\Psi_N(\beta,x,\zeta)$ with the amplitude for the basic brane (Dirichlet or Neumann, depending on the Stokes half--plane \cite{Cecotti:2013mba}). The brane amplitude is a solution to the $tt^*$ Lax equations which satisfies specific boundary conditions. We use these conditions to uniquely define the full $tt^*$ geometry, that is, to determine the unknown Fourier coefficients $A_m(N)$ or, equivalently, the magnetic charge function $F_N(z)$, eqn.\eqref{magneticchargefunction}. 
 $F_N(z)$ is uniquely determined \cite{Cecotti:2013mba} by the condition that $\Psi_N(\beta,x,\zeta)$ reproduces the correct brane amplitude in the asymmetric limit
$\bar\beta\to 0$ while keeping fixed $\beta$ and $x$.
In this limit $U_N$ remains the same, $\overline{U}_N\to 0$, and
\begin{equation}
\Phi_N\to \Phi_N^\mathrm{as}\equiv\frac{1}{2\pi i}\int_\ell \frac{dt}{t-\zeta^{-1}}\,F_N(\beta t-ix)-\frac{1}{2\pi i}\int_\ell \frac{dt}{t+\zeta^{-1}}\,F_N(\beta t+i x).
\end{equation}
In particular, at $x=0$ and assuming $\mathrm{Im}(\beta/\zeta)>0$,
\begin{equation}\label{xxxy1}
\begin{split}
\log\Psi_N(x=0)^\mathrm{as}= &\frac{i\,U_N}{\zeta}+\frac{1}{\pi i\, \zeta}\int_\ell \frac{dt}{t^2+(i/\zeta)^2}\;F_N(\beta t)=\\
=&\frac{i\,U_N}{\zeta}-
\frac{1}{\pi}\int_0^\infty \frac{ds}{s^2+1}\;F_N\big(i\beta\,s/\zeta\big).
\end{split}
\end{equation}
To get the coefficients $A_m(N)$ we compare this expression with the
asymmetric limit of the amplitude $\Psi_N(x=0)^\mathrm{as}$ computed directly. 

\subsubsection{The Selberg integral}

The asymmetric limit is just the normalized holomorphic period, which may be computed using the Selberg (Metha) integral
\cite{selberg1,selberg2,selberg3}. After shifting $x\to x+1/2$ (to compensate for the Jacobian of $X_i\to e_k$), we have\footnote{ The fields $Y_i$ are related to the $X_i$ by the rescaling
$Y_i= \sqrt{i/\zeta}X_i$. The bizarre--looking factors $i$ arise from the replacement $\zeta\to i/\zeta$ with respect to the usual 2d conventions.}
\begin{equation}\label{xxxy2}\begin{split}
\Psi_N(x)^\mathrm{as}
&\sim\frac{1}{(2\pi)^{N/2}N!}\;d_N(\beta)^{-x}\, (\zeta/i)^{N(N-1) i\beta/2\zeta}\,\int dY\,e^{-\sum_i Y_i^2/2}\,\prod_{i<j}(Y_i-Y_j)^{2(i\beta/\zeta+x)}=\\
&=d_N(\beta)^{-x}\,(\zeta/i)^{N(N-1) i\beta/2\zeta}\,\prod_{k=1}^N
\frac{\Gamma\big(k \,i\beta/\zeta+k\, x\big)}{\Gamma\big(i\beta/\zeta+x\big)},
\end{split}
\end{equation}
where $d_N(\beta)$ is the value of the Szeg\"o discriminant of the vacuum configuration, eqn.\eqref{szego}, and the factor
$(\zeta/i)^{N(N-1) i\beta/2\zeta}$ arises from the field rescaling  $X_i\to Y_i\equiv\sqrt{i/\zeta}\,X_i$.
 Using the following variant of Binet formula \cite{gammafun}
\begin{equation}
\log\Gamma(z)=(z-1/2)\log z-z+\frac{1}{2}\log2\pi-\frac{1}{\pi}\int_0^\infty \frac{ds}{s^2+1}\,\log(1-e^{-2\pi z t}),
\end{equation}
we rewrite
\begin{multline}\label{xxxy3}
\log \Psi_N(x=0)^\mathrm{as}=
\sum_{k=1}^N \Big[\log\Gamma(k i\beta/\zeta) - k\frac{i\beta}{\zeta} \log(i/\zeta)-\log\Gamma( i\beta/\zeta) +\frac{i\beta}{\zeta} \log(i/\zeta)\Big]\equiv\\
\equiv\frac{i}{\zeta}\,U_N-
\frac{1}{\pi}\int_0^\infty \frac{ds}{s^2+1}\left\{\sum_{k=1}^N\log\!\big(1-e^{-2\pi k (i\beta/\zeta)s}\big)-N\log\!\big(1-e^{-2\pi (i\beta/\zeta) s}\big)\right\}+\mathrm{const},
\end{multline}
where the constant\footnote{ The constant is related to the overall normalization of the amplitude $\Psi_N$; fixing the normalizations in the standard way, also this constant matches in the asymmetric limit.} depends only on $N$.
Comparing eqns.\eqref{xxxy1} and \eqref{xxxy3} yields the identification
\begin{gather}\label{wwwcv}
\begin{aligned}
F_N(z)&=\sum_{k=1}^N\log\!\Big(1-e^{-2\pi k\,z}\Big)-N \log\!\big(1-e^{-2\pi z}\big)=\\
&=\int \log\!\big(1-e^{-2\pi y\,z}\big)\,\omega_N(y)\,dy,
\end{aligned}\\
\omega_N(y)\equiv\sum_{k=1}^N\delta(y-k)-N\,\delta(y-1),
\end{gather}
corresponding to a linear charge distribution
\begin{equation}
\rho_N(x)=N\,\delta_\mathbb{Z}(x)-\sum_{k=1}^N \delta_\mathbb{Z}(k x)\equiv -\int \delta_\mathrm{Z}(y\,x)\,\omega_N(y)\,dy,
\end{equation}
that is, to a superposition of point Abelian monopoles at the 
points $(0,0, j/k)\in\mathbb{R}^3$ of charge $-1/k$.

We note that a crucial ingredient in matching the asymmetric limit of the brane amplitude with the period integral was the identity
\begin{equation}
U_N(\beta)= \int\limits_0^1 (y\beta)\Big(\log(y\beta)-1\Big)\,\omega_N(y)\,dy,
\end{equation}
which allows to read the function $F_N(z)$
directly from the chiral ring $\mathcal{R}$ and viceversa.

\subsubsection{The brane amplitude}

Using \eqref{wwwcv} the full brane amplitude for the Gaussian matrix LG model becomes
\begin{multline}\label{fullGaussianamplitude}
\log\Psi_N= \frac{i}{\zeta}\,U_N-i\,\zeta\,\overline{U}_N-\\
-\frac{1}{2\pi i}\int_\ell \frac{ds}{s}\,\frac{\zeta}{s-\zeta}
\left\{\sum_{k=1}^N\log\!\Big(1-e^{-2\pi k(\beta/s-ix+\bar\beta s)}\Big)-N \log\!\big(1-e^{-2\pi (\beta/s-ix+\bar\beta s)}\big)\right\} -\\
-\frac{1}{2\pi i}\int_\ell \frac{ds}{s}\,\frac{\zeta}{s+\zeta}
\left\{\sum_{k=1}^N\log\!\Big(1-e^{-2\pi k(\beta/s+ix+\bar\beta s)}\Big)-N \log\!\big(1-e^{-2\pi (\beta/s+ix+\bar\beta s)}\big)\right\} 
\end{multline}
This expression is reminiscent of the TBA equations of \cite{Gaiotto:2008cd}. The precise relation of the brane amplitude
\eqref{fullGaussianamplitude} with the TBA equations for 4d SQED will be discussed in section
\ref{s:twistopstrings} below. 

\subsubsection{Interlude: The twistorial Gamma function}\label{twistgammafun}

The brane amplitudes for an interesting class of $tt^*$ geometries may all be expressed in terms of a single new transcendent function which we call the \emph{twistorial} Gamma function.
The twistorial Gamma function, $\Gamma(\mu,\bar\mu,x)$, is
defined as the analytic continuation to $\mu$, $\bar\mu$ independent complex variables of the $\zeta=i$ brane amplitude for the Abelian $tt^*$ geometry associated to the charge distribution
$\rho(x)=1-\delta_\mathbb{Z}(x)$, i.e.\! to the periodic linear array of Abelian monopoles, eqn.\eqref{wellknownide}. Explicitly (assuming $\mathrm{Re}\,\mu>0$ and $\mathrm{Re}\,\bar\mu>0$)
\begin{multline}\label{twistgamma}
\log\!\left(\frac{\Gamma(\mu,\bar\mu,x)}{\sqrt{2\pi}}\right)=\mu(\log\mu-1)+\bar\mu(\log\bar\mu-1)-\\
-\frac{1}{2\pi}\int\limits_0^\infty \frac{dt}{t(t-i)}\log\!\Big(1-e^{-2\pi(\mu/t-ix+\bar\mu t)}\Big)
- \frac{1}{2\pi}\int\limits_0^\infty \frac{dt}{t(t+i)}\log\!\Big(1-e^{-2\pi(\mu/t+ix+\bar\mu t)}\Big)
\end{multline}
$\Gamma(\mu,\bar\mu,x)$ is a three--variable extension of the Euler Gamma function. It reduces to it in various limits; moreover,
for each functional equation satisfied by the classical Gamma function, there is an analogue identity for its twistorial extension. 
These identities are most conveniently expressed in term of the related function
\begin{equation}\label{whatLambda}
\Lambda(\mu,\bar\mu,x)= \exp\!\Big[-(1/2-x)\log\mu\Big]\;\Gamma(\mu,\bar\mu,x).
\end{equation} 
To the difference equation $\Gamma(z+1)=z\,\Gamma(z)$ it corresponds the identity
\begin{equation}
\Lambda(\mu,\bar\mu, x+1)= \mu\, \Lambda(\mu,\bar\mu,x).
\end{equation}
The reflection property $\Gamma(z)\,\Gamma(1-z)= \pi/\sin(\pi z)$
generalizes to
\begin{equation}
\Lambda(\mu,\bar\mu, x)\; \Lambda(-\mu,-\bar\mu,1-x) = \frac{\pi}{\sin\!\big(\pi (\mu+x-\bar\mu)\big)}.
\end{equation}
Finally, the Gauss product formula ($n$ is any natural number)
\begin{equation}
\Gamma(n z)=(2\pi)^{(1-n)/2}\, n^{nz-1/2}\,\prod_{k=0}^{n-1}\Gamma(z+k/n)
\end{equation}
extends to the product formula
\begin{equation}
\Lambda(n\mu,n\bar\mu,n x)=(2\pi)^{(1-n)/2}\; n^{n(\mu+x+\bar\mu)-1/2}\;\prod_{k=1}^{n-1}\Lambda(\mu,\bar\mu,
x+k/n).
\end{equation}
The relation of $\Gamma(\mu,\bar\mu,x)$ with the classical Gamma function is manifest in its two asymmetric limits
(we assume $0<x<1$)
\begin{align}
\Lambda(\mu,0,x)&=\Gamma\big(\mu+x\big)\\
\phantom{\bigg|}\Lambda(0,\bar\mu,x)&=\frac{\Gamma(x)}{\Gamma(1-x)}\;\Gamma(1-x+\bar\mu).
\end{align}
The Weierstrass infinite product representation of the Gamma function
\begin{equation}
\label{weirss2}
 \Gamma(z)^{-1}=z\, e^{\gamma\,z}\prod^{\infty}_{k=1}
\left( 1+\frac{z}{k} \right)e^{-z/k}
\end{equation}
has the twistorial extension (for $\mathrm{Re}\,\mu>0$, $\mathrm{Re}\,\bar\mu>0$ and $0<x<1$)
\begin{equation}\label{theorem}
\begin{split}
 \Gamma(\mu,\bar\mu,x)=&\mu^{1/2-x}\;\frac{2\,\Gamma(1-x-\mu+\bar\mu)}{x+2\mu+\sqrt{x^2+2\mu\bar\mu}}\;e^{-2\gamma(x+\mu)}\;\times\\
&\times\prod_{k\geq 1}\left\{\frac{1-\frac{x+2\mu}{k}+\sqrt{\left(1-\frac{x}{k}\right)^2+
4\,\frac{\mu\bar\mu}{k^2}}}{1+\frac{x+2\mu}{k}+\sqrt{\left(1+\frac{x}{k}\right)^2+4\,\frac{\mu\bar\mu}{k^2}}}\;e^{2(x+\mu)/k}\right\},
\end{split}
\end{equation}
which may be taken as the definition of the function. The properties listed above easily follow from this representation.

\subsubsection{The brane amplitude in terms of twistorial Gamma functions}\label{s:thumbrule}

Our findings for the $tt^*$ Lax equations of the Gaussian model may be summarized in a simple `rule of thumb': to get the exact brane amplitude, take the
period integral expressing the asymmetric limit of the brane amplitude, eqn.\eqref{xxxy2}, and simply replace each Euler Gamma function in the \textsc{rhs} by its twistorial version\footnote{ \label{fooor}For $\zeta\neq i$ the twistorial amplitude with the normalization in eqn.\eqref{fullGaussianamplitude} differs from the `rule of thumb' one by the trivial factor $\exp[iU_N(\beta)/\zeta-U_N(i\beta/\zeta)-i\overline{U_N}(\bar\beta)\zeta-\overline{U_N}(-i\bar\beta\zeta)]$, cfr.\! eqn.\eqref{xxxy3}. }
\begin{equation}
\Gamma(k i\beta/\zeta+k x)\rightsquigarrow \Gamma(k i\beta/\zeta,-k i \zeta\,\bar\beta, k x).
\end{equation}
We shall see in the remaining part of this section that this `rule of thumb' works for a larger class of $tt^*$ geometries.

\subsection{Abelian $tt^*$ geometries in $(\mathbb{R}^2\times S^1)^r$}\label{ss:abelian}

The Gaussian matrix model of section \ref{s:gaussianmodel}
is only the first instance in a large class of multi--matrix $(2,2)$ models for which the exact $tt^*$ brane amplitudes may be computed in closed form. In this and the next sections we give some further examples,
deferring to appendix \ref{appendixmultimatrix} the discussion of an even larger class of solvable $tt^*$ geometries describing twistorial extensions of Toda theories.  All these models are characterized
by the condition that they have no magnetic BPS states.

$tt^*$ geometry \cite{Cecotti:1991me,Cecotti:2013mba} states that, for a four--supercharge theory with $m$ susy vacua, the vacuum Berry connection is an $U(m)$ 
hyperholomorphic connection
on some hyperK\"ahler manifold, possibly dimensionally reduced along the orbits of an isometry group. 
For the present class of LG models, eqn.\eqref{whichmodels}, freezing all couplings in $W(z)$,
the $tt^*$ Berry connection, as a function of the coupling $\beta$ in front the log of the Vandermonde determinant
and its associated vacuum angle $\theta$ \cite{Cecotti:1991me,Cecotti:2010fi,Cecotti:2013mba},
is an $U(m)$ Bogomolnyi monopole in $\mathbb{R}^2\times S^1$ \cite{Cecotti:2013mba}. Taking into account the dependence on the couplings in $W(z)$ (and related angles),
we extend the Bogomolyi monopole to a higher dimensional hyperholomorphic $U(m)$ `monopole' \cite{Cecotti:2013mba}.
For the models with $m=1$ the $tt^*$
monopole is Abelian, and the $tt^*$ equations become linear, hence explicitly solvable. 

In view of this fact, it is interesting to classify all functions $W(z)$ such that the corresponding model
\eqref{whichmodels} has a single susy vacuum \textit{for all} $N\in\mathbb{N}$.
Limiting ourselves to the case of $W^\prime(z)$ rational, we see from the  discussion in \S.\ref{HSandvanvleck} that 
this requires the unique van Vleck polynomial $f(z)$ to have degree zero, and hence to be the eigenvalue of
a Schroedinger Hamiltonian. Then the requirement that there is a unique susy vacuum for all $N$ is equivalent to the
requirement that the Schroedinger Hamiltonian has a complete system of polynomial eigenfunctions. 
By a theorem of Bochner \cite{bochner} there are just three such Schoedinger operators, whose Heine polynomials are respectively the Hermite, the Laguerre, and Jacobi polynomials\footnote{ Other classical orthogonal polynomials, such as the Chebyshev, Legendre and Gegenbauer ones, are special instances of the Jacobi polynomials.}. 
Although this result is classical, it is instructive to look at it from the viewpoint of
the class $\mathcal{S}[A_1]$ theories \cite{Gaiotto}. To each LG model \eqref{whichmodels} with $W^\prime(z)$ rational we may associate the
$\mathcal{S}[A_1]$ theory enginereed on the sphere
by the quadratic differential
\begin{equation}
\phi_2(z)=\Big({W^\prime}^{\,2}+\text{lower order}\Big)dz^2,
\end{equation}
 (see also \cite{Gaiotto:2014bza}). The three Abelian $tt^*$ geometries then correspond to the three \emph{free} $\mathcal{S}[A_1]$ theories.
In the language of complete theories \cite{Cecotti:2011rv}, and ideal triangulations of marked surfaces \cite{fomin}, they correspond to\footnote{ Note that all three models may be obtained from $T_2$ by taking suitable limits. Colliding two of the three regular singularities
of $T_2$ we get a degree 4 irregular singularity \textit{i.e.}\! the $D_2$ model; colliding all three regular singularities we get a degree 6 irregular one, \textit{i.e.}\! the $A_1$ model. The Laguerre and Hermite polynomials are obtained from the Jacobi ones in the corresponding limits.}
\begin{itemize}
 \item[a)] a disk with four marked points on the boundary, whose triangulation quiver is the $A_1$ Dynkin graph,
which corresponds to a free hypermultiplet with flavor symmetry $SU(2)$;
\item[b)] a disk with one regular puncture and two marked points on the boundary, whose triangulation quiver is the $D_2$ (disconnected)
Dynkin graph, corresponding to a free hypermultiplet doublet with flavor symmetry $SU(2)\times SU(2)$;
\item[c)] a sphere with three regular punctures (the $T_2$ theory) which corresponds to a free \emph{half}--hypermultiplet
in the $\tfrac{1}{2}(\mathbf{2},\mathbf{2},\mathbf{2})$ representation of the flavor $SU(2)\times SU(2)\times SU(2)$ group.
\end{itemize}

The $tt^*$ Abelian geometries associated to the free $\mathcal{S}[A_1]$ theories
are higher--dimensional hyperholomorphic `monopoles' in the space $(\mathbb{R}^2\times S^1)^r$,
where $r$ is the number of $SU(2)$ factors in their flavor group. The three real coordinates associated to each $SU(2)$ factor
are given by the complex coupling $\mu_i$ in front of the corresponding logarithmic term in $\mathcal{W}$ and its associated vacuum angle
$\theta_i$.   

To each one of the three Abelian $tt^*$ geometries there is associated yet another remarkable mathematical structure,
namely a Selberg integral whose evaluation yields a product of Gamma functions \cite{selberg1,selberg2,selberg3}. While deep relations between all these structures are well known in the mathematical literature (see \textit{e.g.}\! chapter 8 of \cite{selberg1}), $tt^*$ geometry makes their mutual connections more transparent and natural.

We summarize the several structures in the table: 
\begin{center}
\vskip-12pt
\begin{tabular}{c|c|c|c|c|c}\hline
model & $W(z)$ & polyn. & $r$ & $\mathcal{S}[A_1]$ & Selberg int.\\\hline\hline
Gaussian & $-z^2/2$ & Hermite & 1 & $A_1$ & $A_{N-1}$ Metha\\
Generalized Penner & $\mu \log z-\lambda z$ & Laguerre & 2 
& $D_2$ & $BC_N$ Metha\\
Double Penner & $\mu_1\log z+\mu_2\log(z-\rho)$ & Jacobi & 3
& $T_2$ & Morris\\\hline\hline
\end{tabular}
\end{center}

\subsubsection{Solving Abelian $tt^*$ equations in $(\mathbb{R}^2\times S^1)^r$}\label{SS:solvingabelian}

We consider a general Abelian $tt^*$ geometry in $(\mathbb{R}^2\times S^1)^r$. To fix the ideas we  focus on a LG model defined by a family of rational one--forms $\lambda(\mu_a)$ in $\mathbb{C}^N$, which are
invariant under a discrete group $\Gamma$ of symmetries of $\mathbb{C}^N$, and whose zeros form a \emph{single} orbit of $\Gamma$.
The form $\lambda(\mu_a)$ are parametrized by the independent
periods $2\pi i \mu_1,2\pi i \mu_2,\dots, 2\pi i \mu_r$ of $\lambda(\mu_a)$ (cfr.\! eqn.\eqref{genericWm}). Then the LG model with superpotential the
multi--valued function 
\begin{equation}
\mathcal{W}(X_i,\mu_a)= \int^{X_i} \lambda(\mu_a),\end{equation} 
restricted to the $\Gamma$--invariant sector,
defines an Abelian $tt^*$ geometry on the space $(\mathbb{R}^2\times S^1)^r$ coordinatized by the reduced periods $\boldsymbol{\mu}\equiv(\mu_1,\cdots,\mu_r)$ and angles $\boldsymbol{\theta}/2\pi \overset{\text{def}}{=}\boldsymbol{x}\equiv(x_1,\dots, x_r)$. 
The rescaled coordinates $4\pi i\, \mu_a$ are just the central charges of the BPS solitons wrapping the generators $\{\gamma_a\}$ of the field space homology group
$$H_1\!\big((\mathbb{C}^N\setminus\{\text{polar locus}\})/\Gamma, \mathbb{Z}\big).$$
The $tt^*$ metric $G(\boldsymbol{\mu}, \boldsymbol{\bar\mu},\boldsymbol{x})$
satisfies the equations
\begin{gather}\label{hh1}
\partial_{\mu_a}\partial_{x_b}\log G(\boldsymbol{\mu}, \boldsymbol{\bar\mu},\boldsymbol{x})=
\partial_{\mu_b}\partial_{x_a}\log G(\boldsymbol{\mu}, \boldsymbol{\bar\mu},\boldsymbol{x})\\
\label{hh2}\partial_{\bar\mu_a}\partial_{x_b}\log G(\boldsymbol{\mu}, \boldsymbol{\bar\mu},\boldsymbol{x})=
\partial_{\bar\mu_b}\partial_{x_a}\log G(\boldsymbol{\mu}, \boldsymbol{\bar\mu},\boldsymbol{x})\\
\label{hh3}\Big(\partial_{\bar\mu_a}\partial_{\mu_b}+\partial_{x_a}\partial_{x_b}\Big)\log G(\boldsymbol{\mu}, \boldsymbol{\bar\mu},\boldsymbol{x})=0,
\end{gather}
for all $a,b=1,2,\dots,r$.
 Using periodicity in the $x_a$, invariance under overall $U(1)$ rotations of the $\mu_a$, and (generic) decoupling at infinite mass $|\mu_a|$, we end up with solutions of the form
\begin{equation}\label{rrraq}
\log G(\boldsymbol{\mu}, \boldsymbol{\bar\mu},\boldsymbol{x})=
\sum_{\boldsymbol{m}\in\mathbb{Z}^r\atop \boldsymbol{m}\neq 0} 
A(\boldsymbol{m})\; K_0\Big(4\pi \sqrt{(\boldsymbol{m}\cdot \boldsymbol{\mu})(\boldsymbol{m}\cdot\boldsymbol{\bar\mu})}\,\Big)\, \exp\!\big(2\pi i\, \boldsymbol{m}\cdot\boldsymbol{x}\big),
\end{equation}
for some numerical coefficients $A(\boldsymbol{m})$ which are further restricted by the $tt^*$ reality condition \cite{Cecotti:1991me}
\begin{equation}
A(-\boldsymbol{m})=-A(\boldsymbol{m}),\qquad A(\boldsymbol{m})\equiv  i\,a(\boldsymbol{m})\in 
i\,\mathbb{R}.
\end{equation}
\paragraph{Sign--coherence.} In all known examples, the coefficients $A(\boldsymbol{m})$
enjoy a sign--coherence property:  there exists a natural basis of periods, $\{\gamma_a\}$, such that $A(\boldsymbol{m})=0$ unless the components of the $r$--vector $\boldsymbol{m}$ are all non--negative integers or all non--positive integers. As we shall argue in section \ref{s:twistopstrings}, this property arises from 4d: it is related to sign--coherence of BPS spectra of 4d $\mathcal{N}=2$ theories which have a quiver description \cite{Cecotti:2011rv}. In the sign--coherent case we may summarize the real coefficients $a(\boldsymbol{m})$ in a
magnetic charge function $F(\boldsymbol{z})$ which generalizes the one defined in eqn.\eqref{magneticchargefunction} for the Gaussian model, i.e.
\begin{equation}\label{genmagnfunction}
F(\boldsymbol{z})=-\pi \sum_{\boldsymbol{m}\in \mathbb{Z}_+^r} a(\boldsymbol{m})\;e^{-2\pi \,\boldsymbol{m}\cdot\boldsymbol{z}}.
\end{equation} 
We may repeat word--for--word the analysis for the Gaussian case, with the result that the exact brane amplitude $\Psi(\boldsymbol{\mu},\boldsymbol{\bar\mu}, \boldsymbol{x},\zeta)$ (a section of the $tt^*$ hyperholomorphic line bundle $\mathcal{L}$ which is holomorphic in complex structure $\zeta$)  has the form\footnote{ This is correct in a region where the complex numbers $\mu_a$, $\bar\mu_a$ all belong to the same half--plane, so that the integral of all terms in the sum defining $F(\boldsymbol{z})$ converge for the \emph{same} choice of ray $\ell\subset\mathbb{C}$. In the general case one needs to choose \emph{different} rays $\ell$ for each term in the sum \cite{Cecotti:2013mba}. The implied deformation of the contours introduces various Stokes factors from the residues at poles (cfr.\! eqn.\eqref{stokesgauss}). For simplicity we write expressions valid for $\mathrm{Re}\,\mu_a>0$, $\mathrm{Re}\,\bar\mu_a>0$, the general case being obtained by analytic continuation and multiplication by the appropriate Stokes factors.  }
\begin{equation}
\begin{split}\label{aaaab1}
\log \Psi(\boldsymbol{\mu},\boldsymbol{\bar\mu}, \boldsymbol{x},\zeta)=
\frac{i}{\zeta}\,U(\boldsymbol{\mu})&-i\zeta\, \overline{U}(\boldsymbol{\bar\mu})-\frac{1}{2\pi i}\int_\ell  \frac{ds}{s}\,
\frac{\zeta}{s-\zeta}\, F\big(\boldsymbol{\mu}/s- i\boldsymbol{x}+\boldsymbol{\bar\mu}\,s\big)-\\
&-\frac{1}{2\pi i}\int_\ell  \frac{ds}{s}\,
\frac{\zeta}{s+\zeta}\, F\big(\boldsymbol{\mu}/s+ i\boldsymbol{x}+\boldsymbol{\bar\mu}\,s\big),
\end{split}
\end{equation}
where $U(\boldsymbol{\mu})$ is the value\footnote{ The value is not uniquely defined; varying $U(\boldsymbol{\mu})$ is equivalent to changing the trivialization of $\mathcal{L}$. However there is a physically natural trivialization, and hence a natural $U(\boldsymbol{\mu})$.} of the superpotential $\mathcal{W}(X_i,\boldsymbol{\mu})$ at the zero of $\lambda(\boldsymbol{\mu})$. 
We also write the $tt^*$ metric $G(\boldsymbol{\mu},\boldsymbol{\bar\mu}, \boldsymbol{x})$, eqn.\eqref{rrraq}, and the CFIV index
$Q(\boldsymbol{\mu},\boldsymbol{\bar\mu}, \boldsymbol{x})$
\cite{Cecotti:1992qh}
of a general Abelian $tt^*$ geometry in terms of
the magnetic charge function $F(\boldsymbol{z})$
\begin{gather}\label{aaaab2}
\log G(\boldsymbol{\mu},\boldsymbol{\bar\mu}, \boldsymbol{x})=
\frac{1}{2\pi i}\int_\ell \frac{ds}{s}\,\Big[F\big(\boldsymbol{\mu}/s- i\boldsymbol{x}+\boldsymbol{\bar\mu}\,s\big)-F\big(\boldsymbol{\mu}/s+ i\boldsymbol{x}+\boldsymbol{\bar\mu}\,s\big)\Big]\\
Q(\boldsymbol{\mu},\boldsymbol{\bar\mu}, \boldsymbol{x})=
-\frac{1}{2\pi i}\int_\ell \frac{ds}{s^2}\;\mu_a\Big[ F_a\big(\boldsymbol{\mu}/s- i\boldsymbol{x}+\boldsymbol{\bar\mu}\,s\big)-F_a\big(\boldsymbol{\mu}/s+ i\boldsymbol{x}+\boldsymbol{\bar\mu}\,s\big)\Big]\label{aaaab3}\\
\text{where}\qquad F_a(\boldsymbol{z})= 2\pi^2\,\sum_{\boldsymbol{m}\in \mathbb{Z}_+^r} m_a\,a(\boldsymbol{m})\;e^{-2\pi \,\boldsymbol{m}\cdot\boldsymbol{z}} 
\end{gather}
These quantities satisfy the two basic $tt^*$
bilinear relations (cfr.\! eqn.(4.19) of ref.\cite{Cecotti:1992rm}) which in the present situation reduce to
\begin{gather}\label{bil1}
\Psi(\boldsymbol{x},\zeta)\;\Psi(-\boldsymbol{x},-\zeta)=1\\
\overline{\Psi(-\boldsymbol{x},1/\bar\zeta)}=G(-\boldsymbol{x})\,\Psi(\boldsymbol{x},\zeta).\label{bil2}
\end{gather}

It remains to determine the Fourier coefficients $a(\boldsymbol{m})$, or equivalently the magnetic function $F(\boldsymbol{z})$. $tt^*$ theory gives at least four different ways to fix them:
\begin{itemize}
\item[a)] from the IR asymptotics of the CFIV index \cite{Cecotti:1992qh}. The real number $-\pi\,a(\boldsymbol{m})$ gets  identified with the (net signed) number of BPS solitons with central charge
$4\pi i\, \boldsymbol{m}\cdot\boldsymbol{\mu}$, i.e.\! solitons wrapping the homological cycle $\boldsymbol{m}\cdot\boldsymbol{\gamma}$. One should count BPS solitons with collinear central charges with the appropriate weight (see \cite{Cecotti:2010qn} appendix). In particular, $\pi\gcd(\boldsymbol{m})\,a(\boldsymbol{m})$ should be an integer.
\item[b)] from the UV limit of the CFIV index which expresses
the UV dimensions $h(\boldsymbol{x})$ of the chiral primaries \cite{Cecotti:1991me,Cecotti:1992qh}. These dimensions are constrained by the chiral ring $\mathcal{R}$ to be piecewise linear functions of the
$\boldsymbol{x}$'s. This implies $a(\boldsymbol{m})=O(1/|\boldsymbol{m}|)$ in agreement with the IR predictions.
The discontinuities of $h(\boldsymbol{x})$ should be integers \cite{Cecotti:1992rm}, implying specific integrality properties of the $a(\boldsymbol{m})$'s.
\item[c)] matching the asymmetric limit of the brane amplitude
with the period integral, as we did for the Gaussian model. For instance, if we have
\begin{equation}\label{genasymBC}
\int e^{i\mathcal{W}/\zeta}\; d^N\!X= \begin{pmatrix}\text{elementary}\\
\text{function}\end{pmatrix}\frac{\prod_{k=1}^{n} \Gamma(i\ell_{k,a} \mu_a/\zeta)}{\prod_{h=1}^{n^\prime} \Gamma(i\tilde\ell_{h,a} \mu_a/\zeta)},
\end{equation}
we immediately conclude that the model has $n+n^\prime$
primitive BPS solitons of central charges $4\pi i\,\ell_{k,a}\mu_a$
and $-4\pi i\, \tilde\ell_{h,a}\mu_a$, respectively. In particular, 
$\ell_{k,a}$ and $\tilde \ell_{h,a}$ should be \emph{non--negative} integers corresponding to the cycles $\ell_{k,a}\gamma_a$, $-\tilde\ell_{h,a}\gamma_a$ wrapped by the \emph{primitive} BPS solitons. The BPS solitons have the sign--coherence property 
discussed around eqn.\eqref{genmagnfunction}, and the magnetic function may be read directly from the \textsc{rhs} of eqn.\eqref{genasymBC}
\begin{equation}\label{whichmagnfun}
F(\boldsymbol{z})= \sum_{k=1}^n \log(1-e^{-2\pi\, \boldsymbol{l}_k\cdot \boldsymbol{z}})-\sum_{h=1}^{n^\prime}\log(1-e^{-2\pi\, \boldsymbol{\tilde l}_h\cdot \boldsymbol{z}}).
\end{equation}
\item[d)] from the holomorphic function $U(\boldsymbol{\mu})$.
Indeed, the very fact that our $tt^*$ geometry is Abelian means that the BPS solitons do not interact, and then $U(\boldsymbol{\mu})$
should be identified with the effective twisted superpotential of a $(2,2)$ $U(1)^r$ gauge theory coupled to $n+n^\prime$ chiral superfields of
charges $\ell_{k,a}$ and $-\tilde\ell_{h,a}$, respectively. This gives
\begin{equation}\label{exxpectform}
U(\boldsymbol{\mu})=
\sum_{k=1}^n (\boldsymbol{l}_k\cdot \boldsymbol{\mu})\Big(\log (\boldsymbol{l}_k\cdot \boldsymbol{\mu})-1\Big)-
\sum_{h=1}^{n^\prime} (\boldsymbol{\tilde l}_h\cdot \boldsymbol{\mu})\Big(\log (\boldsymbol{\tilde l}_h\cdot \boldsymbol{\mu})-1\Big).
\end{equation}
Thus we may extract the charge vectors $\boldsymbol{l}_k$,
$-\boldsymbol{\tilde l}_h$ directly from the critical value of the (canonical) superpotential $U(\boldsymbol{\mu})$. $F(\boldsymbol{z})$ is then given by eqn.\eqref{whichmagnfun}.
More generally, as in eqn.\eqref{wwwcv}, we may
introduce a BPS density $\omega(\boldsymbol{y})$ such that
\begin{align}\label{bpsden1}
F(\boldsymbol{z})&=\int \log\!\big(1-e^{-2\pi \boldsymbol{y}\cdot \boldsymbol{z}}\big)\,\omega(\boldsymbol{y})\,d\boldsymbol{y},\\
U(\boldsymbol{\mu})&=\int (\boldsymbol{y}\cdot\boldsymbol{\mu})\Big(\!\log(\boldsymbol{y}\cdot\boldsymbol{\mu})-1\Big)\,\omega(\boldsymbol{y})\,d\boldsymbol{y}.\label{bpsden2}
\end{align}
\end{itemize}

The fact that one gets the same coefficients $a(\boldsymbol{m})$ by using any one of the above four methods is a non--trivial check of the correctness of the procedure.

The `rule of thumb' of \S.\ref{s:thumbrule} gives the \emph{exact} brane amplitude corresponding to the boundary condition \eqref{genasymBC} as a product of twistorial Gamma functions\footnote{ Again, this expression differs from the brane amplitude in the standard normalization, eqn.\eqref{aaaab1}, by the trivial factors discussed in footnote \ref{fooor}.}
\begin{equation}\label{braneamtwist}
\Psi(\boldsymbol{\mu},\boldsymbol{\bar\mu}, \boldsymbol{x},\zeta)=
\prod_{k=1}^n \Gamma\big(i \boldsymbol{\ell}_k\cdot\boldsymbol{\mu}/\zeta,-i \boldsymbol{\ell}_k\cdot\boldsymbol{\bar\mu}\,\zeta,
\boldsymbol{\ell}_k\cdot\boldsymbol{x}\big)
\prod_{h=1}^{n^\prime} \Gamma\big(i \boldsymbol{\tilde\ell}_h\cdot\boldsymbol{\mu}/\zeta,-i \boldsymbol{\tilde\ell}_h\cdot\boldsymbol{\bar\mu}\,\zeta,
\boldsymbol{\tilde\ell}_h\cdot\boldsymbol{x}\big)^{-1},
\end{equation}
(modulo, possibly, rational shifts of $\boldsymbol{x}$, which
correspond to insertions of chiral primaries in the period integral).

A priori, extracting the soliton charges $\boldsymbol{l}_k$,
$-\boldsymbol{\tilde l}_h$ from the integral \eqref{genasymBC} may be a bit ambiguous since, using the functional identities for the Euler
Gamma function, we may write the \textsc{rhs} as a product of Gamma functions in many different ways (allowing also for rational shifts of the arguments). However, since the twistorial Gamma function satisfies the same identities as the Euler one, the brane amplitude is well defined, independently of how we write the \textsc{rhs} of \eqref{genasymBC}. 

\subsubsection{Constructing new Abelian $tt^*$ geometries from old ones}\label{s:fromoldtonew}

Our explicit computations of twistorial brane amplitudes
in the examples of section \ref{s:twistopstrings} consist of several steps in which one constructs a sequence of $tt^*$ geometries one after another. For instance, to get the brane amplitudes for SQED, we first solve the $tt^*$ equations for the Gaussian model at finite $N$; for each $N$ we get a one--dimensional\footnote{ In the quaternionic sense.} Abelian $tt^*$ geometry, which depends on the coordinates $(\epsilon_1,\theta)$. This yields the SQED amplitude at special discrete loci in its parameter space
\begin{equation}\label{discreteloci}a_e=(N+1/2)\epsilon_1\quad \text{and}\quad \theta_e=(N+1/2)\theta.\end{equation} 
The next task is to extend this discrete family of one--dimensional $tt^*$ geometries to a sound $tt^*$ geometry in one more quaternionic dimension, depending on continuous coordinates $(a_e,\theta_e, \epsilon_1,\theta)$, such that its restriction to the locus \eqref{discreteloci} reproduce the $tt^*$ geometry of the Gaussian model of size $N$. Here one needs appropriate techniques to construct the higher--dimensional  $tt^*$ geometry from the 
smaller ones. Having done that, the next task is to consider various physically interesting limits of the result. These limits should produce full (regular) limiting $tt^*$ geometries with sound metrics, indices, and brane amplitudes.
Again, we need appropriate techniques.

Constructing new $tt^*$ geometries out of old ones to serve as physically natural generalizations/limits is a formidable task.
However, if the $tt^*$ geometries are Abelian, it can be done explicitly with the help of the magnetic charge function $F(\boldsymbol{z})$ introduced in the previous subsection.
The magnetic charge function $F(\boldsymbol{z})$, being directly related to the charge distribution of the $tt^*$ higher--dimensional monopole \cite{Cecotti:2013mba}, is
a gauge--invariant datum of the $tt^*$ geometry, i.e.\! a totally unambiguous way of describing all $tt^*$ quantities. 
$F(\boldsymbol{z})$ is also the datum on which we have the best physical control: from eqn.\eqref{aaaab1} we see that $F(\boldsymbol{z})$ is simply the
logarithm of the Stokes jump of the (Abelian) $tt^*$
amplitude which is given by well understood wall crossing formulae.  The datum $F(\boldsymbol{z})$ determines  the $tt^*$ metric and CFIV index by eqns.\eqref{aaaab2} and \eqref{aaaab3}.
In addition, from eqns.\eqref{whichmagnfun} and \eqref{exxpectform} we see that from $F(\boldsymbol{z})$ we may also reconstruct the
`semi--flat' holomorphic function $U(\boldsymbol{\mu})$. Concretely,
to give $F(\boldsymbol{z})$ is equivalent to specifying the 2d BPS charge distribution $\omega(\boldsymbol{y})$ which describes the 2d Stokes jumps, cfr.\! eqn.\eqref{bpsden1}
which determines the function $U(\boldsymbol{\mu})$
\begin{equation}
U(\boldsymbol{\mu})= \int d\boldsymbol{w}\; \omega(\boldsymbol{w})\, \big(\boldsymbol{w}\cdot\boldsymbol{\mu}\big)\Big(\!\log\big(\boldsymbol{w}\cdot\boldsymbol{\mu}\big)-1 \Big).
\end{equation} 
Plugging this expression in eqn.\eqref{aaaab1}, we write the (Abelian) twistorial brane amplitude corresponding to the given datum $F(\boldsymbol{z})$. 

In section \ref{s:twistopstrings} we shall use this strategy to get the general SQED $tt^*$ geometry away from the special locus \eqref{discreteloci} and to define some of its interesting limits.

\subsection{Example: the generalized Penner model}

The $(2,2)$ generalized Penner model is given by the superpotential
\begin{equation}
\mathcal{W}(e_k)=\sum_{i=1}^N\Big(-X_i+\mu_1\log X_i\Big)+\mu_2\,\log\!\!\!\!\prod_{1\leq i<j\leq N}(X_i-X_j)^2
\end{equation}
where the basic chiral fields are the elementary symmetric
polynomials $e_k$ and $\mu_1\neq 0$. The two complex couplings, $\mu_1$ and $\mu_2$ are the
independent residues of the rational one--form $\lambda(\boldsymbol{\mu})=d\mathcal{W}$. 
From the discussion in \S\S.\ref{chiringva}, \ref{HSandvanvleck} and \ref{ss:abelian}
we know that $\lambda(\boldsymbol{\mu})$ has a unique simple zero (modulo $\mathfrak{S}_N$) specified by the $N$--th associated Laguerre polynomial
\begin{equation}
P_N(z)=(-1)^N\, N!\, \mu_2^N\; L_N^{(\mu_1/\mu_2-1)}\!(z/\mu_2).
\end{equation}
The generalized Penner model corresponds to an Abelian
$tt^*$ geometry in $(\mathbb{R}^2\times S^1)^2$ to which our results apply. To get the explicit expressions one has only to specify the magnetic function $F_N(\mu_1,\mu_2)$ which may be determined by any one of the four methods of the previous section. Conceptually, the simplest one is from the effective twisted superpotential $U(\mu_1,\mu_2)$ which is defined by the two equations
\begin{align}
\partial_{\mu_1}U(\mu_1,\mu_2)&= \log\!\big[ (-1)^N P_N(0)\big]
\equiv \sum_{k=0}^{N-1}\log(\mu_1+ k\mu_2) \\
\partial_{\mu_2}U(\mu_1,\mu_2)&= \log \mathrm{Discr}\, P_N(z)\equiv
\sum_{k=1}^N\Big(k \log(k \mu_2)-\log\mu_2\Big)+\sum_{k=1}^{N-1} k\,\log(\mu_1+k \mu_2),
\end{align}
where we used known properties of the Laguerre polynomials \cite{szego}.
$U(\mu_1,\mu_2)$ has the expected form, eqn.\eqref{exxpectform}, to be interpreted as the effective twisted 
superpotential of an Abelian gauge theory 
\begin{multline}
U(\mu_1,\mu_2)=\sum_{k=1}^N\Big\{(k\mu_2)\big(\log(k\mu_2)-1\big)-\mu_2\big(\log\mu_2-1\big)\Big\}+\\
+\sum_{k=0}^{N-1}(\mu_1+k\mu_2)\big(\log(\mu_1+k\mu_2)-1\big).
\end{multline}
Therefore $U(\mu_1,\mu_2)$ defines a magnetic function
\begin{equation}
F(z_1,z_2)=\sum_{k=1}^N\Big\{\!\log\!\Big(1-e^{-2\pi k z_2}\Big)-\log\!\Big(1-e^{-2\pi z_2}\Big)\!\Big\}
+\sum_{k=0}^{N-1}\log\!\Big(1-e^{-2\pi(z_1+k z_2)}\Big).
\end{equation}
Replacing this magnetic function in eqns.\eqref{aaaab1},
\eqref{aaaab2}, and \eqref{aaaab3}, we get the various $tt^*$
quantities for the generalized Penner model. Using eqn.\eqref{braneamtwist}, we may write the brane amplitude in closed form as a product of twistorial Gamma functions
\begin{equation}\label{eeqqrry}
\Psi=\prod_{k=1}^N \frac{\Gamma(k\mu_2,k\bar\mu_2, kx_2)}{\Gamma(\mu_2,\bar\mu_2,x_2)}\prod_{k=0}^{N-1}\Gamma\big(\mu_1+k\mu_2,\bar\mu_1+k\bar\mu_2,x_1+kx_2\big).
\end{equation}
Alternatively, we arrive at the same conclusion using the
period integral which in this case is the $BC_N$ Metha integral \cite{selberg2,selberg3}
\begin{equation}\label{wwww123}
\begin{split}
\frac{1}{N!}\int \prod_i&\left(\frac{dx_i}{\sqrt{2\pi}}e^{-x_i^2/2}\big(2|x_i|^2\big)^{\mu_1}\right)\prod_{i<j}|x_i^2-x_j^2|^{2\mu_2}=\\
&\qquad\quad=\prod_{k=1}^N\frac{\Gamma(k \mu_2)}{\Gamma(\mu_2)}
\;\prod_{k=1}^{N-1}
\frac{\Gamma(1+2\mu_1+2k \mu_2)}{\Gamma(1+\mu_1+k \mu_2)},
\end{split}
\end{equation}
which corresponds to \eqref{eeqqrry} after keeping into account the  Jacobian factor (which shifts $x_1$ by $1/2$ and $x_2$ by an integer). 

\subsection{Example: the Double Penner model}\label{s:doublepenner}

The superpotential for the double Penner (matrix) LG model
 is
\begin{equation}
\mathcal{W}(X_i,\boldsymbol{\mu})=\mu_1\sum_{i=1}^N \log X_i+\mu_2\sum_{i=1}^N \log(1-X_i)+\mu_3\!\!\!\!\sum_{1\leq i<j\leq N}\log(X_i-X_j)^2,
\end{equation}
where field configurations are identified modulo permutations of the $X_i$'s. The polynomial $P(z)$ describing the vacuum configuration satisfies the hypergeometric
ODE
\begin{equation}
\mu_3\,P^{\prime\prime}+\left(\frac{\mu_1}{z}+\frac{\mu_2}{z-1}\right)P^\prime-\frac{N(\mu_1+\mu_2+(N-1)\mu_3)}{z(z-1)}P=0,
\end{equation}
whose solution is the $N$--th Jacobi polynomial of argument $1-2z$ and parameters
$\alpha=\mu_1/\mu_3-1$, $\beta=\mu_2/\mu_3-1$, normalized 
to be monic, that is, (\cite{selberg1},\S.8.5)
\begin{equation}
P(z)=\frac{(-1)^N (\mu_1/\mu_3)_N}{(N+\mu_1/\mu_3+\mu_2/\mu_3-1)_N}\;\, P^{(\mu_1/\mu_3-1,\,\mu_2/\mu_3-1)}_N(1-2 z)
\end{equation}
where $(a)_N$ is the Pochhammer symbol
\begin{equation}
(a)_N=a(a+1)(a+2)\cdots(a+N-1)=
\frac{\Gamma(a+N)}{\Gamma(a)}.
\end{equation}
By definition one has
\begin{equation}
U(\boldsymbol{\mu})=\mu_1\log\!\big[(-1)^N P(0)\big]
+\mu_2\log\!\big[P(1)\big]+\mu_3\log\mathrm{Discr}\,P(z).
\end{equation}
Using the properties of the Jacobi polynomials 
(cfr.\cite{selberg1}, \textbf{Theorem 8.5.2}) we get 
\begin{equation}\begin{split}
U(\boldsymbol{\mu})&=\sum_{k=1}^N \bigg\{\big(\mu_1+(k-1)\mu_3\big)\Big(\!\log\!\big(\mu_1+(k-1)\mu_3\big)-1\Big)+\\
&+\big(\mu_2+(k-1)\mu_3\big)\Big(\!\log\!\big(\mu_2+(k-1)\mu_3\big)-1\Big)-\\
&-\big(\mu_1+\mu_2+(N+k-2)\mu_3\big)\Big(\!\log\!\big(\mu_1+\mu_2+(N+k-2)\mu_3\big)-1\Big)+\\
&+
(k\mu_3)\Big(\!\log(k\mu_3)-1\Big)-\mu_3\Big(\!\log\mu_3-1\Big)\bigg\}
\end{split}\end{equation}
which again has the proper form to be interpreted as an effective twisted superpotential. The corresponding magnetic function $F(\boldsymbol{z})$ is
\begin{multline}
F(\boldsymbol{z})=\sum_{k=1}^N\bigg\{\log\!\big(1-e^{-2\pi(z_1+(k-1)z_3)}\big)+\log\!\big(1-e^{-2\pi(z_2+(k-1)z_3)}\big)+\\
+\log\!\big(1-e^{-2\pi k z_3}\big)
-\log\!\big(1-e^{-2\pi(z_1+z_2+(N+k-2)z_3)}\big)-
\log\!\big(1-e^{-2\pi z_3}\big)\bigg\},
\end{multline}
from which we compute all $tt^*$ quantities.
In particular,
the brane amplitude is written as a product of twistorial
Gamma functions (for brevity we omit to write the obvious second argument of the various functions)
\begin{equation}\label{qqqert}\begin{split}
\Psi=& \prod_{k=1}^N\bigg\{\Gamma\Big(i\big(\mu_1+(k-1)\mu_3\big)/\zeta, x_1+(k-1)x_3\Big)\;\Gamma\Big(i k\mu_3/\zeta, k x_3\Big)\times\\
&\times \Gamma\Big(i\big(\mu_2+(k-1)\mu_3\big)/\zeta,x_2+(k-1)x_3\Big)\;
\Gamma\Big(i\mu_3/\zeta,x_3\Big)^{\!-1}
\times\\
&\times \Gamma\Big(i\big(\mu_1+\mu_2+(N+k-2)\mu_3\big)/\zeta,x_1+x_2+(N+k-2)x_3\Big)^{\!-1}\bigg\}.
\end{split}\end{equation}
Again, the same result could have been obtained by matching the asymmetric limit of the brane amplitude with the period integral which in this case is the Selberg integral \cite{selberg1,selberg2,selberg3}.

\section{Twistorial Topological Strings for the ${\cal N}=2$  SQED\\ (conifold B-model)} \label{s:twistopstrings}

In this section we show how the large $N$ behavior of the 
$tt^*$ geometries studied in section \ref{s:lgmatrixmodels} produces the 
twistorial topological string amplitudes of section \ref{s:twiststringgs}.
In particular, the large $N$ asymptotics of 
the Gaussian model reproduce the $\mathcal{N}=2$ SQED amplitudes with the Coulomb branch parameter $a$ identified with
$N\epsilon_1$. The simplest way to connect the Gaussian $tt^*$ geometry  
with $\mathcal{N}=2$ SQED is to compare the $tt^*$ metric $G_N$ and brane amplitude $\Psi_N$ computed in section \ref{s:lgmatrixmodels} with the hyperK\"ahler geometry of SQED compactified on a circle studied in \cite{Gaiotto:2008cd}.
The relation between $tt^*$ and the TBA--like equations of \cite{Gaiotto:2008cd} was already outlined in \cite{Cecotti:2013mba}; we start by making the dictionary between the two geometries more precise.

\subsection{The dictionary between $tt^*$
and GMN hyperK\"ahler geometries}\label{ss:dictionary}

Both $tt^*$ Lax equations and the TBA equations of GMN \cite{Gaiotto:2008cd} determine a hyperK\"ahler geometry; moreover they both 
encode a Riemann--Hilbert (RH) problem. From the discussion in section \ref{ss:thetalim} we expect the RH problem of  \cite{Gaiotto:2008cd} to be a certain classical limit of
the $tt^*$ one 
for the large $N$ $(2,2)$ model dual to the given $\mathcal{N}=2$ theory. 
Both RH problems are formulated as integral equations with singular kernels. The TBA equations are based on the kernel \cite{Gaiotto:2008cd}
\begin{equation}\label{kernel1a}
\frac{1}{4\pi i}\,\frac{d\zeta^\prime}{\zeta^\prime}\,\frac{\zeta+\zeta^\prime}{\zeta-\zeta^\prime}
\end{equation}
while in $tt^*$ one uses the
simpler  kernel
\begin{equation}\label{kernel2a}
\frac{1}{2\pi i}\,\frac{d\zeta^\prime}{\zeta^\prime-\zeta}\quad\text{which under }\zeta^\prime,\zeta\to 1/\zeta^\prime,1/\zeta\ \text{becomes}\quad \frac{1}{2\pi i}\,\frac{d\zeta^\prime}{\zeta^\prime}\,\frac{\zeta}{\zeta^\prime-\zeta}.
\end{equation}
To reconcile the two kernels, we notice that the
basic $tt^*$ relations, eqns.\eqref{bil1}\eqref{bil2},
imply the reality condition
\begin{equation}\label{real2}
\overline{\big(G(\boldsymbol{x})^{-1/2}\,\Psi(\boldsymbol{x},-1/\bar\zeta)\big)}=\big(G(\boldsymbol{x})^{-1/2}\,\Psi(\boldsymbol{x},\zeta)\big)^{-1},
\end{equation}
which has the same form as the GMN one
$\overline{X_\gamma(-1/\bar\zeta)}=X_\gamma(\zeta)^{-1}$. Then the $tt^*$ amplitudes should be identified with the corresponding GMN objects times $G^{1/2}$.
More conceptually, the $tt^*$ amplitude $\Psi(\zeta)$ is a section of the hyperholomorphic vector bundle $\mathcal{V}\to\mathcal{M}$ written in
a holomorphic trivialization (in complex structure $I$, corresponding to $\zeta=0$). To get the section in a unitary trivialization we have to perform a complex gauge transformation by $G^{-1/2}$. 
The GMN quantities correspond to unitary gauge $tt^*$ quantities, and satisfy the reality condition in the form \eqref{real2}. 
The change of kernels from \eqref{kernel2a} to \eqref{kernel1a}
just implements the change of gauge. 
Indeed, 
\begin{align}
\frac{1}{2\pi i}\frac{d\zeta^\prime}{\zeta^\prime-\zeta}&=
\frac{1}{4\pi i}\frac{d\zeta^\prime}{\zeta^\prime}\;\frac{\zeta^\prime+\zeta}{\zeta^\prime-\zeta}+\frac{1}{4\pi i}\,\frac{d\zeta^\prime}{\zeta^\prime}\\
\frac{1}{2\pi i}\frac{d\zeta^\prime}{\zeta^\prime}\,\frac{\zeta}{\zeta^\prime-\zeta}&=
\frac{1}{4\pi i}\frac{d\zeta^\prime}{\zeta^\prime}\;\frac{\zeta^\prime+\zeta}{\zeta^\prime-\zeta}-\frac{1}{4\pi i}\,\frac{d\zeta^\prime}{\zeta^\prime}
\end{align}
Using these identities and eqn.\eqref{aaaab2},
the Abelian $tt^*$ brane amplitude \eqref{aaaab1} may be rewritten as
\begin{equation}
\begin{split}\label{aaaab1y}
\log \Psi(\zeta)^{tt^*}=
\frac{i}{\zeta}\,U&-i\zeta\, \overline{U}-\frac{1}{4\pi i}\int_\ell  \frac{d\zeta^\prime}{\zeta^\prime}\,
\frac{\zeta^\prime+\zeta}{\zeta^\prime-\zeta}\, F\big(\boldsymbol{\mu}/\zeta^\prime- i\boldsymbol{x}+\boldsymbol{\bar\mu}\,\zeta^\prime\big)-\\
&+\frac{1}{4\pi i}\int_\ell  \frac{d\zeta^\prime}{\zeta^\prime}\,
\frac{\zeta^\prime-\zeta}{\zeta^\prime+\zeta}\, F\big(\boldsymbol{\mu}/s+ i\boldsymbol{x}+\boldsymbol{\bar\mu}\,s\big)+\frac{1}{2}\log G,
\end{split}
\end{equation}
which shows that
\begin{equation}
\log \Psi(\zeta)^\text{GMN}= \log\!\Big(G^{-1/2}\,\Psi(\zeta)\Big)^{\!tt^*}.
\end{equation}
In particular, as $\zeta\to 0$ (resp.\! $\zeta\to\infty$)
the GMN amplitude becomes $G^{-1/2}$, (resp.\! $G^{1/2}$) in agreement with general $tt^*$ geometry \cite{Cecotti:1992rm}.

\subsection{Twistorial Double Gamma Function}

The analysis of \S.\,\ref{s:gaussianmodel} gives the brane amplitudes for the Gaussian matrix model in the form of a product of twistorial Gamma functions
(up to elementary normalization factors)
\begin{equation}\label{gaugaubrane}
\Psi_N(\beta,\bar\beta,x, \zeta)=\prod_{k=1}^N \frac{\Gamma(i k\beta/\zeta, -i k \bar\beta\zeta, kx)}{\Gamma(i\beta/\zeta, -i\bar\beta \zeta,x)},
\end{equation}
where $\beta=\epsilon_1/\tilde\epsilon_2$ and $x=\theta/2\pi$.
To compare with $\mathcal{N}=2$ SQED or the B-model for the conifold,  it is convenient to introduce the twistorial version of the double Gamma function,
$\Gamma_2(\mu,\bar\mu, y\,|\,\epsilon,\bar\epsilon,x)$ whose arguments are four complex variables $\mu, \bar\mu, \epsilon, \bar \epsilon$ and two period real ones $y$, $x$. 
The function $\Gamma_2(\mu,\bar\mu, y\,|\,\epsilon,\bar\epsilon,x)$ allows us to write in a closed form the
brane amplitude for \emph{all} Abelian $tt^*$ geometries. 

For $\mathrm{Re}\,\mu>0$, $\mathrm{Re}\,\epsilon>0$ we define the function\footnote{ All multivalued functions are assumed to have their normal values (i.e.\! real on the real axis).}
\begin{equation}\label{whatfunU}
 U(\mu\,|\,\epsilon)= \epsilon \,\psi^{(-2)}\!\big(\mu/\epsilon\big)+\frac{\mu^2}{2\epsilon}\,\log\epsilon -\frac{\mu}{2}\,\log(2\pi \epsilon),
\end{equation}
where $\psi^{(-2)}(z)$ is the polygamma function of order $-2$,
\begin{equation}
\begin{split}
 \psi^{(-2)}(z)&\equiv \int_0^z \log\Gamma(s)\,ds=\\
 &= \frac{1}{2}\,z\log2\pi -\frac{1}{2}\,z(z-1)+ z\log\Gamma(z)-\log G(z+1),
 \end{split}
\end{equation}
$G(z)$ being the Barnes $G$--function.
From the identity $G(z+1)=\Gamma(z)\, G(z)$
we see that $\psi^{(-2)}(z)$
satisfies the difference equation
\begin{equation}\label{qqqwa}
 \psi^{(-2)}(z+1)-\psi^{(-2)}(z)= z(\log z-1)+\frac{1}{2}\log2\pi.
\end{equation}
The function $U(\mu\,|\,\epsilon)$ may be defined as the \emph{unique} solution to the difference equation
\begin{equation}\label{udiffXX}
U(\mu+\epsilon\,|\,\epsilon)=U(\mu\,|\,\epsilon)+\mu\Big(\log \mu-1\Big),
\end{equation}
normalized as $U(0\,|\,\epsilon)=0$
and satisfying
\begin{equation}\label{hyperconv}
\frac{d^3}{ds^3} \frac{U(s\,\epsilon\,|\,\epsilon)}{\epsilon}>0,\quad \text{for all }s\in\mathbb{R}_+.
\end{equation}
\medskip

For $\mathrm{Re}\,\mu>0$, $\mathrm{Re}\,\epsilon>0$, the twistorial double Gamma function
$\Gamma_2(\mu,\bar\mu,y\,|\,\epsilon,\bar\epsilon,x)$ is defined as
\begin{multline}\label{doublegamma}
\log\Gamma_2(\mu,\bar\mu,y\,|\,\epsilon,\bar\epsilon,x)= U(\mu+\epsilon/2\,|\,\epsilon)+U(\bar \mu+\bar\epsilon/2\,|\,\bar\epsilon)-\\
-\frac{1}{2\pi}\int\limits_0^\infty \frac{ds}{s(s-i)} 
\log\boldsymbol{\Psi}\big(X(s);q(s)\big)+\frac{1}{2\pi}
\int\limits_0^{-\infty} \frac{ds}{s(s-i)}
\log\boldsymbol{\Psi}\big(X(s)^{-1};q(s)^{-1}\big)\end{multline}
where
\begin{equation}
X(\zeta)=e^{-2\pi(\mu/\zeta-i y+\bar\mu\zeta)},\qquad
\quad q(\zeta)=e^{-2\pi (\epsilon/\zeta-ix+\bar\epsilon\zeta)},
\end{equation}
and $\boldsymbol{\Psi}(x;q)$ is the quantum dilogarithm \cite{Faddeev:1993rs}
defined for $|q|<1$ (i.e.\! for $\mathrm{Re}\,\epsilon>0$)
as
\begin{equation}
 \boldsymbol{\Psi}(x;q)=(xq^{1/2};q)^{-1}_\infty=\prod_{k=0}^\infty (1-x\,q^{k+1/2})^{-1}.
\end{equation}
$\boldsymbol{\Psi}(x;q)$ satisfies the recursion relation
\begin{equation}\label{xxx456}
\boldsymbol{\Psi}(xq;q)=(1-x q^{1/2})\;\boldsymbol{\Psi}(x;q).
\end{equation}
Eqns.\eqref{udiffXX}, \eqref{xxx456} and
\eqref{twistgamma} imply that $\Gamma_2(\mu,\bar\mu,y\,|\,\epsilon,\bar\epsilon,x)$ satisfies the difference equation
\begin{equation}\label{hammatwodif}
\begin{split}
 \log\Gamma_2(\mu+\epsilon,\bar\mu+&\bar\epsilon,y+x\,|\,\epsilon,\bar\epsilon,x)-
\log\Gamma_2(\mu,\bar\mu,y\,|\,\epsilon,\bar\epsilon,x)=\\
&=\log\!\Big[\Gamma(\mu+\epsilon/2,\bar \mu+\bar\epsilon/2, y+x/2)\big/\sqrt{2\pi}\Big].
\end{split}
\end{equation}

Then the Gaussian brane amplitude, eqn.\eqref{gaugaubrane}, may be written in the compact form
(absorbing $\zeta$ in $\beta$, $\bar\beta$, an neglecting the trivial factors in footnote \ref{fooor})
\begin{equation}\label{guasbrane}
\Psi_N(\beta,\bar\beta,x,\zeta=i)=\frac{\Gamma_2\big((N+\tfrac{1}{2})\beta,(N+\tfrac{1}{2})\bar\beta, (N+\tfrac{1}{2})x\,|\,\beta,\bar\beta,x\big)}{\Gamma_2(\beta/2,\bar\beta/2, x/2,\,|\beta,\bar\beta,x)\; \Gamma(\beta,\bar\beta,x)^N}
\end{equation}

From eqn.\eqref{hammatwodif} it is clear that in the
asymmetric limit $\bar\mu\to 0$, $\bar\epsilon\to 0$
\begin{equation}
\Gamma_2(\mu,\bar\mu,y\,|\,\epsilon,\bar\epsilon,x)^\text{asym}= \Gamma_2\big(\mu+y-\tfrac{1}{2}\epsilon-\tfrac{1}{2}x\,\big|\, \epsilon+\tfrac{1}{2}x\big)
\end{equation}
where $\Gamma_2(z\,|\,\omega)$ is the (ordinary) double Gamma function defined by the recursion relation
\begin{equation}
\Gamma_2(z+\omega\,|\,\omega)= \Gamma_1(z)\, \Gamma_2(z\,|\,\omega),
\end{equation}
where $\Gamma_1(z)\equiv \Gamma(z)/\sqrt{2\pi}$.

\subsection{Relation with SQED amplitudes in $\tfrac{1}{2}\,\Omega$--background}

In the \textsc{rhs} of eqn.\eqref{guasbrane} we may neglect the denominator which has trivial dependence of $N$ (and may argued to arise from the normalization of the measure in the LG model). Then,
keeping into account the factor $G_N^{-1/2}$ as required by the dictionary of \S.\ref{ss:dictionary} (and restoring all elementary factors) we may rewrite \eqref{guasbrane} in the form
\begin{equation}\label{brannne}
\begin{split}
&\log(G^{-1/2}_N\Psi_N)=i\,\frac{U\big(a_e+\epsilon_1/2)\,\big|\, \epsilon_1\big)}{\tilde\epsilon_2\,\zeta}-i\,\frac{U\big(\bar a_e+\bar\epsilon_1/2)\zeta\,\big|\,\bar\epsilon_1\big)}{\tilde\epsilon_2}\,\zeta-\\
&-
\frac{1}{4\pi i}\int_\ell \frac{d\zeta^\prime}{\zeta^\prime}\,\frac{\zeta^\prime+\zeta}{\zeta^\prime-\zeta}\,
\log\boldsymbol{\Psi}\big(X_e(\zeta^\prime);q(\zeta)\big)+\frac{1}{4\pi i}\int_{-\ell} \frac{d\zeta^\prime}{\zeta^\prime}\,\frac{\zeta^\prime+\zeta}{\zeta^\prime-\zeta}\,
\log\boldsymbol{\Psi}\big(X_e(\zeta^\prime)^{-1};q(\zeta)^{-1}\big)
\end{split}
\end{equation}
 where 
 \begin{equation}\label{defdedff}
 X_e(\zeta)=e^{-2\pi R\, a_e/\zeta+i\theta_e-2\pi R\, \bar a_e\zeta}
 \end{equation}
is the (exact) GMN electric line for SQED, and 
 \begin{equation}q(\zeta)=e^{-2\pi( \epsilon_1/\tilde\epsilon_2\zeta-i x+
\bar\epsilon_1\zeta/\overline{\tilde\epsilon}_2)}.
\end{equation}
Eqn.\eqref{brannne} coincides with the original Gaussian model brane amplitude, 
eqn.\eqref{guasbrane}, provided the Coulomb branch parameter $a_e$ is equal to
\begin{equation}
R\,a_e=\big(N+\tfrac{1}{2}\big)\frac{\epsilon_1}{\tilde\epsilon_2}\equiv \big(N+\tfrac{1}{2}\big)\beta,
\end{equation}
as predicted by the large $N$ duality, \S.\,\ref{s:largeNduality}. As anticipated in \S.\ref{ss:thetaquant}, in the present set--up
this relation is enhanced to a fully twistorial relation
between the Coulomb branch and $\tfrac{1}{2}\,\Omega$--background parameters
\begin{equation}\label{twiscorre}
\begin{pmatrix}R\,a_e\\
\theta_e\\
R\,\bar a_e\end{pmatrix}= \big(N+\tfrac{1}{2}\big)\!\!
\begin{pmatrix}\epsilon_1/\tilde\epsilon_2\\
2\pi\, x\\
\bar\epsilon_1/\,\overline{\tilde\epsilon}_2\end{pmatrix}.
\end{equation}
Note that, apart from the first two `elementary' terms\footnote{ In the language of \cite{Gaiotto:2008cd}
these terms are the `semi--flat' part of the amplitude.} in the \textsc{rhs} of \eqref{brannne}, which we shall discuss momentarily, the amplitude depends on $a_e$, $\theta_e$ and $\bar a_e$ only through the electric line $X_e(\zeta)$. 

The exact GMN magnetic line for SQED, $X_m(\zeta)$, may also be recovered from the Gaussian matrix model;
indeed
\begin{equation}
\frac{G_N^{-1/2}\,\Psi_N(\zeta)}{G_{N-1}^{-1/2}\,\Psi_{N-1}(\zeta)}= X_m(\zeta)\Big|_{\theta_m=0}.
\end{equation}
This result is consistent with the physical idea that
the magnetic line is the cost in energy for carrying a matrix eigenvalue from infinity to the vacuum configuration. This result agrees with the physical picture of \S.\,\ref{ss:genpicctu}, where we think of the one--vacuum Gaussian model as a $n=2$ model in the limit of infinite separation between the vacua.
In the \textsc{rhs} we have set the magnetic angle
$\theta_m$ to zero. In SQED the dependence on $\theta_m$ is trivial, and hence $\theta_m$ is just part of the overall normalization of the Gaussian amplitude.

\subsection{The higher--dimensional $tt^*$ geometry}\label{s:hfher}

As discussed in \S.\ref{s:fromoldtonew}, we promote the LG model amplitudes \eqref{brannne} to the brane amplitudes for a higher--dimensional Abelian $tt^*$ geometry on $(\mathbb{R}^2\times S^1)^2$. We simply do this by declaring $(a_e,\theta_e,\epsilon,\theta\equiv 2\pi x)$ to be independent coordinates of $(\mathbb{R}^2\times S^1)^2$. The resulting geometry is best described in the general framework of section  \ref{ss:abelian};
it is the Abelian $tt^*$ geometry defined by the magnetic function
\begin{equation}\label{sqedmagnffunc}
F(z_1,z_2)= -\sum_{k=0}^\infty \log\!\big(1-e^{-2\pi[z_1+(k+1/2)z_2]}\big)
\end{equation}
which yields the holomorphic function
\begin{equation}\label{ffffunc}
U(a_2,\epsilon_1)=-\sum_{k=0}^\infty \big(a_e+(k+\tfrac{1}{2})\epsilon_1\big)\Big(\log\big(a_e+(k+\tfrac{1}{2})\epsilon_1\big)-1\Big)
\end{equation}
where we used the standard notation for the two periods of the $tt^*$ geometry $\mu_1=a_e$ and $\mu_2=\epsilon_1$. The function \eqref{ffffunc}
 satisfies the recursion relation
\begin{equation}
U(a_e+\epsilon_1,\epsilon_1)=U(a_e,\epsilon_2)+
\big(a_e+\tfrac{1}{2}\epsilon_1\big)\Big(\log\!\big(a_e+\tfrac{1}{2}\epsilon_1\big)-1\Big),\end{equation}
as well as the analogue of eqn.\eqref{hyperconv}, and hence 
\begin{equation}
U(a_e,\epsilon_1)\equiv U(a_e+\tfrac{1}{2}\epsilon_1\,|\,\epsilon_1)+\text{irrelevant constant,}
\end{equation}
so that the brane amplitudes of the higher dimensional $tt^*$ geometry, restricted to the 
$\mathbb{R}^2\times S^1$ loci \eqref{twiscorre}, reproduce the LG amplitudes
\eqref{brannne}.
In \S.\ref{ss:nssuperW} we show that the holomorphic function $U(a_e+\tfrac{1}{2}\epsilon_1\,|\,\epsilon_1)$ is equal to the
Nekrasov--Shatashvili effective twisted superpotential for $\mathcal{N}=2$ SQCD in $\tfrac{1}{2}\Omega$ background.

The above procedure gives the $tt^*$ geometry for SQED in half--Omega background without any restriction on the parameters nor problems with the absolute normalization of the functional measure. The corresponding brane amplitude in unitary gauge is given by eqn.\eqref{brannne}, where now $a_e$, $\bar a_e$, $\theta_e$, $\epsilon_1$, $\bar\epsilon_1$, $x\equiv \theta/2\pi$ are arbitrary parameters. In the holomorphic gauge (in complex structure $I$) it is the twistorial double Gamma function itself (setting, for simplicity, $\zeta=i$, $\tilde\epsilon_2=1$)
\begin{equation}\label{fulllam}
\Psi(a_e,\bar a_e, \theta_e\, |\,\epsilon_1,\bar\epsilon_1,x)^\mathrm{hol}= \Gamma_2(a_e,\bar a_e,\theta_e/2\pi\,|\,\epsilon_1,\bar\epsilon_1,x).
\end{equation}
From eqn.\eqref{hammatwodif} we see that
under the shift
\begin{equation}
\Big(a_e,\;\bar a_e,\; \theta_e\Big)\longmapsto \Big(a_e+\epsilon_1,\; \bar a_e+\bar\epsilon_1,\;
\theta_e+2\pi x\Big),
\end{equation}
the amplitude gets multiplied by the twistorial Gamma function of arguments $a_e+\epsilon_1/2$, $\theta_e/2\pi+x/2$, i.e.\! by the GMN magnetic line at $\theta_m=0$
evaluated at the mid point in the Coulomb branch
\begin{equation}\label{doublegammarecrelationX}
\begin{split}
\Psi(&a_e+\epsilon_1,\bar a_e+\bar\epsilon_1, \theta_e+2\pi x\, |\,\epsilon_1,\bar\epsilon_1,x)^\mathrm{hol}
=\\
&=X_m(a_e+\epsilon_1/2,\bar a_e+\bar\epsilon_1/2, \theta_e+\pi x, \theta_m=0)\;\Psi(a_e,\bar a_e, \theta_e\, |\,\epsilon_1,\bar\epsilon_1,x)^\mathrm{hol}.
\end{split}
\end{equation}

The $tt^*$ metric and CFIV index may be obtained by plugging in the magnetic function \eqref{sqedmagnffunc} in the general expressions
\eqref{aaaab2} and \eqref{aaaab3}. The $tt^*$ metric is 
\begin{equation}\label{metricXXs}
G(a_e,\bar a_e, \theta_e\, |\, \epsilon_1,\bar\epsilon_1, x)=\exp\!\left(\frac{1}{\pi}\; \mathrm{Im}\int\limits_0^\infty \frac{ds}{s}\; \log\boldsymbol{\Psi}\!\big(X_{e}(s); q(s)\big)\right).
\end{equation}

\subsection{SQED twistorial amplitudes and the $\tfrac{1}{2}\Omega$--background $\widetilde{\mathcal{W}}_\mathrm{eff}(a_e,\epsilon_1)$}\label{ss:nssuperW}

To properly identify the brane amplitude \eqref{brannne} with the partition function of
$\mathcal{N}=2$ SQED in $\tfrac{1}{2}\,\Omega$--background, it remains to consider the holomorphic term
\begin{equation}
U\big(a_e+\tfrac{1}{2}\epsilon_1\,\big|\, \epsilon_1\big)
\end{equation}
 and its antiholomorphic counterpart. 
 
In the discussion around eqn.\eqref{exxpectform} we saw that, in an Abelian $tt^*$ geometry, the holomorphic function $U(\boldsymbol{\mu})$ is identified with the effective
twisted superpotential of the corresponding $(2,2)$ model. In the case of the large--$N$ Gaussian amplitude the corresponding $(2,2)$ model is expected to be $\mathcal{N}=2$ SQED in
$\tfrac{1}{2}\Omega$--background, so
consistency requires the holomorphic function $U$
to be identical to the Nekrasov--Shatashvili
effective twisted superpotential $\widetilde{\mathcal{W}}_\mathrm{eff}(a_e,\epsilon_1)$ \cite{Nekrasov:2009rc}.  In general,   
the effective superpotential consists of a
perturbative part plus an instanton part. In theories like SQED only the perturbative part is present. The perturbative part may be formally written in terms of a
Schwinger proper--time integral which requires $\zeta$--regularization; then 
\begin{equation}
\widetilde{\mathcal{W}}_\mathrm{eff}(a_e,\epsilon_1)= \epsilon_1\,\lim_{s\to 0}\frac{d}{ds}\!\left\{\frac{(\Lambda/\epsilon_1)^s}{\Gamma(s)} \int_0^\infty 
\frac{t^s\, dt}{t^2}\;
\frac{e^{-t\,a_e/\epsilon_1}}{2\,\sinh(t/2)}\right\}
\end{equation}
The integral may be written in terms of the Hurwitz zeta--function $\zeta(s,z)$. 
In particular,
\begin{equation}
\begin{split}
-\partial_{a_e}&\widetilde{\mathcal{W}}_\mathrm{eff}(a_e,\epsilon_1)= \lim_{s\to 0}\frac{d}{ds}\!\Big\{(\Lambda/\epsilon_1)^s\;\zeta(s,a_e/\epsilon_1+\tfrac{1}{2})\Big\}\equiv\\
&\equiv-\frac{a_e}{\epsilon_1}\log(\Lambda/\epsilon_1)+\log\Gamma\big(\tfrac{1}{2}+a_e/\epsilon_1\big)-
\frac{1}{2}\log2\pi.
\end{split}
\end{equation}
Now, from eqn.\eqref{whatfunU}
\begin{equation}
\partial_{a_e} U\big(a_e+\tfrac{1}{2}\epsilon_1\,\big|\,\epsilon_1\big)= \log\Gamma\big(\tfrac{1}{2}+a_e/\epsilon_1\big)
+\frac{a_e}{\epsilon_1}\,\log\epsilon_1-\frac{1}{2}\log2\pi,
\end{equation}
so (we have set $\Lambda\equiv\tilde\epsilon_2$ to $1$) the two twisted superpotentials\footnote{ The overall sign just reflects a different convention on the sign of the twistor parameter $\zeta$.}
\begin{equation}
-\widetilde{W}_\mathrm{eff}(a_e,\epsilon_1)\quad \text{and}\quad
U(a_e+\tfrac{1}{2}\epsilon_1\,|\,\epsilon_1)
\end{equation}
 are equal (up to an irrelevant additive constant). This completes the proof that the Gaussian matrix LG model brane amplitudes correspond to SQED in $\tfrac{1}{2}\Omega$--background under the identification \eqref{twiscorre}.

\subsection{The $\theta$--limit $tt^*$ geometry}\label{s:thetal}

Consider the metric of the higher--dimensional SQED $tt^*$ geometry, eqn.\eqref{metricXXs}.
The second argument of the quantum dilogarithm
\begin{equation}\label{nnome}
q(\zeta)=e^{-2\pi\epsilon_1/\tilde\epsilon_2\zeta+i \theta-2\pi
\bar\epsilon_1\zeta/\overline{\tilde\epsilon}_2}, 
\end{equation}
is a non--trivial function of the twistor parameter $\zeta$ instead of a fixed elliptic nome $q$. This corresponds to the fact that $(\epsilon_1,\theta,\bar\epsilon_1)$ are not fixed parameters of the geometry, but rather coordinates of the higher--dimensional $tt^*$ manifold $(\mathbb{R}^2\times S^1)^2$.

To make contact with more standard (and limited)
approaches, one would like to reduce the $tt^*$ geometry to a simpler hyperK\"ahler manifold of quaternionic dimension $1$ of coordinates $a_e,\theta_e,\bar a_e$ (and $\theta_m$) with fixed $q$. This would correspond to a family of Abelian $tt^*$ geometries on $\mathbb{R}^2\times S^1$ depending on the parameter $q$. 
In view of eqn.\eqref{nnome}, this is roughly equivalent to taking $\epsilon_1\to 0$ while keeping fixed $q=e^{i\theta}$. This direct limit is however not well--defined, and the geometric construction gives a precise meaning to the $\theta$--limit.

The family of reduced $tt^*$ geometries is specified, according to section 
\ref{ss:abelian}, by a magnetic function $F(z;q)$.
One takes
\begin{equation}
F(z;q)=\log\boldsymbol{\Psi}(e^{-2\pi z};q)\equiv-\sum_{m\geq 1}\frac{1}{m}\;\frac{e^{-2\pi m z}}{q^{m/2}-q^{-m/2}},
\end{equation}
with the proviso that $q$ is morally a phase, $e^{i\theta}$, and hence we must state the additional rule that under complex conjugation\footnote{ Note that this is exactly the action of `complex conjugation' on the quantum torus algebra in the related subject of quantum cluster algebras \cite{Fock031,FockX2,FockX3}.} $q\leftrightarrow q^{-1}$
so that
\begin{equation}
\overline{F(z;q)}=\sum_{m\geq 1}\frac{1}{m}\;\frac{e^{-2\pi m \bar z}}{q^{m/2}-q^{-m/2}}=-\log\boldsymbol{\Psi}(\bar z;q),
\end{equation}
with a sign flip. Hence in the $\theta$--limit
the SQED $tt^*$ metric \eqref{metricXXs} becomes
\begin{equation}\label{thetattmetric}
\log G_\theta=\frac{1}{2\pi i}\int_\ell \frac{ds}{s}\,\log\boldsymbol{\Psi}(X_e(s);q)+
\frac{1}{2\pi i}\int_\ell \frac{ds}{s}\,\log\boldsymbol{\Psi}(\overline{X_e(s)};q).
\end{equation}
To the function
\begin{equation}
F(z;q)\equiv -\sum_{k\geq 0} \log\!\big(1-e^{-2\pi[z+(k+1/2)\log q]}\big)
\end{equation}
there corresponds the holomorphic function
(cfr.\! eqn.\eqref{ffffunc})
\begin{equation}
U(\mu)=-\sum_{k\geq 0}\big(\mu+(k+1/2)\log q\big)\Big(\log\!\big(\mu+(k+1/2)\log q\big)-1\Big)
\equiv U(\mu,\log q).
\end{equation}
Hence, reintroducing all factors,
the $\theta$--limit brane amplitude for SQED is
\begin{multline}\label{thetalimitfin}
\log\psi^\theta= -\frac{i}{\tilde\epsilon_2 \,\zeta} \,\mathcal{W}^{NS}\!\Big(a_e,-i \frac{\zeta\tilde\epsilon_2 \log q}{2\pi}\Big)+\frac{i\zeta}{\tilde\epsilon_2}\,\mathcal{W}^{NS}\!\Big(\bar a_e, -i\frac{\tilde\epsilon_2\log q}{2\pi\,\zeta}\Big)-\\
-\frac{1}{4\pi i}\int_\ell \frac{d\zeta^\prime}{\zeta^\prime}\, \frac{\zeta^\prime+\zeta}{\zeta^\prime-\zeta}\;\log\boldsymbol{\Psi}(X_e(\zeta^\prime);q)-
\frac{1}{4\pi i}\int_{-\ell} \frac{d\zeta^\prime}{\zeta^\prime}\, \frac{\zeta^\prime+\zeta}{\zeta^\prime-\zeta}\;\log\boldsymbol{\Psi}(X_e(\zeta^\prime)^{-1};q).
\end{multline}

\subsection{$\theta$--limit vs.\! the quantum KS wall crossing formula}\label{st:thetaaali}

The Stokes jumps of the brane amplitude are intrinsically defined, independently of a choice of trivialization.  From eqn.\eqref{thetalimitfin} we see that, in the
 $\theta$--limit, the SQED brane amplitude $\Psi(\zeta)_\theta$ jumps at the BPS rays in $\zeta$--plane as
 \begin{equation}
 \Psi(e^{+i0}\zeta)_\theta=\boldsymbol{\Psi}\!\big(X_\gamma(\zeta)^{\pm 1};q\big)\,\Psi(e^{-i0}\zeta)_\theta.
 \end{equation} 
 While the explicit computation above holds for  Abelian $tt^*$ geometries, we argued in section \ref{ss:thetalim},
 that this should be true for all (reduced)
 $tt^*$ geometries defined by the $\theta$--limit of a 4d $\mathcal{N}=2$ theory on $\tfrac{1}{2}\Omega$ background.

Composing all jumps at the several BPS phases
one would get an ordered product
 \begin{equation}
 \Psi(e^{2\pi i}\zeta)_\theta=\left(\prod^\curvearrowleft_{\text{BPS}\atop \text{phases}} \boldsymbol{\Psi}\!\big(X_\gamma(\zeta);q\big)\right)\Psi(\zeta)_\theta,
 \end{equation} 
which looks like the 4d $\mathcal{N}=2$ quantum monodromy $\mathbb{M}(q)$ \cite{Cecotti:2010fi,Cecotti:2009uf} whose invariance (up to conjugacy) under arbitrary changes of parameters is equivalent to the refined
version \cite{Cecotti:2009uf,Dimofte:2009bv, 
Dimofte:2009tm}
of the
 Kontsevich--Soibelman wall crossing formula \cite{Kontsevich:2008fj}.  However, for this identification to work we also need that in the
 $\theta$--limit line operators $X_\gamma(\zeta)$ satisfy the quantum torus algebra
 \cite{Fock031,FockX2,FockX3}
 \begin{equation}\label{quantumtorusalgebra}
 X_\gamma(\zeta)\,X_{\gamma^\prime}(\zeta)= q^{\langle\gamma,\gamma^\prime\rangle}\,
 X_{\gamma^\prime}(\zeta)\,X_{\gamma}(\zeta),
 \end{equation}
where $\langle\cdot,\cdot\rangle\colon\Gamma\times \Gamma\to \mathbb{Z}$ is the Dirac electro--magnetic pairing. If this relation holds, then the adjoint action of the
operators $\boldsymbol{\Psi}\!\big(X_\gamma(\zeta);q\big)$ on the quantum torus algebra
generate a quantum cluster algebra action 
\cite{Fock031,FockX2,FockX3} which produces the correct action of the quantum monodromy \cite{Cecotti:2010fi,Cecotti:2014zga}.

Thus, to conclude that the twistorial brane amplitudes in the $\theta$--limit correspond to the \emph{refined} version of the GMN quantities, it remains to show that the quantum torus algebra commutation relations
\eqref{quantumtorusalgebra} are satisfied in this limit.
In particular, in SQED case the $\theta$--limit quantum torus algebra would read 
\begin{equation}\label{qqqasxl}
X_m(\zeta)\,X_e(\zeta)= q\,X_e(\zeta)\,X_m(\zeta),
\end{equation}
where $X_e(\zeta)$ and $X_m(\zeta)$ are the electric and magnetic line operators, respectively. 
The validity of \eqref{qqqasxl} is equivalent to
the $\theta$--limit of the functional equation for the
twistorial double Gamma function $\Gamma_2$.
Indeed, the $\theta$--limit of eqn.\eqref{doublegammarecrelationX} is
\begin{equation}
X_m(\zeta)\,\Psi(a_e,\bar a_e, \theta_e;\zeta)_\theta=\Psi(a_e,\bar a_e, \theta_e+\theta;\zeta)_\theta,
\end{equation}  
which says that, in this limit, $X_m(\zeta)$ may be
represented by the operator
\begin{equation}
\exp\!\left(\theta\,\frac{\partial}{\partial\theta_e}\right).
\end{equation}
Since $X_e(\zeta)=e^{-2\pi a_e/\zeta+i\theta_e-2\pi\bar a_e\zeta}$, this implies
\begin{equation}
X_m(\zeta)\,X_e(\zeta)=e^{i\theta}\,X_e(\zeta)\,X_m(\zeta),
\end{equation}
which is the relation we wanted to check, eqn.\eqref{qqqasxl}.

\section{Twistorial invariant aspects of the $tt^*$ geometry}\label{s:other}

We have defined the twistorial topological string as the D-brane wave function
for the ${1\over 2} \Omega$ background.  In this section we briefly discuss other aspects of $tt^*$ geometry
for the ${1\over 2} \Omega$ background.  In particular we focus on aspects where
the twistor parameter $\zeta$ disappears from the computations.
We will focus on three aspects: The $tt^*$ metric, the $Q$-function
(CFIV index) and the monodromy operator.  

The $tt^*$ metric does not depend on the twistor parameter because there are no boundaries involved.  We will mainly focus on the metric
for the ground state.  This involves roughly speaking $\sum_{\vec k}|\psi_{\vec k} |^2$ by gluing two hemispheres
to get a sphere.  This will only depend on the external parameters defining the theory (i.e.\! the triplet of masses) as
well as the $\epsilon_1,{\tilde \epsilon_2}$.  We will explicitly compute it for the case of the twistorial version
of 2 M5 branes on a sphere with 3-punctures.  By the AGT relation this is related to the three point
function of the Liouville theory.  Here we will be able to compute a twistorial version of the DOZZ formula.
Its interpretation as an amplitude of some integrable 2d system remains to be seen.

Another thing we can do is to study the 2d monodromy.  This was a powerful tool in classification
of the 2d ${\cal N}=2$ theories \cite{Cecotti:1992rm}.  In fact the computations done in \cite{Cecotti:2010fi} can
now be interpreted as computing the trace of the monodromy for the ${1\over 2}\Omega $ background in the
$\theta$-limit.  In particular it corresponds to considering the 2d theory on a torus, where as we go around
the temporal circle we do an R-twist. 

Another aspect of the $tt^*$ geometry is the CFIV index \cite{Cecotti:1992qh}, which at the conformal
point measures the central charge of the theory.  It is natural to ask what is the interpretation of this in the 4d theory.
In this section we argue that the computation of AMNP {\cite{Alexandrov:2014wca} can be interpreted
as computing the CFIV index of the ${1\over 2}\Omega$ background in the $C$-limit.  We provide
evidence for this by computing the two indices for the case of SQED.  
The fact that both CFIV index and the AMNP index
do not depend on $\zeta$ and are continuous, and that they agree in various limits (e.g.\! when SQED dominates)
strongly suggests they are the same in general.

In the rest of this section we discuss the relation between CFIV index and the AMNP index as well as the $tt^*$ metric for the 2 M5 brane theory
on three times punctured sphere, as examples of these other aspects of the $tt^*$ geometry.

\subsection{$C$--limit: the $4d$ AMNP index vs. the CFIV index}

The  CFIV index of the 2d (2,2) theory \cite{Cecotti:1992qh} is clearly an interesting quantity to compute.  It encodes
the amount of degrees of freedom in the 2d theory (and at 2d conformal points measures the central charge and the dimension of the chiral primary operators \cite{Cecotti:1991me,Cecotti:1992qh}). When the 2d (2,2) theory arises from the ${1\over 2} \Omega$
background of a 4d $\mathcal{N}=2$ theory, its CFIV index computes also a four--dimensional
susy--protected physical quantity. More generally,
as discussed in the context of theories with 2d LG duals \S\S.\,\ref{s:fromoldtonew}, \ref{s:hfher}, and \ref{s:thetal}, the CFIV index is defined for all $tt^*$ geometries, including the higher--dimensional analytically--continued one (\S.\ref{s:hfher}),
the $\theta$--limit one (\S.\ref{s:thetal}), and its $C$--limit geometry
(to be discussed in detail in section \ref{s:c-limit}).
Then, in the present setup, the CFIV indices of these diverse $tt^*$ geometries yield
an  increasingly--refined sequence of susy--protected quantities
 for the 
4d $\mathcal{N}=2$ theory
\begin{equation}
Q_C,\quad Q(\theta)_\theta,\quad Q(\theta,\epsilon_1).
\end{equation} 
In the previous sections we wrote the exact form of all these indices under the assumption that the higher--dimensional $tt^*$ geometry is Abelian. In particular $Q_C$, $Q(\theta)_\theta$ and $Q(\theta,\epsilon_1)$ are explicitly known for $\mathcal{N}=2$ SQED.
All these indices, being susy--protected, count 4d
multi--BPS states with some very special weight
so that the full index is wall crossing invariant and smooth in parameter space.

In this section we focus on the coarser version of the index,
the $C$--limit one $Q_C$. In \S.\ref{eee45cq} below we show that the exact
$Q_C$ index for $\mathcal{N}=2$ SQED is equal (up to overall normalization) to the 4d wall-crossing invariant quantity $\mathcal{I}$, proposed by
AMNP \cite{Alexandrov:2014wca}.
Thus $Q_C$ and $\mathcal{I}$ are two smooth, wall crossing invariant, susy--protected quantities, both counting multi--BPS states with weights which happen to agree for all mutually--local multi--hypermultiplet states. Then the (properly normalized)  indices
$Q_C$ and $\mathcal{I}$ are expected to be equal in general.

It would be desirable to have a direct four--dimensional definition of $Q_C$ (and its refined cousins $Q(\theta)_\theta$, $Q(\theta,\epsilon_1)$).
This in particular allows us to directly compare it with the index proposed in AMNP.  Therefore we start with a direct 4d discussion of $Q_C$
based on the target space interpretation of the twistorial topological string in \S.\ref{s:target}. 
There are two aspects to the AMNP work:  One is the identification of a wall-crossing invariant constructed out of the classical hyperK\"ahler geometry of the circle compactification of 4d ${\cal N}=2$ supersymmetric theories.  The second aspect is to identify it with
a particular 4d index computation.  While we find that computation of CFIV index agrees with the object introduced in AMNP,  the definition of the
4d index we find is distinct from the 4d index of AMNP, even though there are some formal similarities.

\subsubsection{4d interpretation of $Q_C$}\label{4dQC}

We consider our 4d $\mathcal{N}=2$ theory quantized in
the space $S^1\times I_L \times \mathbb{R}^2_{(34)}$, where $S^1$ is a circle of length $R$
viewed as periodic Euclidean time $t$, $I_L$ is a segment of lenght $L$
(we shall take $L\to\infty$ at the end of the computation), and $\mathbb{R}^2_{(34)}$
is the orthogonal plane on which we switch on the $\tfrac{1}{2}\Omega$ background.
Having non--zero $\theta$ means that as $t\to t+R$ we rotate the 3--4 plane
$\mathbb{R}^2_{(34)}$ by an angle proportional to $\theta$ and make a compensating $SU(2)_R$ rotation
to preserve half the supercharges. One also introduces fugacities $\theta_a$ for the various conserved charges $q_a$.
Then the fully--refined twistorial CFIV index, before taking any limits, is
\begin{equation}\label{qqindex}
 i\,Q(\theta,\epsilon_1)= \lim_{L\to\infty} \frac{R}{L} \;
\mathrm{Tr}_{\frac{\Omega}{2}}\Big[(-1)^F\, J\, e^{i\theta(J-J_{34})+i\sum_a \theta_a q_a}\, e^{-RH}\Big],
\end{equation}
 where $J_{34}$ is the generator of rotations in the 3--4 plane and $J$ is the Cartan generator of
$SU(2)_R$ which, from the 2d viewpoint, is identified with $F/2$, so that \eqref{qqindex} coincides with the
standard definition
of the 2d CFIV index \cite{Cecotti:1992qh} for the 2d theory on
$S^1\times I_L$, the insertion of the twisting operator $e^{i\theta(J-J_{34})+i\sum_a \theta_a q_a}$
implementing the reduction from a $8$--supercharge theory to a $(2,2)$ one. The \textsc{rhs} of eqn.\eqref{qqindex}
has an obvious representation in terms of path integrals of the 4d $\mathcal{N}=2$ theory with periodic boundary conditions in Euclidean time.

The $C$--limit index is defined by first
turning off the ${1\over 2} \Omega$ background (i.e.\! sending $\eps_1\rightarrow 0$) and then
$\theta\rightarrow 0$
\begin{equation}\label{zzza12m}
\begin{split}
 Q_C&= \lim_{\theta\to 0} \left(\frac{\theta}{2\pi}\; \lim_{\epsilon_1\to 0} Q(\theta,\epsilon_1)\right)=\\
 &=\lim_{\theta\to 0\atop L\to \infty} \left(\frac{\theta R}{2\pi L}\;   
\mathrm{Tr}\Big[(-1)^F\, J\, e^{i\theta(J-J_{34})+i\sum_a \theta_a q_a}\, e^{-RH}\Big]\right).
\end{split}
\end{equation}

We would like to consider the $C$--limit CFIV index $Q_C$ to compare with the AMNP index $\mathcal{I}$.
As a check, let us compute the \emph{one} BPS half--hypermultiplet contribution
to the \textsc{rhs} of \eqref{qqindex} in the $C$--limit. 
We first note that the hypermultiplet fermions are $SU(2)_R$
invariant, and hence do not contribute to the trace in
\eqref{qqindex} because of the insertion of the $SU(2)_R$ generator $J$. The scalars
are in the fundamental of $SU(2)_R$ and hence have $J=\pm 1/2$. Then the $SU(2)_R$ representation content produces an
overall factor $i\sin(\theta/2)$ in the one--particle trace.
Therefore the contribution to
the \textsc{rhs} of \eqref{qqindex} from one half--hyper of mass $M$ and charges $q_a$ is
\begin{equation}\label{999qw}
\begin{split}
i&\frac{R\, \sin(\theta/2)}{L}\,  e^{i\sum_a \theta_a q_a}\; \mathrm{Tr}^{(1)}_{\frac{\Omega}{2}}\!\Big[e^{-R H-i\theta J_{34}}\Big]=\\
&=i \frac{R\,\sin(\theta/2)}{L}\,  e^{i\sum_a \theta_a q_a}\; \frac{R}{2\sqrt{\pi}}\int_0^\infty \frac{dt}{t^{3/2}}\,
e^{-M^2 t-R^2/4t}\;\; \mathrm{Tr}^{(1)}_{\frac{\Omega}{2}}\!\Big[e^{t\,\Delta_{\epsilon_1}-i\theta J_{34}}\Big]
\end{split}
\end{equation}
 where $\mathrm{Tr}^{(1)}_{\frac{\Omega}{2}}$ stands for the trace over the one--scalar Hilbert space as regularized by
$\tfrac{1}{2}\Omega$ background, and $-\Delta_{\epsilon_1}$ is the $\Omega$--regularized version of the free bosonic  `Schwinger time' 
Hamiltonian $-\Delta\equiv p^2$.
From \cite{Nekrasov:2002qd,Nakayama:2011be} we know that on a free hyper a $\tfrac{1}{2}\Omega$ background
has the same effect as a constant background electromagnetic field
\begin{equation}
 F\propto \epsilon_1\, dx^3\wedge dx^4,
\end{equation}
plus a compensating $SU(2)_R$--twist of order $O(\epsilon_1)$ to preserve half--supersymmetry.  If we are interested only in the 
limit $\epsilon_1\to 0$, we may neglect the twist, and work
simply with the magnetic background (which provides an effective IR regulation). Hence, in this limit, $-\Delta_{\epsilon_1}$ may be replaced by
\begin{equation}
\begin{split}
-\Delta_{\epsilon_1}&= -\partial_{x^2}^2+\Big(-i\partial_{x^3}-\tfrac{1}{2}\epsilon_1\,x^4\Big)^{\!2}+\Big(-i\partial_{x^4}+\tfrac{1}{2}\epsilon_1\,x^3\Big)^{\!2}\equiv\\
&\equiv  p^2_2 +2\left\{\frac{1}{2}\,p_3^2+\frac{1}{2}\,p_4^2+\frac{1}{2} \left(\frac{\epsilon_1}{2}\right)^{\!2}\Big((x^3)^2+(x^4)^2\Big)+\frac{\epsilon_1}{2}\, J_{34}\right\}
\end{split}
\end{equation}
i.e.\! as a free particle moving in the segment $I_L$ times twice the harmonic oscillator in the $3$-$4$ directions of frequency $\epsilon_1/2$ shifted by $\epsilon_1\, J_{34}/2$. 
Then
\begin{equation}\label{qqq1234}
\begin{split}
\mathrm{Tr}^{(1)}_{\frac{\Omega}{2}}\!\Big[e^{t\,\Delta_{\epsilon_1}-i\theta J_{34}}\Big]& =
\left(\frac{1}{4\pi t}\right)^{\!1/2}\, L\, \sum_{k\geq 0} e^{-(k+1/2)\epsilon_1\, t}\;\,\frac{\sin[(k+1)(\theta-i\epsilon_1 t)]}{\sin (\theta-i\epsilon_1 t)}
\equiv\\
&\equiv \left(\frac{1}{4\pi t}\right)^{\!1/2}\, \frac{ L\,e^{-\epsilon_1 t/2}}{1-2\, e^{-\epsilon_1 t}\,\cos (\theta-i\epsilon_1 t)+ e^{-2\epsilon_1 t}}\\ 
&\xrightarrow{\ \ \ \epsilon_1\,\to\; 0\ \ \ }\ 
\left(\frac{1}{4\pi t}\right)^{\!1/2}\, \frac{ L}{4\,\sin^2(\theta/2)},
\end{split}
\end{equation}
where we used the explicit form of the generating function for the Chebyshev polynomials of the second kind $U_{k-1}(\cos\theta)=\sin(k\theta)/\sin(\theta)$.

Inserting back \eqref{qqq1234} into \eqref{999qw},
we get the one half--hyper contribution to
$Q(\theta)_\theta$
\begin{equation}\label{eeq123v}
Q(\theta)_\theta^{\text{(1/2-hy)}}=\frac{R^2}{16\pi} \frac{e^{i\sum_a \theta_a q_a}}{\sin(\theta/2)}\int_0^\infty \frac{dt}{t^2}\,
e^{-M^2 t-R^2/4t}\equiv \frac{e^{i\sum_a \theta_a q_a}}{4\pi} \frac{MR}{\sin(\theta/2)}\,K_1(MR).
\end{equation}
This 4d result should be compared with the half--hyper contribution to the $\theta$--limit index as predicted by the large--$N$ 2d dual $tt^*$ geometry. Since one--particle contributions to the CFIV index are universal, we may compute it by expanding the
SQED one and keeping only the first term. Using the $\theta$--limit $tt^*$ metric, eqn.\eqref{thetattmetric}, we get for the half--hyper contribution to $Q(\theta)_\theta|_\text{2d dual}$ \footnote{ For simplicity we take the central charge of the hyper to be real positive, hence equal $M$.}
\begin{multline}
Q(\theta)_\theta^{\text{(1/2-hy)}}\Big|_\text{2d dual}\equiv
-\frac{1}{2}R\,\frac{\partial}{\partial R} \left(\frac{1}{2\pi i}
\int_0^\infty \frac{ds}{s} \Big(\log(X_e(s);q)\Big)^{\text{(1/2-hy)}}\right)=\\
=\frac{1}{4\pi i}\,\frac{e^{i\theta_e}}{q^{1/2}-q^{-1/2}}\;
R\frac{\partial}{\partial R}\int_0^\infty \frac{ds}{s}\, e^{-MR(s+1/s)/2}=\frac{e^{i\theta_e}\, MR\; K_1(MR)}{4\pi\,\sin(\theta/2)},
\end{multline}
in full agreement with \eqref{eeq123v}.

From the definition, eqn.\eqref{zzza12m},  the one half--hyper contribution to $Q_C$ then is
\begin{equation}
Q_C^{\text{(1/2-hy)}}= 
\frac{R}{4\pi^2}\,\; e^{i\sum_a\theta_a q_a}\, M\, K_1(MR),
\end{equation}
which is the standard one particle contribution to the CFIV index \cite{Cecotti:1991me,Cecotti:1992qh} as well
as the regularization prescription proposed for AMNP index \cite{Alexandrov:2014wca} (up to an extra factor of $2\pi$, cfr.\! eqn.\eqref{eee123}).

\subsubsection{$Q_C$ for $\mathcal{N}=2$ SQED}\label{eee45cq}

Next, we compute CFIV index for SQED to check its relation with AMNP index.
The $C$--limit consists in taking $q\to 1$ in the $\theta$--limit. 
The $C$--limit of the metric is\footnote{ Recall that $\log q$ is formally purely imaginary.}
\begin{equation}\label{cmetric}
\begin{split}
\log G_C&\overset{\text{def}}{=} \lim_{q\to 1}\left(\frac{-i\,\log q}{2\pi}\,\log G_\theta\right)
\equiv \\
&\equiv\frac{1}{4\pi^2}\int_{\ell_e} \frac{ds}{s}\, \mathrm{Li}_2(X_e(s))+\frac{1}{4\pi^2}\int_{-\ell_e} \frac{ds}{s}\, \mathrm{Li}_2(X_e(s)^{-1}).
\end{split}
\end{equation}

The CFIV index $Q$ of a $tt^*$ geometry is the component of the
 Berry connection\footnote{ Written in the normalized `point' holomorphic gauge \cite{Cecotti:1992rm}.} in the direction of the RG flow
 \cite{Cecotti:1991me,Cecotti:1992qh}. For the SQED metric this is
 \begin{equation}
Q(a_e,\bar a_e, \theta_e\, |\, \epsilon_1,\bar\epsilon_1, \theta; \tilde\epsilon_2)=\frac{1}{2}\,\tilde\epsilon_2\,\frac{\partial}{\partial \tilde\epsilon_2}
\log G(a_e,\bar a_e, \theta_e\, |\, \epsilon_1,\bar\epsilon_1, \theta; \tilde\epsilon_2).
 \end{equation}
Replacing in this formula the general metric with the $\theta$--limit one, eqn.\eqref{thetattmetric},
we get the $\theta$--limit CFIV index.
In the same vein, replacing $G$ by the $C$--limit $tt^*$ metric $G_C$,
eqn.\eqref{cmetric}, we get the $C$--limit CFIV index $Q_C$
\begin{equation}\label{thetaindex}
\begin{split}
Q(a_e,\bar a_e,\theta_e; \tilde\epsilon_2)_C=&\frac{1}{2}\,\tilde\epsilon_2\,\frac{\partial}{\partial\tilde\epsilon_2} \log G_C=\\
=&-\frac{1}{4\pi\,\tilde\epsilon_2}\,
\int_{\ell_e} \!\frac{ds}{s}\left(\frac{a_e}{s}+\bar a_e\,s\right)\log\!\big(1-X_e(s)\big)-\\
&-\frac{1}{4\pi\,\tilde\epsilon_2}
\,
\int_{\ell_{-e}} \!\frac{ds}{s}\left(\frac{a_{-e}}{s}+\bar a_{-e}\,s\right)\log\!\big(1-X_{-e}(s)\big)
\end{split}
\end{equation}
(here $a_{-e}=-a_e$, $X_{-e}(s)=X_e(s)^{-1}$, and
$\ell_{-e}=-\ell_e$).
 This expression should be compared with the 4d BPS index $\mathcal{I}_\text{SQED}$ for $\mathcal{N}=2$ SQED introduced in ref.\cite{Alexandrov:2014wca} (which can also be identified with the TBA free energy \cite{Alexandrov:2010pp,Alexandrov:2014wca}).
Comparing eqn.\eqref{thetaindex} with
 eqn.(11) of
 ref.\cite{Alexandrov:2014wca}, and taking into account the identifications $i Z_\gamma= 2 a_\gamma$ and
 $R=1/\tilde\epsilon_2$, we see that the $C$--limit 2d index is related to the 4d one as
 \begin{equation}\label{eee123}
Q(a_e,\bar a_e, \theta_e; \tilde\epsilon_2)_C=2\pi\;\mathcal{I}(a_e,\bar a_e, \theta_e; \tilde\epsilon_2)_\text{SQED}.
 \end{equation}
 As we have already argued, even if this relation has been shown  for SQED,
 we expect it to hold for all 4d $\mathcal{N}=2$ theories.

\subsection{Twistorial Liouville amplitudes}

The analysis in section \ref{s:twistopstrings} of the twistorial brane amplitudes
for the Gaussian model applies, with minor modifications, to any theory whose $tt^*$ geometry is Abelian, in particular to the examples in section
\ref{s:lgmatrixmodels} and those in appendix \ref{appendixmultimatrix} which correspond to $ADE$ Toda amplitudes. Up to an elementary pre--factor, for all models in this class the brane amplitude is obtained by the following
`rule of thumb': one starts from the matrix
period integral $\int e^{\mathcal{W}/\zeta}\,dX$ written
as a product of Euler Gamma functions, and replaces
each product of $\Gamma$'s of the form\footnote{ Here and below we take $\zeta=i$.}
\begin{equation}
\prod_{j=0}^N \Gamma(\alpha+\beta j)
\rightsquigarrow \frac{\Gamma_2\big(\alpha+(N+\tfrac{1}{2})\beta, \bar\alpha+(N+\tfrac{1}{2})\bar\beta,
y+(N+\tfrac{1}{2})x\,|\, \beta,\bar\beta,x\big)}{\Gamma_2\big(\alpha+\tfrac{1}{2}\beta, \bar\alpha+\tfrac{1}{2}\bar\beta,
y+\tfrac{1}{2}x\,|\, \beta,\bar\beta,x\big)},
\end{equation} 
where $\Gamma_2$ is the twistorial double Gamma function,  $2\pi y$ the angle associated to the coupling $\alpha=\mu/\tilde\epsilon_2$, and $2\pi x$ the angle associated to $\beta=\epsilon_1/\tilde\epsilon_2$. Finally, one has to identity the 't Hooft coupling to be kept fixed while sending $N\to\infty$ with the correct physical quantity, and then analytically continue to arbitrary values of this quantity.
 
In the particular case of the double Penner model of section \ref{s:doublepenner}, the large-$N$ limit of the brane amplitude
gives the twistorial version of the Liouville three--point holomorphic block; indeed, in the asymmetric limit the
amplitude reduces to the period integral which is identified
  \cite{Cheng:2010yw}  with the ordinary three--point conformal block of the Liouville.
  
The correct identification of the physical parameters in
may be read from ref.\cite{Cheng:2010yw}   
\begin{gather}\label{dict1}
(N-1)\beta=-\frac{1}{2}(\mu_1+\mu_2+\mu_3)\ \mod 1,\\
b\equiv\sqrt{-\beta}\\
\beta=-b \,Q \ \mod 1.
\end{gather}
We complete these expressions to twistorial triplets
by setting also
\begin{gather}
(N-1)\bar\beta=-\frac{1}{2}(\bar\mu_1+\bar\mu_2+\bar\mu_3)\\
(N-1)x= -\frac{1}{2}(y_1+y_2+y_3)\\
\bar\beta=-\bar b\, \bar Q\ \mod 1.
\end{gather}
Moreover, we set  \cite{Cheng:2010yw} 
\begin{equation}
\mu_i=-2b\,\alpha_i,\qquad \bar\mu_i=-2\bar b\,\bar\alpha_i,\quad i=1,2,3\label{dictn}
\end{equation}
where the  $\alpha_i$ are the external Liouville momenta. We define
\begin{equation}
\Gamma_{b,\bar b,x}(\alpha, \bar\alpha,y)\equiv \Gamma_2(b\,(\alpha-b/2), \bar b\,(\bar\alpha-\bar b/2), y+x/2\,|\, -b^2,-\bar b^2,x).
\end{equation}
With these conventions, the $tt^*$ brane amplitude in \eqref{qqqert} may be rewritten as:
\begin{equation}\label{twisFalpha123}
\begin{split}
&\frac{\Gamma_{b,\bar b,x}\big(\alpha_1+\alpha_2+\alpha_3-Q,\bar\alpha_1+\bar\alpha_2+\bar\alpha_3-\bar Q, x-(y_1+y_2+y_3)/2\big)}{\Gamma_{b,\bar b,x}(0,0,0)}\times\\
&\ \ \times 
\frac{\Gamma_{b,\bar b,x}\big(\alpha_2+\alpha_3-\alpha_1,\bar\alpha_2+\bar\alpha_3-\bar \alpha_1, (y_1-y_2-y_3)/2\big)}{\Gamma_{b,\bar b,x}(Q-2\alpha_1,\bar Q-2\bar\alpha_1,y_1-x)}\times\\
&\ \ \times
\frac{\Gamma_{b,\bar b,x}\big(\alpha_1+\alpha_3-\alpha_2,\bar\alpha_1+\bar\alpha_3-\bar \alpha_2, (y_2-y_1-y_3)/2\big)}{\Gamma_{b,\bar b,x}(Q-2\alpha_2,\bar Q-2\bar\alpha_2,y_2-x)}\times\\
&\ \ \times
\frac{\Gamma_{b,\bar b,x}\big(Q+\alpha_3-\alpha_1-\alpha_2,
\bar Q+\bar\alpha_3-\bar\alpha_1-\bar \alpha_2, (y_1+y_2-y_3)/2-x\big)}{\Gamma_{b,\bar b,x}(2\alpha_3,2\bar\alpha_3,-y_3)}.
\end{split}
\end{equation}
This twistorial amplitude is obtained from the ordinary Liouville $3$--point chiral block 
 $\mathcal{F}_{\alpha_1,\alpha_2,\alpha_3}$ \cite{Cheng:2010yw}  
by replacing each double Gamma factor with the corresponding twistorial double Gamma function according to the dictionary
\begin{equation}
\Gamma_b(a_i\,\alpha_i+c\, Q) \rightsquigarrow
\Gamma_{b,\bar b,x}( a_i\,\alpha_i+c\,Q,
a_i\,\bar\alpha_i+c\,\bar Q, -a_i\,y_i/2-c\,x),
\end{equation} 
where the coefficient $a_i=0,\pm 1,\pm 2$ and $c=0,\pm1$ are the same as in ref.\cite{Cheng:2010yw}.
In this expression the twistorial parameter $\zeta$ was absorbed in the couplings. By construction, the above expression reduces to the usual Liouville $3$--point chiral block 
 $\mathcal{F}_{\alpha_1,\alpha_2,\alpha_3}$ in the asymmetric limit. To restore the $\zeta$ dependence one replaces each factor of 
\eqref{twisFalpha123} with the rule
\begin{equation}\label{eeeqp}
\Gamma_{b,\bar b,x}(\xi,\bar \xi, y)\rightsquigarrow
\text{(elementary factor)}\;\Gamma_{\sqrt{i/\zeta}\,b, \sqrt{-i\zeta}\,\bar b,x}(i \xi/\zeta,-i\zeta\bar\xi,y).
\end{equation}

The extension of this result to certain 3--point blocks for $ADE$ Toda theories is described in appendix \ref{appendixmultimatrix}.

\subsubsection{The $tt^*$ metric: the twistorial $\Upsilon$--function}

While the brane amplitude of the double Penner model is the twistorial extension of the three--point chiral block, its $tt^*$ metric, which has the form $\Psi\,\Psi^\dagger$
(for an appropriate notion of Hermitian conjugation) should be thought of as the twistorial extension of the full three--point amplitude. 

To write the $tt^*$ in a nice form, one needs the twistorial extension of the standard $\Upsilon$--function
\begin{align}\label{upsordinary}
&\Upsilon_b(x)= \frac{1}{\Gamma_b(x)\, \Gamma_b(Q-x)}\\
& \text{where } \Gamma_b(x)=\Gamma_2(b\,x\,|\,b,b^{-1})\ \text{and }\; Q=b+b^{-1},
\end{align}
which is obviously symmetric under the reflection $x\leftrightarrow Q-x$
\begin{equation}\label{reflsssym}
\Upsilon_b(Q-x)=\Upsilon_b(x).
\end{equation}
In the non--twistorial set up,
the arguments of the various $\Upsilon_b$--functions entering in the Liouville amplitude are shifted by integral multiples of $Q/2$ which may be seen as a shift by $k b/2$ followed by a shift by $k b^{-1}/2$ (the second one being a half--integral shift for the argument of the double Gamma function).

The twistorial $\Upsilon$--function is defined as\begin{equation}
\Upsilon_{b,\bar b, \theta}(\alpha,\bar\alpha, \phi)=\exp\!\left(\frac{1}{\pi}\,\mathrm{Im}\int_0^\infty\,\frac{ds}{s}\,
\log\boldsymbol{\Psi}\big(Z(s);q(s)\big)\right)
\end{equation}
where
\begin{align}
q(s)&=\exp\!\Big[-2\pi b^2/s+i\theta-2\pi \bar b^2\,s\Big]\\
Z(s)&= \exp\!\left[-\frac{2\pi}{s}b\!\left(\alpha-\frac{b}{2}\right)+i\phi-2\pi s\,\bar b\!\left(\bar \alpha-\frac{\bar b}{2}\right)\right].
\end{align}
The shift by $kQ/2$ of the argument of the classical $\Upsilon$--function is enhanced to the multiple argument shift
\begin{equation}\label{twisrorialshift}
(\alpha,\bar\alpha, \phi)\longmapsto (\alpha+kb/2,
\bar\alpha+k\bar b/2, \phi+k\pi)\qquad k\in\mathbb{Z}.
\end{equation}
Note that in the asymmetric limit the chiral block 
\eqref{twisFalpha123} becomes a function of $\alpha^\text{asym}\equiv \alpha+\phi/(2\pi b)$ only, so that the twistorial shift
\eqref{twisrorialshift} reproduces the usual shift $\alpha^\text{asym}\to \alpha^\text{asym}+k Q/2$ in this limit. The twistor version of reflection symmetry 
\eqref{reflsssym} is\footnote{ The two sides of
\eqref{reftwis} differ by a factor which cancels in the ratios of twistorial $\Upsilon$--functions which express the metric of any Abelian $tt^*$ geometry.}
\begin{equation}\label{reftwis}
\Upsilon_{b,\bar b,\theta}(b-\alpha,\bar b-\bar\alpha, 2\pi-\phi)\longleftrightarrow \Upsilon_{b,\bar b,\theta}(\alpha,\bar\alpha, \phi),
\end{equation}
which is again the (twistor version of) sign flip followed by a shift by $Q$.

From section \ref{s:doublepenner}, the $tt^*$ metric of the double Penner model may be written as (we neglect an elementary factor which may be absorbed in the normalization, and omit writing the barred variables)
\begin{equation}
G=\frac{\Upsilon_{b,\bar b,\theta}(0,0)}{\Upsilon_{b,\bar b,\theta}(\sum_j\alpha_j-b, \sum_j\phi_j-2\pi)}
\prod_{i=1}^3\frac{\Upsilon_{b,\bar b,\theta}(2\alpha_i,2\phi_i)}{\Upsilon_{b,\bar b,\theta}(\sum_j\alpha_j-2\alpha_i,\sum_j\phi_j-2\phi_i)}
\end{equation}
where the various parameters are related to the LG model couplings as in eqns.\eqref{dict1}--\eqref{dictn}.
This result
has the same form as the DOZZ expression \cite{Zamolodchikov:1995aa}
for the Liouville $3$--point function with
the ordinary $\Upsilon_b$--functions replaced by their twistorial counterparts using the dictionary
(here $n_i$, $k$ are arbitrary integral coefficients)
\begin{equation}
\Upsilon_b\big(n_i\alpha_i+k Q/2\big)\rightarrow\Upsilon_{b,\bar b,\theta}\big(n_i\alpha_i+k b/2,
n_i\bar\alpha_i+k \bar b/2, n_i\phi_i+k\pi\big).
\end{equation} 
As in the case of the twistorial $\Gamma$--function, \S.\ref{twistgammafun}, the above `twistorial extension map' preserves the functional identities. 

The twistorial $\Upsilon$--function may be written as a product of two twistorial double Gamma functions, generalizing eqn.\eqref{upsordinary} to the twistorial set up. Of course, the metric of all
Abelian $tt^*$ geometries may be written in terms of twistorial $\Upsilon$--functions.  

Again the extension to Toda $3$--point function is straightforward in view of appendix \ref{appendixmultimatrix}.

\section{The $C$-limit} \label{s:c-limit}

In this section we consider the $C$-limit which we first defined in \S\ref{s:c-limit-def} above, 
by first taking the $\theta$-limit $\eps_1 \to 0$ and then
taking the further limit $\theta \to 0$.  We provide 
evidence that in this limit the partition function
becomes computable in terms of a pure classical geometric object developed in \cite{Neitzke:2011za}, 
which in turn was based on the hyperK\"ahler geometry studied in \cite{Gaiotto:2008cd}.
We also give evidence that, if we take the asymmetric limit of the $C$-limit,
then we recover the NS limit of the topological string partition function,
as claimed in \S\ref{s:corztons} above.

\subsection{The $C$--limit amplitude in SQED} \label{s:c-limit-sqed}

Consider again the SQED brane amplitude in the $\theta$--limit, which is given by
\begin{multline}
\log\psi^\theta= {{W^{NS}(a,\theta {\tilde \epsilon_2}\zeta)\over \zeta {\tilde \epsilon_2}}}+{{\zeta {\overline W}^{NS}({\overline a},\theta {\tilde \epsilon_2}/\zeta)\over {\tilde \epsilon_2}}}
+\frac{1}{2\pi i}\int_{\ell_e} \frac{d\zeta^\prime}{\zeta^\prime}\;\frac{\zeta^\prime+\zeta}{\zeta-\zeta^\prime}\log\boldsymbol{\Psi}\!\big(X_e(\zeta^\prime);q\big)+\cdots
\end{multline}
and define the $C$--limit amplitude as 
\begin{equation}
\begin{split}
&\log \psi^{C}= \lim_{\theta \to 0} \left(\frac{\theta }{2\pi}\;\log \psi^\theta \right)=\\
&=\frac{R^2 \mathcal{F}(a_e)}{\zeta^2}+R^2\overline{\mathcal{F}(a_e)}\,\zeta^2+\frac{1}{8\pi^2}\int_{\ell_e}
\frac{d\zeta^\prime}{\zeta^\prime}\;\frac{\zeta^\prime+\zeta}{\zeta-\zeta^\prime}\,\mathrm{Li}_2\big(X_e(\zeta^\prime)\big)+\cdots
\end{split}\label{nsndampl}
\end{equation}
Here $\mathcal{F}(a_e)$ is the prepotential of the ${\cal N}=2$ theory in the electric basis.

On the hyperK\"ahler manifold $\mathcal{M}$
describing SQED compactified on a circle, let us consider the one--form 
\begin{equation}
\eta(\zeta):=\frac{1}{4\pi^2}\,\log X_m(\zeta)\, \frac{d X_e(\zeta)}{X_e(\zeta)}-d\log \psi^C(\zeta),
\end{equation}
which is a primitive for the canonical hyperK\"ahler symplectic form
\begin{equation}
\Omega(\zeta)\equiv \frac{1}{4\pi^2}\,\frac{dX_m(\zeta)}{X_m(\zeta)}\wedge \frac{d X_e(\zeta)}{X_e(\zeta)}= d\eta(\zeta).
\end{equation}
Since $\Omega(\zeta)$ is invariant under all symplectomorphisms, under a KS symplectomorphism associated to a BPS state $\eta(\zeta)$ may change only by a closed form.
In fact, we claim that $\eta(\zeta)$ is smooth across the BPS rays. Indeed at the BPS ray $\ell_e$
\begin{align}
\log X_m(\zeta) &\longrightarrow \log X_m(\zeta)-\log\!\big(1-X_e(\zeta)\big)\\
d\log\psi^C(\zeta)&\longrightarrow
d\log\psi^C(\zeta)-\frac{1}{4\pi^2}\,\log\!\big(1-X_e(\zeta)\big)\,\frac{dX_e(\zeta)}{X_e(\zeta)}.
\end{align}
The property that $\eta(\zeta)$ is globally holomorphic as a function of $\zeta$ determines $\log\psi^C(\zeta)$ up to a globally defined function on $\mathbb{C}^\times$ which is easily fixed using the behavior at the North and South poles. Therefore,
this property, together with the prescribed behavior for $\zeta\to 0,\infty$, may be taken to be \emph{the definition} of the brane amplitude in the $C$-limit. Then, if we are interested only in the $C$--limit, we may dispense ourselves of all the intricacies of the twistorial $tt^*$ geometry and focus on the simpler hyperholomorphic geometry of $\mathcal{M}$.
This reinterpretation of the $C$-limit brane amplitude
in terms of hyperK\"ahler geometry
allows us go beyond the simple $\mathcal{N}=2$ model which have an Abelian $tt^*$ geometry, and study the amplitudes of more interesting 4d theories directly in the $C$-limit.

The study of the $C$-limit of brane amplitudes for general 4d $\mathcal{N}=2$ from the viewpoint of hyperK\"ahler geometry is the main goal of the rest of this section.
We close this introductory section with an elementary comment.
The amplitude $\log\psi^C(\zeta)$ is fixed by the requirement that its discontinuities across the BPS rays are the Hamilton--Jacobi generating function for the corresponding Kontsevich--Soibelman symplectomorphism. As it is well--known, the Hamilton--Jacobi function depends on
the choice of the contact form. In SQED there are only electrically charged BPS particles, and it is natural to choose a one--form $\eta(\zeta)$ proportional to $dX_e(\zeta)$ (up to exact terms). 
In the general case, it is more natural to make a choice which is symmetric between electric and magnetic
\begin{equation}
\varpi(\zeta)^\mathrm{sym}= \frac{1}{8\pi^2}\left(\log X_m(\zeta)\,\frac{dX_e(\zeta)}{X_e(\zeta)}-\log X_e(\zeta)\, \frac{dX_m(\zeta)}{X_m(\zeta)}\right)-d\log \psi^C(\zeta)^\text{sym}
\end{equation} which is also a primitive of the hyperK\"ahler symplectic form $\Omega(\zeta)=d\varpi(\zeta)^\text{sym}$. Clearly, the symmetric amplitude $\psi^C(\zeta)^\text{sym}$ is related to the original one by a trivial transformation
\begin{equation}
\log \psi^C(\zeta)^\text{sym}=\log\psi^C(\zeta)-\frac{1}{8\pi^2}\log X_e(\zeta)\;\log X_m(\zeta).
\end{equation}
Across the BPS line $\ell_e$ then
\begin{equation}
\begin{split}
&d\log \psi^C(\zeta)^\text{sym}\longrightarrow\\
&\rightarrow d\log \psi^C(\zeta)^\text{sym}-\frac{1}{4\pi^2}\,\log\!\big(1-X_e(\zeta)\big)\,\frac{dX_e(\zeta)}{X_e(\zeta)}
+\frac{1}{8\pi^2}d\Big(\log X_e(\zeta)\;\log(1- X_e(\zeta))\Big)=\\
&=d\log \psi^C(\zeta)^\text{sym}+\frac{1}{4\pi^2} d \,L(X_e(\zeta)),
\end{split}
\end{equation}
where $L(z)=\mathrm{Li}_2(z)+\tfrac{1}{2} \log z \log(1-z)$ is the Roger dilogarithm. Hence the symmetric version of the brane amplitude has the form as the \textsc{rhs} of eqn.\eqref{nsndampl} with the Euler dilogarithms $\mathrm{Li}_2(X_e)$ replaced by the Rogers ones $L(X_e)$.

\subsection{Review on the $X_\gamma$}

Let us now briefly review the relevant facts about the hyperK\"ahler geometry of $\cM$;
some of them have been used earlier in this paper, and in the previous section 
in the special case of SQED, but now we want to formulate things in a way that 
goes beyond the SQED example.

Being hyperK\"ahler, $\cM$ carries in particular a family of holomorphic symplectic structures
$\varpi_\zeta$, labeled by the twistor parameter $\zeta \in \C^\times$.
Around any generic point of $\cM$, for any generic $\zeta \in \C^\times$,
$\cM$ also carries canonical local holomorphic coordinates.
These coordinates, denoted $X_\gamma(\zeta)$, are 
labeled by $\gamma$ in the IR charge lattice $\Gamma$.  Physically, these coordinates
can be thought of as the vevs of IR line defects wrapped on $S^1$.  
Their analytic properties as functions of $\zeta$ 
are somewhat subtle.  Asymptotically as $\zeta \to 0$
or $\zeta \to \infty$ they behave as 
\begin{equation} \label{x-asymptotics}
X_\gamma \sim c_\gamma \exp \left( \pi R \zeta^{-1} Z_\gamma + \I \theta_\gamma + \pi R \zeta \bar{Z}_\gamma \right)
\end{equation}
where $c_\gamma$ is real and independent of $\zeta$.  Despite the fact that these asymptotics are continuous,
the actual functions $X_\gamma(\zeta)$ are not:  rather, they are \ti{piecewise} continuous.
Their discontinuities occur at the ``BPS rays,'' defined as the loci where there exists a BPS state
of charge $\gamma$, and $Z_\gamma / \zeta$ is a negative real number.  When $\zeta$ crosses such a locus
in the clockwise direction, the functions $X_\gamma$ jump by the transformation
\begin{equation} \label{x-jumps}
X'_\mu = X_\mu (1 - X_\gamma)^{\Omega(\gamma) \IP{\mu,\gamma}}
\end{equation}
where $\Omega(\gamma)$ is the BPS index counting states of charge $\gamma$ (second helicity 
supertrace).
The asymptotics \eqref{x-asymptotics} and jumps \eqref{x-jumps} are actually sufficient
to characterize $X_\gamma$, and even give a useful scheme for \ti{computing} $X_\gamma$ 
in practice.
Namely, $X_\gamma(\zeta)$ can be written in the form
\begin{equation}
X_\gamma(\zeta) = X_\gamma^\sf(\zeta) X_\gamma^\inst(\zeta)
\end{equation}
where $X_\gamma^\sf$ is given by an explicit formula
\begin{equation}
X^\sf_\gamma(\zeta) = \exp \left[ \frac{\pi R}{\zeta} Z_\gamma + \I \theta_\gamma + \pi R \zeta \bar{Z}_\gamma \right]
\end{equation}
and the ``instanton corrections'' $X_\gamma^\inst$ 
are determined by the TBA-like integral equation
\begin{equation} \label{x-tba}
X_\gamma^\inst(\zeta) = \exp\!\!\left[ -\frac{1}{4
\pi \I} \sum_{\gamma'} \Omega(\gamma') \langle \gamma,\gamma'
\rangle \int_{\ell_{\gamma'}} \frac{\de \zeta'}{\zeta'} \frac{\zeta' +
\zeta}{\zeta' - \zeta} \log (1 - X_{\gamma'}(\zeta'))\right],
\end{equation}
where the $\Omega(\gamma')$ are the BPS degeneracies.

The usefulness of the $X_\gamma$ comes partly from the fact that they are Darboux coordinates.
Indeed, suppose we introduce the (multivalued) logarithms
\begin{equation}
x_\gamma = \frac{1}{2 \pi \I} \log X_\gamma.
\end{equation}
Then the holomorphic symplectic form on $\cM$ takes the simple shape
\begin{equation}
\Omega_\zeta = - \frac{1}{2 R} \IP{\de x, \de x}
\end{equation}
or more concretely, choosing a basis $\{\gamma_i\}$ for $\Gamma$ and setting $\eps_{ij} = \IP{\gamma_i, \gamma_j}$, $x_{i} = x_{\gamma_i}$,
\begin{equation}
\Omega_\zeta = - \frac{1}{2R} \sum_{i,j} \eps_{ij} \de x_i \wedge \de x_j.
\end{equation}

\subsection{Prequantization and $\Psi$}

Now let us restrict attention to theories which in the IR do not have any continuous flavor symmetries
(so e.g.\! pure $\N=2$ super Yang-Mills would qualify, but not $\N=2$ super Yang-Mills coupled to matter.)
The inclusion of flavor symmetries would require a more elaborate formalism.

Under this restriction, 
the hyperK\"ahler moduli space 
$\cM$ of the theory carries a distinguished line bundle $V$, which is closely related to our
interpretation of $\Psi$.  We will not review the full 
explicit construction of $V$ here; for that see \cite{Neitzke:2011za}.
For our purposes in this section, we want to emphasize a feature which was not discussed explicitly 
there, namely, $V$ is a kind of ``prequantum line bundle'' for the holomorphic symplectic
structures $\Omega_\zeta$.  Indeed, $V$ carries a family of holomorphic connections
$\nabla(\zeta)$, such that the curvature of $\nabla(\zeta)$ is
\begin{equation}
F_{\nabla(\zeta)} = - 2 \pi \I R \Omega_\zeta = \pi \I \IP{\de x, \de x}.
\end{equation}
Moreover, like the holomorphic symplectic structures, the connections $\nabla(\zeta)$ can also 
be put into a simple canonical form:  indeed there exist canonical local sections $\Psi$
of $V$ such that $\nabla(\zeta)$ is represented by the 1-form
\begin{equation} \label{nabla-gauge}
\cA(\zeta) = \pi \I \IP{x,\de x} = \pi \I \sum_{i,j} \eps_{ij} x_i  \de x_j.
\end{equation}
Said otherwise, we have
\begin{equation}
\nabla(\zeta) \Psi = \pi \I \IP{x,\de x} \Psi.
\end{equation}
Equivalently, $\nabla(\zeta)$ is represented as
\begin{equation}
\nabla(\zeta) = \de + \eta(\zeta),
\end{equation}
where
\begin{equation}
\eta(\zeta) = \pi \I \IP{x(\zeta), \de x(\zeta)} - \de \log \Psi(\zeta).
\end{equation}

Let us now describe $\Psi$ more concretely.  For this we must 
first explain how $V$ is defined.  A function on $\cM$
can be represented as $f(u^i, \theta_j)$ where $u^i$ are local coordinates on the Coulomb branch,
$\theta_j$ are linear coordinates on the torus fibers of $\cM$ (with respect to some fixed basis $\gamma_j$
of $\Gamma$), and $f$ is \ti{periodic} under shifts
of the $\theta_j$ by multiples of $2\pi$.  Similarly, a section of $V$ over $\cM$ may be represented
as a function $s(u^i, \theta_j)$ 
which obeys\footnote{ We are suppressing a ``twisting'' discussed in \cite{Neitzke:2011za}, which
introduces some extra minus signs into the story, but will play no important role here.}
\begin{equation} \label{v-periodicity}
s(u^i, \theta_j + 2\pi) = e^{\I \eps_{ij} \theta_i / 2} s(u^i, \theta_j).
\end{equation}
In other words, we consider $s$ to be an honest periodic section of $V$ if it is represented
by a function with this twisted periodicity.  (In particular, it follows that 
$V$ is a topologically nontrivial bundle:  it admits 
no global nonvanishing section, even on a single torus fiber of $\cM$.)

Now $\Psi$ is given by a formula
\begin{equation} \label{psi-explicit}
 \Psi = \Psi^\sf \Psi^{\inst,1} \Psi^{\inst,2}
\end{equation}
where
\begin{equation} \label{eq:psi-sf}
 \Psi^{\sf} = \exp \left[\frac{\I \pi R^2}{4} \left( \zeta^{-2} U + \zeta^2 \bar{U} \right) - \frac{R}{4} \left(\zeta^{-1} C + \zeta \bar{C} \right) \right],
\end{equation}
with
\begin{align}
U &= \int \IP{Z, \de Z}, \label{def-U} \\
C &= \IP{Z, \theta},
\end{align}
and
\begin{align} \label{eq:psiinst1}
 \Psi^{\inst,1} &= \exp \left[ - \sum_\gamma \frac{\Omega(\gamma)}{16 \pi^2} \int_{\ell_\gamma} \frac{\de \zeta'}{\zeta'} \frac{\zeta'+\zeta}{\zeta'-\zeta}
\left[ 2 \Li_2(X_\gamma(\zeta')) + \log X_\gamma(\zeta') \log (1 - X_\gamma(\zeta')) \right]\right],\\ \label{eq:psiinst2}
 \Psi^{\inst,2} &= \exp \left[ - \sum_\gamma \frac{\Omega(\gamma)}{16 \pi^2} \int_{\ell_\gamma} \frac{\de \zeta'}{\zeta'} \frac{\zeta'+\zeta}{\zeta'-\zeta}
\left[ (\log X^\sf_\gamma(\zeta') - \log X^\sf_\gamma(\zeta)) \log (1 - X_\gamma(\zeta')) \right]\right].
\end{align}
As written, $\Psi$ is only a local section of $V$, 
because it does not obey the periodicity \eqref{v-periodicity}.
Nevertheless we can use it to define a local gauge for $V$, and relative to this local gauge,
$\nabla(\zeta)$ is given by \eqref{nabla-gauge}.

In this construction it is important that, as $\zeta$ crosses a BPS ray, $\Psi(\zeta)$ transforms
by
\begin{equation}
\Psi' = \Psi \exp \left( \frac{1}{2 \pi \I} (\Li_2(X_\gamma) + \frac{1}{2} \log(X_\gamma) \log(1 - X_\gamma)) \right)
\end{equation}
Indeed, this transformation law, combined with \eqref{x-jumps} above, guarantees that the
connection form $\eta(\zeta)$ is continuous --- the jumps
of $X$ and $\Psi$ cancel one another.

A second important property of the construction is that the connections $\nabla(\zeta)$ are
well behaved in the limit $\zeta \to 0, \infty$; this gives a further constraint on the
form of $\Psi(\zeta)$.  What ``well behaved'' means precisely is
that the $(0,1)_\zeta$ part of $\nabla(\zeta)$ has a finite limit as $\zeta \to 0$ or $\zeta \to \infty$.
This property was used
in \cite{Neitzke:2011za} to build a \ti{hyperholomorphic} structure on $V$, i.e.\! a single unitary connection
$D$ in $V$ such that the curvature $F_D$ is of type $(1,1)_\zeta$ for all $\zeta \in \C{\mathbb{P}}^1$.
Indeed, for every $\zeta$, the $(0,1)_\zeta$ part of the connection $\nabla(\zeta)$ agrees with the 
$(0,1)_\zeta$ part of $D$; this is one way of characterizing $D$.

Much as with our previous discussion of $X_\gamma$, 
these two properties of $\Psi$ --- its asymptotics and its jumps at the BPS rays 
--- are sufficient to determine
$\Psi$ completely.  They were what motivated the complicated explicit formula \eqref{psi-explicit} 
above.

\subsection{Comparing $\Psi$ with $\psi^C$}

Now, in the special case of SQED, we would like to connect $\Psi$ with the object $\psi^C$ which we obtained 
as the $C$-limit of the twistorial topological string.
Roughly the relation is that $\Psi$ is the ``symmetric'' version $({\Psi^C})^{sym}$
which we introduced at the end of \S\ref{s:c-limit-sqed}. 

Now let us say this more precisely.
Given an electric-magnetic splitting, we can define an unsymmetrized object by
\begin{equation}
 \tilde\Psi = \Psi \exp \left( - \frac{\I \pi R^2}{2} \left( \zeta^{-2} U + \zeta^2 \bar{U} \right) - \pi \I \IP{x^e, x^m} \right)
\end{equation}
If there are no magnetically charged states, then a short computation shows that 
$\tilde\Psi$ can be given more directly as
\begin{equation}
 \tilde\Psi = \tilde\Psi^\sf \tilde\Psi^{\inst,1} \tilde\Psi^{\inst,2}
\end{equation}
where
\begin{equation}
 \tilde \Psi^{\sf} = \exp \left[\frac{\I \pi R^2}{2} \left( \zeta^{-2} {\mathcal F} + \zeta^2 \bar{{\mathcal F}} \right) - \frac{R}{4} \left(\zeta^{-1} \tilde W + \zeta \bar{\tilde W} \right) \right],
\end{equation}
with
\begin{align}
\tilde W &= 2 \IP{Z^e, \theta^m},
\end{align}
and
\begin{align}
 \tilde\Psi^{\inst,1} &= \exp \left[ - \sum_\gamma \frac{\Omega(\gamma)}{4 \pi^2} \int_{\ell_\gamma} \frac{\de \zeta'}{\zeta'} \frac{\zeta'+\zeta}{\zeta'-\zeta}
 \Li_2(X_\gamma(\zeta')) \right], \\
 \tilde \Psi^{\inst,2} &= \exp \left[ - \sum_\gamma \frac{\Omega(\gamma)}{8 \pi^2} \int_{\ell_\gamma} \frac{\de \zeta'}{\zeta'} \frac{\zeta'+\zeta}{\zeta'-\zeta}
\left[ (\log X^\sf_\gamma(\zeta') - \log X^\sf_\gamma(\zeta)) \log (1 - X_\gamma(\zeta')) \right]\right].
\end{align}
In particular, the Rogers dilogarithm which appeared in $\Psi^{\inst,1}$
has been replaced by the ordinary $\Li_2$ in $\tilde\Psi^{\inst,1}$.

Now, in the special case of SQED,
we would like to compare this unsymmetrized $\tilde\Psi$ with the
$C$-limit amplitude $\psi^C$ given in \eqref{nsndampl}.
Naively the two cannot match since $\tilde\Psi$ is a function depending on $(a, \bar{a}, \theta^e, \theta^m)$
while $\psi^C$ does not involve $\theta^m$. Nevertheless the two are very similar.  Indeed,
the part of $\tilde\Psi^\sf$ involving ${\mathcal F}$ and $\bar{\mathcal F}$, 
and the instanton terms $\tilde\Psi^{\inst,1}$, match with corresponding terms in $\psi^C$.  
The remaining terms in $\tilde\Psi$
can be rewritten in the form
\begin{equation}
\exp \left[ - \frac{R}{2} \left(\zeta^{-1} \IP{Z^e, \hat \theta^m_+} + \zeta \IP{\bar{Z}^e, {\hat{\theta}}^m_-} \right) \right],
\end{equation}
where $\hat \theta^m_\pm$ is the quantum-corrected version of $\theta^m$ (called $\Upsilon_{0/\infty}$ in \cite{Gaiotto:2008cd}),
\begin{equation}
\hat \theta^m_\pm = \theta^m \pm \sum_\gamma \frac{\Omega(\gamma)}{8 \pi^2} \int_{\ell_\gamma} \frac{\de \zeta'}{\zeta'} \log (1 - X_\gamma(\zeta')).
\end{equation}
Thus, if we formally set these to zero, $\hat \theta^m_{\pm} = 0$, then we get agreement 
\begin{equation}
 \tilde\Psi = \psi^C.
\end{equation}

Since the $C$-limit of the twistorial topological string arises from the $\theta$-limit and the associate
quantum Riemann-Hilbert problem fixing it is symmetric between electric and magentic degrees of freedom,
makes us believe that it should possible also to recover the magnetic angles in the classical limit
making the above equality more general.  Moreover we expect this to extend to arbitrary theories,
and not just SQED.  Again this is natural because the classical limit of the quantum Riemann Hilbert
problem and what chracterizes the topological string wave function seem to formally reduce
to the above partition function in the $C$-limit.

\subsection{$\Psi$ as a generating function}

In \cite{Neitzke:2011za} the object $\Psi$ played a sort of auxiliary role; it was key for the construction of
$D$, but it was not given a direct 
physical interpretation.  Now we want to explain one place where $\Psi$ appears more directly.  

We consider the asymmetric limit
\begin{equation} \label{conformal-limit}
R \to 0, \qquad \zeta \to 0, \qquad \eps = \zeta / R \text{ fixed}.
\end{equation}
and specialize to the subset $L \subset \cM$ given by
\begin{equation} \label{zero-section}
\theta_\gamma = 0.
\end{equation}
Before taking the limit \eqref{conformal-limit}, the locus \eqref{zero-section} is not geometrically
distinguished as a subset of the complex manifold $\cM(R, \zeta)$:  in particular it is not a 
complex submanifold.  However, after taking this limit,
it was proposed in \cite{Gaiotto:2014bza} that $L$ becomes a complex Lagrangian submanifold of the limiting
manifold $\cM(\eps)$.
In this section we will explain that, in the limit \eqref{conformal-limit}, 
$\log \Psi$ has an interpretation as \ti{generating function}
for this Lagrangian submanifold, in the coordinates $x_\gamma$.
(To be precise, we will show this only in some special theories such as Argyres-Douglas
theories and $U(1)$ SQED, where we have sufficiently good understanding of how the $X_\gamma$
behave in the asymmetric limit; but we believe it should hold more generally.)

Let us say more precisely what we mean by ``generating function.''  We must 
first choose an electric-magnetic splitting.
Using this splitting we can define the unsymmetrized $\tilde \Psi$ which appeared in the last section.
Our generating function will be a slightly modified version of $\log \tilde \Psi$.  Indeed, in the asymmetric limit
$\log \tilde \Psi$ diverges; fortunately this
divergence can be removed by subtracting a function of $R$ alone; call
the result $\log \tilde \Psi^{reg}$.  We will define
\begin{equation} \label{def-W}
\cW = - \frac{1}{2 \pi \I} \left(\log \tilde \Psi^{reg} + G \right)
\end{equation}
where $G$ is an $\eps$-independent function of the Coulomb branch 
parameters, reflecting a kind of ``1-loop holomorphic anomaly'' for $\log \tilde \Psi^{reg}$.
Then what we will show is that $\cW$ is a generating function for $L$, i.e. along $L$ we may
write the $x_{m_i}$ as holomorphic functions of the $x_{e_i}$, and those functions are given by
\begin{equation} \label{generating}
 x_{m_i} = \frac{\partial \cW}{\partial x_{e_i}}.
\end{equation}

The formula \eqref{generating} gives evidence that $\cW$ can be identified with the Nekrasov-Shatashvili
limit $W^{NS}$ of the instanton partition function of the $\N=2$ theory.
Indeed, it was proposed in \cite{Gaiotto:2014bza}
that, in theories of class $S$, 
in the $R \to 0$ limit $L$ can be identified with the locus of opers in $\cM(\eps)$.
On the other hand, it was proposed in \cite{Nekrasov:2011bc} (and verified in some examples) 
that, in theories of class $S[A_1]$, 
the generating function of the locus of opers
should be $W^{NS}$.\footnote{The coordinate system used in \cite{Nekrasov:2011bc} was not identified there with the coordinate system
$(x_{e_i}, x_{m_i})$ which we are using here.  Rather, it was described in geometric language,
in which it appeared as a complexification of a Fenchel-Nielsen-type
coordinate system on a moduli space of flat $SL(2)$-connections.  
However, recently in \cite{Hollands:2013qza} it has been shown that such complexified Fenchel-Nielsen-type
coordinates \ti{do} arise as $(x_{e_i}, x_{m_i})$ in theories of class $S[A_1]$!
(More precisely, if we evaluate $(x_{e_i}, x_{m_i})$ on the distinguished ``real'' locus of 
the Coulomb branch, where the periods $a_i / \zeta$ are real and negative,
then on this locus they agree with complexified Fenchel-Nielsen-type coordinates.)}
Thus, in theories of class $S[A_1]$, we conclude that 
$\cW = W^{NS}$ up to a constant shift, since they are both giving generating functions 
for the same locus $L$ in $\cM(\eps)$.  This is in accord with our general expectations about 
twistorial topological strings as we explained in \S\ref{s:corztons} above.

It is natural to conjecture that the 
same identification holds for a general theory, not only for class $S[A_1]$.
It would be very interesting to verify this identification more directly.

\subsection{Deriving the generating function} \label{deriving-generating}

In this section, we explain how the key relation \eqref{generating} is obtained.
The main player will be the connection 1-form $\eta(\zeta)$ we reviewed above,
\begin{equation} \label{eta-repeat}
\eta(\zeta) = \pi \I \IP{x(\zeta), \de x(\zeta)} - \de \log \Psi(\zeta).
\end{equation}

We begin by noting that
$\eta(\zeta)$ depends holomorphically on $\zeta \in \C^\times$.
As $\zeta \to 0$ or $\zeta \to \infty$ we can study it explicitly,
just because we know the asymptotic behavior of $x_\gamma$ and $\Psi$.
Indeed, we have already written the asymptotics of $x_\gamma$ above in \eqref{x-asymptotics},
and as for $\Psi$, expanding $\Psi^{\inst,1}$ and $\Psi^{\inst,2}$ 
around $\zeta = 0$ reveals that they do not contribute
at leading order:  we just get the asymptotics of $\Psi^\sf$, which gives
\begin{equation} \label{psi-asymptotics}
\Psi = \exp \left[ \frac{\I \pi R^2}{4 \zeta^2} U + \cdots \right].
\end{equation}
Combining \eqref{x-asymptotics} and \eqref{psi-asymptotics} we get directly
\begin{equation}
\eta = - \frac{\I \pi R^2}{2} \IP{Z, \de Z} + \cdots
\end{equation}
and similarly expanding around $\zeta = \infty$
we can complete this expansion to
\begin{equation} \label{eta-expansion}
\eta = - \frac{\I \pi R^2}{2 \zeta^2} \IP{Z, \de Z} + \cdots - \frac{\I \pi R^2 \zeta^2}{2} \IP{\bar Z, \de \bar Z},
\end{equation}
where $\cdots$ represents terms of order $1/\zeta$, $1$, and $\zeta$.
These terms can be written out as well, but they are considerably more complicated, involving the 
BPS degeneracies $\Omega(\gamma)$ and
integrals over the BPS rays $\ell_\gamma$.

When we restrict to $L$, $\eta(\zeta)$ simplifies:  
we obtain an extra symmetry $\zeta \to -\zeta$ which implies 
the terms of order $1/\zeta$ and $\zeta$ drop out, so that we have 
\begin{equation} \label{eta-short}
\eta = - \frac{\I \pi R^2}{2 \zeta^2} \IP{Z, \de Z} + \eta_0 - \frac{\I \pi R^2 \zeta^2}{2} \IP{\bar Z, \de \bar Z},
\end{equation}
for some 1-form $\eta_0$.
In the limit \eqref{conformal-limit}, the expansion further simplifies, to
\begin{equation} \label{eta-reduced}
\eta(\zeta) = - \frac{\I \pi}{2 \eps^2} \IP{Z, \de Z} + \lim_{R \to 0} \eta_0
\end{equation}
so long as $\lim_{R \to 0} \eta_0$ exists.
We expect that this limit does indeed exist and moreover it is a \ti{closed} form,
so that locally we can write
\begin{equation} \label{eta-closed}
 \lim_{R \to 0} \eta_0 = \de G
\end{equation}
for some function $G$; in what follows we assume this is true.
In Appendix \ref{eta-limit} below we show that \eqref{eta-closed} does hold
at least for Argyres-Douglas theories and for $U(1)$ SQED, and incidentally that 
for $U(1)$ SQED we have the explicit formula
\begin{equation}
G = - \frac{1}{48} \log (a/\bar{a}).
\end{equation}

Now combining \eqref{eta-repeat} and \eqref{eta-reduced} we have
\begin{equation}
\de \log \Psi(\zeta) = \pi \I \IP{x(\zeta), \de x(\zeta)} + \frac{\pi \I}{2 \eps^2} \IP{Z, \de Z} - \de G.
\end{equation}
Using the definition of $U$ 
and rearranging, this becomes
\begin{equation} \label{generating-symmetric}
\de \left( \log \Psi(\zeta) - \frac{\pi \I}{2 \eps^2} U + G \right) = \pi \I \IP{x(\zeta), \de x(\zeta)}
\end{equation}
This is essentially the result we want.  To put it in precisely the shape \eqref{generating}
we need to make a further slight adjustment.  The right side of \eqref{generating-symmetric}
can be written explicitly in coordinates as
\begin{equation}
\pi \I (x_{e_i} \de x_{m_i} - x_{m_i} \de x_{e_i})
\end{equation}
Thus adding $- \pi \I \de(x_{e_i} x_{m_i})$ to both sides we obtain
\begin{equation}
\de \left( \log \Psi(\zeta) - \pi \I x_{e_i} x_{m_i} - \frac{\pi \I}{2 \eps^2} U + G \right) = - 2 \pi \I x_{m_i} \de x_{e_i} 
\end{equation}
The right side is just $- 2 \pi \I \cW$ according to \eqref{def-W}, so finally
\begin{equation}
\de \cW = x_{m_i} \de x_{e_i}
\end{equation}
matching the desired \eqref{generating}.

\section{Concluding Remarks}
Studying the vacuum geometry for 4d ${\cal N}=2$ theories on ${1\over 2}\Omega$ background
seems to have unified a number of topics:  Topological strings, hyperK\"ahler geometries associated
to them and their quantization, wall-crossing phenomenon and BPS states, etc.  It is clear that this is just the beginning
of the exploration of this vast topic, as the twistorial topological string seems to be a rather rich object.
Related to this richness, is the complication for explicit computations.  In this paper we have
managed to solve exactly some theories which admit only electric BPS states.  Moreover
we have proposed methods to compute them in the $\theta$-limit using a quantum Riemann-Hilbert
problem.

There are many directions which are naturally suggested by this work.  First of all,
there are a few conjectures in this paper that would be nice to prove.  These include
a proof from first principles that the $\theta$-limit is indeed a solution to the quantum Riemann-Hilbert problem.
The proof that AMNP index is the same as the CFIV index.  Also a better understanding of
how both electric and magnetic angles arise in the $C$-limit would be highly desirable.
It would also be nice to find which 2d system does the twistorial version of AGT (for
which we only have computed the three point Liouville amplitude) relate to.

On another front, this work suggests that one should perhaps study more general
pair of D-brane geometries for ${\cal N}=2$ theories on $T^2\times I$, generalizing
the twistorial topological string where the two D-branes were more or less fixed. This is
very natural from the point of view of a 4d $tt^*$.
  Also
it would be interesting to explore what would happen if the length of the interval $I$ is kept
finite instead of it being infinite.  

Clearly a lot more work remains to be done.  We hope to have conveyed the intrinsic elegance of twistorial
topological strings in its ability to unify a number of different areas. 

\section*{Acknowledgements}

We thank B. Haghighat, N. Nekrasov, V. Schomerus and S. Shatashvili for helpful discussions.
SC and CV would also like to thank the hospitality of the Simons Center for Geometry
and Physics during the 12-th summer workshop on mathematics and physics. SC also thanks the Physics Department of Harvard University for hospitality.

The work of AN is supported by NSF award DMS-1151693 and the work of CV is supported in part by NSF grant PHY-1067976. 

\appendix

\section{On the $R \to 0$ limit of the connection} \label{eta-limit}

Here we tie up a loose end from Section \ref{deriving-generating}:  why does $\lim_{R \to 0} \eta_0$ exist, and why is it a
closed form?

First, $\eta_0$ can be written explicitly, by beginning from the definition
\eqref{eta-repeat} of $\eta$ and expanding the quantities $x(\zeta)$, $\Psi(\zeta)$ 
that appear there around $\zeta = 0$,
then computing the term of order $\zeta^0$.  This leads to the result
\begin{equation} \label{eta0-formula}
 \eta_0 =
 -\frac{\I \pi R^2}{4} \left(\IP{Z, \de \bar Z} + \IP{\bar Z, \de Z}\right)
- \frac{R}{8 \pi} \left[ \sum_\gamma \Omega(\gamma) \int_{\ell_\gamma} \frac{\de \zeta}{\zeta} \left( \frac{Z_\gamma}{\zeta} - \bar{Z}_\gamma \zeta \right) \de \log(1 - X_\gamma(\zeta)) \right].
\end{equation}
The first part evidently vanishes as $R \to 0$, but to understand how the second part behaves,
we need to know something about the behavior of the functions $X_\gamma(\zeta)$
in that limit.

As we have already remarked, these functions
are determined by the TBA-like integral equations \eqref{x-tba}.
The limit $R \to 0$ is the high temperature limit
in the TBA language.  
We have not studied the $R \to 0$ behavior of the $X_\gamma$ in a general $\N=2$ theory;
here we will restrict attention to a 
particular class of simple examples,
studied in \cite{Zamolodchikov:1989cf,Zamolodchikov:1991et}, which correspond 
to taking our $\N=2$ theory to be an Argyres-Douglas theory.
In these theories, as we take $R \to 0$,
the functions $X_\gamma$ restricted to the rays $\ell_\gamma$ develop 
a simple and well-known characteristic profile.  We illustrate that profile in Figure \ref{fig:AD3-x1}.
\insfigscaled{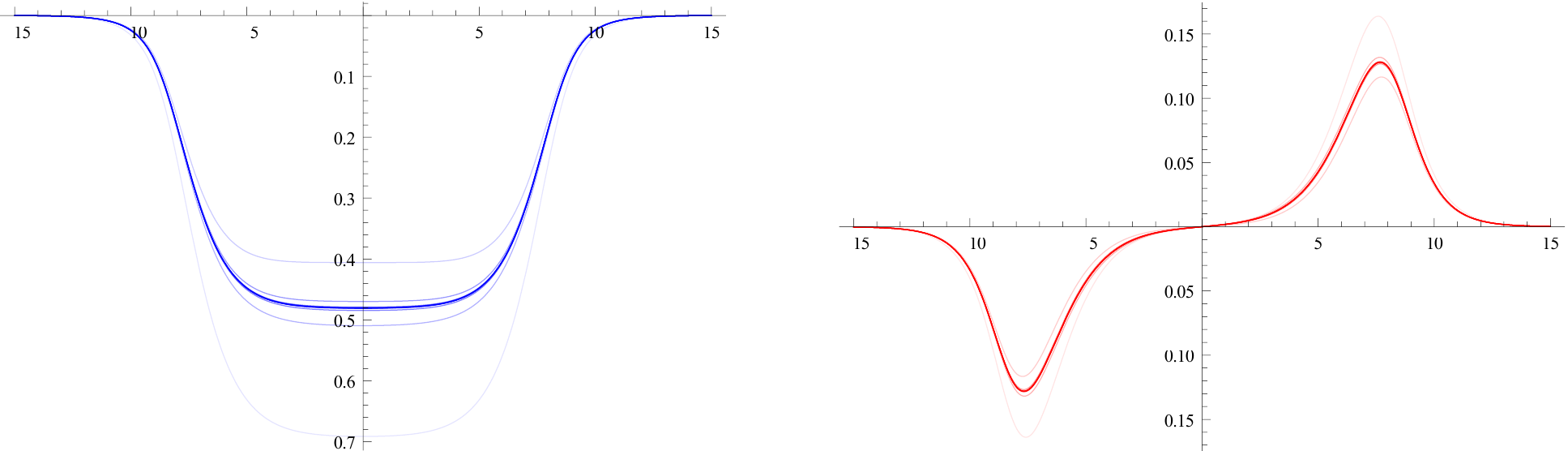}{0.7}{The instanton corrections 
$X^{\inst}_{\gamma_1}$ evaluated along the ray $\ell_{\gamma_1}$, 
at a point in the weak coupling region of the Argyres-Douglas $(A_1, A_2)$ theory, where we took 
$R = 10^{-4}$, $Z_{\gamma_1} = 1 - \frac{\I}{3}$, $Z_{\gamma_2} = 1 + \frac{\I}{2}$, $\theta_{\gamma_1} = \theta_{\gamma_2} = 0$.
The left figure gives $\re \log X^\inst_{\gamma_1}(t)$
and the right is $\im \log X^\inst_{\gamma_1}(t)$, where $t = \log \abs{\zeta}$.  To obtain these 
figures we solved the integral equation by iteration, beginning with $X=X^\sf$, and 
taking $10$ iterations; 
previous iterations are shown as light curves 
on the graph.  At $t = 0$, we get $X_{\gamma_1} \approx -0.61805$,
while the expected plateau value in this example is $\frac{1 - \sqrt{5}}{2} \approx -0.61803$.}

The plateau visible in the middle of the figure reflects the fact that
the $X_\gamma$ become approximately \ti{constant}, independent of 
$\zeta$ and also independent of the Coulomb branch moduli, over a region running
from $\abs{\zeta} \sim R$ to $\abs{\zeta} \sim 1/R$.
To either side of this plateau we see
a characteristic ``kink'' shape:  in the limit $R \to 0$,
the kink on the left depends on $\zeta$ and $R$ only through the combination $\eps = \zeta / R$,
while the one on the right depends only on $\eps' = \zeta R$.
The left kink $X_\gamma(\eps)$ depends \ti{holomorphically} on the Coulomb branch 
moduli, while the right kink depends \ti{antiholomorphically} on them.
Thus $\eta_0$ in \eqref{eta0-formula} splits into two parts:  the $(1,0)$ part receives 
contributions only from the left kink, while the conjugate $(0,1)$ part
comes from the right kink.
Thus we get
\begin{equation}
\lim_{R \to 0} \eta_0 =
- \frac{1}{8 \pi} \re \left[ \sum_\gamma \Omega(\gamma) \int_{\ell_\gamma} \frac{\de \eps}{\eps} \frac{Z_\gamma}{\eps} \de \log(1 - X_\gamma(\eps)) \right].
\end{equation}
The convergence of this integral now follows from the behavior of $X_\gamma(\eps)$
 at small and large $\eps$:
at large $\eps$, $X_\gamma(\eps)$ approaches a constant, so that 
$\de \log(1 - X_\gamma(\eps))$ vanishes; at small $\eps$, we have 
$X_\gamma(\eps) \sim \exp(\pi Z_\gamma / \eps)$,
which is exponentially decaying as $\eps \to 0$ along $\ell_\gamma$, hence so is
$\log ( 1 - X_\gamma(\eps))$.

Finally we would like to see that $\eta_0$ is also \ti{closed} in this limit.
For this we return to the definition \eqref{eta-repeat}.
Using this definition we can write
\begin{equation}
\eta_0 = \oint_{\abs{\zeta}=1} \frac{\de \zeta}{\zeta} (\pi \I \IP{x(\zeta), \de x(\zeta)} - \de \log \Psi(\zeta))
\end{equation}
and thus
\begin{equation}
\de \eta_0 = \pi \I \oint_{\abs{\zeta}=1} \frac{\de \zeta}{\zeta} \IP{\de x(\zeta), \de x(\zeta)}.
\end{equation}
We have already recalled that, 
as $R \to 0$, the functions $X_\gamma(\zeta)$ become approximately constant
along $\ell_\gamma$ in a neighborhood of $\abs{\zeta} = 1$.  To show that $\de \eta_0$
vanishes as $R \to 0$ we need to know more:  we need to know that this behavior extends away 
from the ray $\ell_\gamma$.  Fortunately, it appears (again by numerical experimentation) 
that this is indeed true:  as $R \to 0$, 
the $X_\gamma(\zeta)$ become piecewise constant on a full annulus around $\abs{\zeta} = 1$
(only ``piecewise'' because $X_\gamma(\zeta)$ is discontinuous as a function of $\zeta$ 
when $\zeta$ crosses some of the rays $\ell_{\gamma'}$).
In particular, since this constant is independent of the Coulomb branch moduli, 
$\de x(\zeta)$ approaches $0$ in this limit.
It follows that $\lim_{R \to 0} \eta_0$ is closed as desired.
(This is essentially the statement that as $R \to 0$ the locus $L$ becomes Lagrangian as a subspace
of $\cM(\eps)$, and our argument here is essentially the same as one given in \cite{Gaiotto:2014bza}.)

We now consider one example where the whole story just described becomes completely explicit:
we take our $\N=2$ theory to be the $U(1)$ gauge theory coupled to 1 hypermultiplet with charge
1.  In this theory the charge lattice is spanned by generators $\gamma_e$, $\gamma_m$ (``electric''
and ``magnetic'' respectively), with $\IP{\gamma_e, \gamma_m} = 1$.  The central charges,
as functions of the Coulomb branch modulus $a$, are
\begin{equation}
Z_{\gamma_e} = a, \qquad Z_{\gamma_m} = \frac{1}{2 \pi \I}(a \log (a / \Lambda) - a),
\end{equation}
and the BPS counts are
\begin{equation}
\Omega(\gamma) = \begin{cases} 1 \text{ if } \gamma = \pm \gamma_e, \\
0 \text{ otherwise}.
\end{cases}
\end{equation}

Then \eqref{eta0-formula} specializes to
\begin{equation}
\eta_0 = -\frac{\I \pi R^2}{4} \left(\IP{Z, \de \bar Z} + \IP{\bar Z, \de Z}\right) + \eta_0^\inst
\end{equation}
where
\begin{align}
 \eta_0^\inst =
- \frac{R}{8 \pi} \Bigg[ & \int_{\ell_{\gamma_e}} \frac{\de \zeta}{\zeta} \left( \frac{a}{\zeta} - \bar{a} \zeta \right) \de \log(1 + e^{\pi R a / \zeta + \pi R \bar{a} \zeta})  \\ - &\int_{\ell_{-\gamma_e}} \frac{\de \zeta}{\zeta} \left( \frac{a}{\zeta} - \bar{a} \zeta \right) \de \log(1 + e^{-\pi R a / \zeta - \pi R \bar{a} \zeta}) \Bigg],
\end{align}
and after taking $\zeta \to -\zeta$ in the second integral this becomes
\begin{align}
 \eta^\inst_0 &=
- \frac{R}{4 \pi} \int_{\ell_{\gamma_e}} \frac{\de \zeta}{\zeta} \left( \frac{a}{\zeta} - \bar{a} \zeta \right) \de \log(1 + e^{\pi R a / \zeta + \pi R \bar{a} \zeta}) \\
&= \frac{R^2 \de a}{4} \int_{\ell_{\gamma_e}} \frac{\de \zeta}{\zeta} \left( \frac{a}{\zeta} - \bar{a} \zeta \right) \frac{1}{\zeta} \sum_{n \ge 1} (-1)^n e^{\pi R n(a / \zeta + \bar{a} \zeta)} - c.c. \\
&= \frac{R^2 \de a}{2} \sum_{n \ge 1} (-1)^n \bar{a} (K_2(2 \pi R n \abs{a}) - K_0(2 \pi R n \abs{a})) - c.c. \\
&= \frac{R \bar{a} \de a}{2\pi\abs{a}} \sum_{n \ge 1} \frac{(-1)^n}{n} K_1(2 \pi R n \abs{a}) - c.c.
\end{align}
In the limit $R \to 0$, using $K_1(x) \sim 1/x + O(1)$, this gives finally
\begin{align} \label{eta0-formula-direct}
\lim_{R \to 0} \eta_0 &=
\frac{\de a}{4\pi^2 a} \sum_{n \ge 1} \frac{(-1)^n}{n^2} - c.c.
\\
 &=
-  \frac{1}{48} \Bigg( \frac{\de a}{a} - \frac{\de \bar{a}}{\bar{a}} \Bigg).
\end{align}
In particular, this limit exists and is a \ti{closed} form, as we expected.
The function $G$ appearing in \eqref{eta-closed} can thus be taken to be
\begin{equation}
G = - \frac{1}{48} \log (a/\bar{a}).
\end{equation}
Note that $G$ exists only \ti{locally}, or said otherwise, it suffers from
an ambiguity by shifts in $\frac{\pi \I}{12} {\mathbb Z}$.

\section{Solving the $q$--TBA equation for Argyres-Douglas models}

At the face of it, eqn.\eqref{qTBA} looks quite formidable. However in some very simple cases we may guess its solution. The guessing is based on uniqueness: any operator--valued piecewise holomorphic function with the right asymptotics and discontinuities should be the solution to the integral equation. In this appendix we argue that, in some special cases, to get the solution of the quantum TBA equations it suffices to solve their classical counterpart \cite{Gaiotto:2008cd}.

For simplicity, we focus on an Argyres--Douglas model of type $\mathfrak{g}\in ADE$ in the \emph{minimal} BPS chamber\footnote{ The analysis may extended to more general situations. In particular, the condition of \emph{BPS minimality} may be relaxed; the actual spectral condition depends on the orientation of the Dynkin quiver $Q_\mathfrak{g}$. It suffices that all subquivers which are supports of stable BPS states are $A_r$ quivers with the \emph{linear} orientation. For the region in parameter space covered by the linear $A_n$ quiver \cite{Alim:2011kw}, all BPS chambers satisfy the condition. In particular for the $A_2$ Argyres--Douglas model any BPS chamber will do. Having solved the quantum TBA problem in \emph{one} chamber, one may, in principle, recover the solution in \emph{all} chambers by the appropriate quantum KS jumps.} whose BPS spectrum consists of just $r\equiv \mathrm{rank}\,\mathfrak{g}$ hypermultiplets \cite{Cecotti:2011rv}. We write $\Gamma=\bigoplus_{i=1}^r \mathbb{Z}\alpha_i$ for the charge lattice (isomorphic to the root lattice of $\mathfrak{g}$); its simple--root generators $\{\alpha_i\}_{i=1}^r$ are the charge vectors of the minimal chamber BPS hypermultiplets, and their Dirac pairing
$B_{ij}\equiv \langle \alpha_i,\alpha_j\rangle_{D}$ is the exchange matrix of the corresponding $ADE$ Dynkin quiver $Q_\mathfrak{g}$ \cite{Cecotti:2011rv}.
The datum $Q_\mathfrak{g}$ defines a quantum torus algebra $\mathbb{T}_\Gamma$ \cite{Cecotti:2010fi}
\begin{gather}\label{qtpro}
\mathsf{X}_\gamma\,\mathsf{X}_{\gamma^\prime}=q^{\langle \gamma,\gamma^\prime\rangle_D/2}\;\mathsf{X}_{\gamma+\gamma^\prime}\\
\mathsf{X}_\gamma\,\mathsf{X}_{\gamma^\prime}= q^{\langle \gamma,\gamma^\prime\rangle_D}\;\mathsf{X}_{\gamma^\prime}\,\mathsf{X}_{\gamma},\\
\gamma,\;\gamma^\prime\in\Gamma,\quad q=e^{i\theta},\ \theta\in\mathbb{R},
\end{gather}
whose generators $\mathsf{X}_{\pm\alpha_i}$ may be represented as unitary Weyl operators
\begin{equation}\label{weylop}
\mathsf{X}_{\pm\alpha_i}= e^{\pm i\hat\theta_i},\quad\text{where}\quad \big[\hat\theta_i,\hat\theta_j\big]=-i \theta\, B_{ij}.
\end{equation}
Eqns.\eqref{qtpro}\eqref{weylop} imply that $\mathsf{X}_\gamma$ is unitary for all $\gamma\in\Gamma$, 
\begin{equation}\label{aaX1}
\mathsf{X}_\gamma^\dagger=\mathsf{X}_\gamma^{-1}\overset{\text{def}}{=}\mathsf{X}_{-\gamma},\qquad q^\dagger=q^{-1}.
\end{equation}
To any function $f$ on the classical torus
$(S^1)^r$
\begin{equation}\label{cqtorus1}
f\equiv \sum_{n_i\in\mathbb{Z}^r} f(n_i)\,e^{i n_i \theta_i}\qquad\  f(n_i)\in\mathbb{C},
\end{equation}
there is an associated quantum torus element $\widehat{f}\in\mathbb{T}_\Gamma$ namely
\begin{equation}\label{cqtorus2}
\widehat{f}=\sum_{n_i\in\mathbb{Z}^r} f(n_i)\, \mathsf{X}_{n_i\alpha_i},
\end{equation}
obtained by replacing $e^{i\theta_i}\to e^{i\hat\theta_i}$ in its Fourier expansion
 and taking the operator normal order
 \begin{equation}
 \Big(\mathsf{X}_{\gamma_1}\,\mathsf{X}_{\gamma_2}\cdots \mathsf{X}_{\gamma_s}\Big)_{\text{normal}\atop\text{product}}\equiv 
 \mathsf{X}_{\gamma_1+\gamma_2+\cdots+\gamma_s}.
 \end{equation}

Let 
\begin{gather}
X_{\alpha_i}(a_j,\theta_j;\zeta)=\sum_{n_j\in\mathbb{Z}^r} X_{\alpha_i}(a_j,n_j;\zeta)\, e^{in_j\theta_j},\\
X_{\alpha_i}(a_j,n_j;\zeta)\sim e^{Ra_i/\zeta+R\bar a_i\,\zeta}\,\prod_j\delta_{n_j,\delta_{ij}}\quad \text{as }\ R\to\infty,
\end{gather} 
be the Fourier expansion of the solutions to the \emph{classical} TBA equations of ref.\cite{Gaiotto:2008cd} with $\{a_j\}$ in some domain of the Coulomb branch which belongs to the above \emph{minimal} BPS chamber.

We claim that in this \emph{minimal} case the solution to the quantum TBA equation \eqref{qTBA} is just given by the associated quantum torus elements\footnote{ For convenience, we flip the overall sign of the quantum operator $\widehat{X}_{\alpha_i}(\zeta)$ with respect to the conventions used in section \ref{ss:thetalim}.}
\begin{equation}\label{aaX2}
\widehat{X}_{\alpha_i}(a_j;\zeta)=\sum_{n_j\in\mathbb{Z}^r} X_{\alpha_i}(a_j,n_j;\zeta)\,\mathsf{X}_{n_j \alpha_j}.
\end{equation}
To justify the claim we have to show four facts:
\begin{itemize}
\item[\textit{i)}] $\widehat{X}_{\alpha_i}(a_j;\zeta)$ has the correct quantum KS jumps at all BPS rays $\ell_{\pm\alpha_i}$; 
\item[\textit{ii)}] the $\widehat{X}_{\alpha_i}(a_j;\zeta)$'s satisfy the equal--$\zeta$ canonical commutation relations
\begin{equation}
\widehat{X}_{\alpha_i}(a;\zeta)\,\widehat{X}_{\alpha_j}(a;\zeta)= q^{B_{ij}}\,\widehat{X}_{\alpha_j}(a;\zeta)\,\widehat{X}_{\alpha_i}(a;\zeta);
\end{equation}
\item[\textit{iii)}] the $\widehat{X}_{\alpha_i}(a_j;\zeta)$ satisfy the correct (quantum) reality condition
\begin{equation}
\widehat{X}_{\alpha_i}(a_j;-1/\bar\zeta)^\dagger= \widehat{X}_{-\alpha_i}(a_j;\zeta);
\end{equation}
\item[\textit{iv)}] $\widehat{X}_{\alpha_i}(a_j;\zeta)$ has the correct asymptotics as $R\to\infty$.
\end{itemize}

Fact \textit{iv)} holds by construction: we \emph{chose} the boundary condition of the quantum TBA problem to reproduce the right behavior as $R\to\infty$. For  fact \textit{iii)}, notice that the reality condition on the classical GMN line operators $X_{\alpha_i}(a_j,\theta_j;\zeta)$ \cite{Gaiotto:2008cd} implies for their Fourier coefficients
\begin{equation}
X_{-\alpha_i}(a_j,n_j;\zeta)= X_{\alpha_i}(a_j,-n_j;-1/\bar\zeta)^*,
\end{equation}
while from eqns.\eqref{aaX1}\eqref{aaX2}
\begin{equation}
\begin{split}
\widehat{X}_{\alpha_i}(a_j;-1/\bar\zeta)^\dagger&= \sum_{n_j\in\mathbb{Z}^r} X_{\alpha_i}(a_j,n_j;-1/\bar\zeta)^*\; \mathsf{X}_{-n_j\alpha_j}=\\
&=\sum_{n_j\in\mathbb{Z}^r} {X}_{-\alpha_i}(a_j,-n_j;\zeta)\,\mathsf{X}_{-n_j\alpha_j}\equiv\widehat{X}_{-\alpha_i}(a_j;\zeta).
\end{split}\end{equation}
Note that the normal order prescription is essential for the quantum reality condition.

To argue fact \textit{ii)}, consider the space $\mathcal{M}$ of coordinates $(a_j,\bar a_j, \theta_j)$ endowed with the (degenerate) Poisson bracket
\begin{align}\label{PB1}
\big\{\theta_i,\theta_j\big\}_\mathrm{PB}&=-B_{ij},
\\ 
\big\{a_i,\cdots\big\}_\mathrm{PB}&=\big\{\bar a_i,\cdots\big\}_\mathrm{PB}=0.\label{PB2}
\end{align}
The classical GMN lines $X_{\alpha_i}(a;\zeta)$
satisfy
\begin{equation}\label{PB3}
\big\{\log X_{\alpha_i}(a;\zeta), \log X_{\alpha_j}(a;\zeta)\big\}_\mathrm{PB}= B_{ij}.
\end{equation}
Indeed, this equality is consistent with both the 
$R\to\infty$ asymptotics and the KS jumps.

The $\theta$--limit operator algebra $C(a_j,\bar a_j)\otimes\mathbb{T}_\Gamma$ is just the algebra of functions on $\mathcal{M}$ equipped with the Moyal product $\ast$ induced by the Poisson bracket
\eqref{PB1}\eqref{PB2} \cite{kondef}. Indeed
\begin{equation}
\widehat{f}\cdot \widehat{g}= \widehat{h}\quad\Longleftrightarrow\quad h=f\ast g.
\end{equation}
In this language, the normal product is just the ordinary product of functions
\begin{equation}
\Big(\widehat{f}\;\widehat{g}\Big)_{\text{normal}\atop\text{product}}= \widehat{f\,g\,}.
\end{equation}
Then eqn.\eqref{PB3} yields
\begin{equation}
\big[\log \widehat{X}_{\alpha_i}(a;\zeta), \log \widehat{X}_{\alpha_j}(a;\zeta)\big]= B_{ij},
\end{equation}
which implies fact \textit{ii)}
\begin{equation}
\widehat{X}_{\alpha_i}(a;\zeta)\,\widehat{X}_{\alpha_j}(a;\zeta)= q^{B_{ij}}\,
\widehat{X}_{\alpha_j}(a;\zeta)\,\widehat{X}_{\alpha_i}(a;\zeta),
\end{equation}
and, more generally, the equal--$\zeta$ product rule
\begin{equation}\label{oaprodF}
\widehat{X}_{\gamma}(a;\zeta)\,\widehat{X}_{\gamma^\prime}(a;\zeta)= q^{\langle \gamma,\gamma^\prime\rangle/2}\;
\widehat{X}_{\gamma+\gamma^\prime}(a;\zeta).
\end{equation}

It remains to show fact \textit{i)}, that is, to compare the KS jumps of the classical functions $X_{\alpha_i}(\zeta)$ and of the quantum operators $\widehat{X}_{\alpha_i}(\zeta)$.  Here is where we use our special assumptions that $Q_\mathfrak{g}$ is an $ADE$ Dynkin quiver and the BPS spectrum is minimal. These assumptions entail that the charges of the stable BPS states are $\pm \alpha_j$, while $|B_{ij}|\leq 1$ for all $i,j$. Under these conditions, the classical KS symplectomorphism \cite{Kontsevich:2008fj}
at the BPS ray $\ell_{\pm\alpha_j}$ associated to a hypermultiplet of charge $\pm \alpha_j$ is
\begin{equation}\label{kkks1}
X_{\alpha_i}(a;\zeta)\longrightarrow \big(1+X_{\pm \alpha_j}(a;\zeta)\big)^{\pm B_{ij}}\;X_{\alpha_i}(a;\zeta),
\end{equation}
while the quantum jump is
\begin{equation}\label{kkks2}
\widehat{X}_{\alpha_i}(a;\zeta)\longrightarrow \big(1+q^{\pm B_{ij}/2}\,\widehat{X}_{\pm \alpha_j}(a;\zeta)\big)^{\pm B_{ij}}\;\widehat{X}_{\alpha_i}(a;\zeta).
\end{equation}

To complete the argument, we have to show that for the operators $\widehat{X}_{\alpha_i}(a;\zeta)$ defined in eqn.\eqref{aaX2}, the validity of the classical formula
\eqref{kkks1} implies the validity of the quantum one \eqref{kkks2}.
Since $\pm B_{ij}=1, 0,-1$, we have to consider two cases $\pm B_{ij}=+1$ and $\pm B_{ij}=-1$. In the $+1$ case,
using eqn.\eqref{oaprodF}
the quantum KS formula \eqref{kkks2} may be rewritten as 
\begin{equation}
\widehat{X}_{\alpha_i}(a;\zeta)\longrightarrow \widehat{X}_{\alpha_i}(a;\zeta)+\widehat{X}_{\pm \alpha_j+\alpha_i}(a;\zeta)
\end{equation}
which is precisely the quantum torus operator
corresponding to the \textsc{rhs} of eqn.\eqref{kkks1} under the classical/quantum torus correspondence $f\mapsto \widehat{f}$, eqns.\eqref{cqtorus1}\eqref{cqtorus2}. In the $-1$ case
\begin{equation}
\widehat{X}_{\alpha_i}(a;\zeta)\longrightarrow\sum_{k=0}^\infty (-1)^k q^{-k}\,\widehat{X}_{\pm \alpha_j}(a;\zeta)^k\,\widehat{X}_{\alpha_i}(a;\zeta)
\equiv \sum_{k=0}^\infty (-1)^k \widehat{X}_{\alpha_i\pm k \alpha_j},
\end{equation}
which again is the image of the \textsc{rhs} of eqn.\eqref{kkks1} under the correspondence $f\mapsto \widehat{f}$.

\section{$\beta$--deformed Quiver Matrix LG Models\\
(exact twistorial $ADE$ Toda amplitudes)}\label{appendixmultimatrix}

We saw in section \ref{ss:abelian} that the $tt^*$ equations become linear whenever the vacuum bundle $\mathcal{H}$ over the coupling constant space $\mathcal{K}$ has rank one. In this case the $tt^*$ solution is captured by a pluri--harmonic function $\log G_{tt^*}$ which may be singular only at loci in $\mathcal{K}$ where a new massless sector blows up. This makes possible to construct the $tt^*$ geometry explicitly by the techniques illustrated in \S.\ref{ss:abelian}.  
In view of this fact, one is lead to ask whether there are other Abelian four--supercharge models, besides the ones discussed in \S.\ref{ss:abelian}, which are physically natural, in the sense
that their large--$N$ limit is dual to the topological string in some geometry. In this appendix we consider a class of LG models which may be though of as describing the twistorial extension of a $\beta$--deformed version of the quiver matrix models studied in ref.\cite{Dijkgraaf:2002vw}. 
As discussed in ref.\cite{Dijkgraaf:2009pc}, the large $N$ limit of such quiver matrix models describe the holomorphic blocks of the Toda conformal field theory. Then the results of the present appendix (besides shedding light on some important mathematical conjectures \cite{gaudin3}) may be seen as the \emph{exact} twistorial extension of some \emph{special} three--point amplitudes of $ADE$ Toda conformal field theories, generalizing the Liouville case discussed in section \ref{s:other}.   As already mentioned in \S.\ref{ss:abelian}, the class $\mathcal{S}$ 4d theories associated to these special models via the AGT correspondence have no non--trivial magnetic charges. 

\subsection{The models}

The relevant LG models are labelled by a Lie algebra
$\mathfrak{g}$ which, for simplicity, we take to be  simply--laced, $\mathfrak{g}=ADE$, of rank $r$. To the $\ell$--th node of the 
Dynkin graph $\Gamma_\mathfrak{g}$ of $\mathfrak{g}$
we associate the following data: \textit{i)} a positive integer $N_\ell$,
\textit{ii)} a rational differential $W^\prime_\ell(z)\,dz$, and
\textit{iii)} $N_\ell$ chiral superfields denoted as $X_{\ell,i_\ell}$, $i_\ell=1,2,\dots, N_\ell$. We consider the superpotential
\begin{equation}\label{wwquivermmod}
\begin{split}
\mathcal{W}(e_{\ell,k_\ell})=&\sum_{\ell=1}^r\sum_{i_\ell=1}^{N_\ell}W_\ell(X_{\ell,i_\ell})+\beta\sum_{\ell=1}^r \sum_{1\leq i_\ell <j_\ell\leq N_\ell}\log(X_{\ell,i_\ell}-X_{\ell,j_\ell})^2-\\
&-\beta\sum_{\langle \ell\; \ell^\prime\rangle}\sum_{i_\ell=1}^{N_\ell}\sum_{j_{\ell^\prime}=1}^{N_{\ell^\prime}}\log(X_{\ell,i_\ell}-X_{\ell^\prime,j_{\ell^\prime}}),
\end{split}
\end{equation} 
where $\sum_{\langle \ell\; \ell^\prime\rangle}$ means sum over unordered pairs of nodes $\ell,\ell^\prime\in\Gamma_\mathfrak{g}$ which are connected by a link in
$\Gamma_\mathfrak{g}$. The quiver matrix models
of  \cite{Dijkgraaf:2002vw,Dijkgraaf:2009pc} correspond to the period integrals
\begin{equation}
\int e^\mathcal{W}\bigg|_{\beta=1}\; d^NX,
\end{equation}
where $N=\sum_\ell N_\ell$. The \textsc{rhs} of \eqref{wwquivermmod} is invariant under the product of permutation groups
\begin{equation}\label{premprod}
\mathfrak{S}_{N_1}\times \mathfrak{S}_{N_2}\times \cdots\times \mathfrak{S}_{N_r}\equiv \mathfrak{S}_{\boldsymbol{N}},
\end{equation}
and we identify the field configurations in the same orbit of $\mathfrak{S}_{\boldsymbol{N}}$. Then, as in section \ref{ss:abelian}, the actual chiral superfields are the
elementary symmetric functions $\{e_{\ell,k_\ell}\}$, $\ell=1,\dots,r$, $k_\ell=1,\dots,N_\ell$. In particular,
for $\mathfrak{g}=A_1$ we get back the models studied in section \ref{ss:abelian}.

We assume the rational differentials to be generic, that is, $W^\prime_\ell(z)\,dz$ has only simple poles in $\mathbb{P}^1$ (including the point at infinity). The higher pole models may be  obtained from the generic ones by confluence of ordinary singularities. Then
\begin{equation}\label{ooomn23}
W^\prime_\ell(z)\,dz=\sum_{a=1}^{n_\ell} \frac{\lambda_{\ell,a}}{z-z_{\ell,a}}\;dz.
\end{equation}
The model further simplifies if we assume the position of the poles to be independent of $\ell$, that is, $n_\ell=n$ and $z_{\ell,a}=z_a$, the residues $\lambda_{\ell,a}$  still being  general complex numbers. Notice that in the most relevant case (for our present purposes), \textit{i.e.}\! $n=2$,
this assumption is not at all a limitation of generality, since by field redefinitions we may alway make $z_{\ell,1}=0$ and $z_{\ell,2}=1$. Moreover, the above simplifying assumption automatically holds for LG models describing
$n$--point functions of Toda systems.

Granted this assumption, the residues $\lambda_{\ell,a}$ of the one--form $W^\prime_\ell(z)\,dz$ are most naturally seen as $n$ complex weights $\boldsymbol{\Lambda}_a$ of the Lie algebra $\mathfrak{g}$ under the identification
\begin{equation}
\alpha_\ell(\boldsymbol{\Lambda}_a)=\frac{\lambda_{\ell,a}}{\beta},\qquad a=1,2,\dots,n;\ \ell=1,2,\dots, r,
\end{equation} 
where $\alpha_\ell$ is the simple root associated to the $\ell$--th node of $\Gamma_\mathfrak{g}$. To each weight $\boldsymbol{\Lambda}_a$ there is associated an irreducible highest weight representation $L_{\boldsymbol{\Lambda}_a}$ of $\mathfrak{g}$. Then we think of the rational differential \eqref{ooomn23} as being specified by the collection of highest weight representations $\{L_{\boldsymbol{\Lambda}_1},
L_{\boldsymbol{\Lambda}_2},\cdots,L_{\boldsymbol{\Lambda}_n}\}$ of $\mathfrak{g}$,
$L_{\boldsymbol{\Lambda}_a}$ being attached to the puncture $z_a\in \mathbb{C}$ for $a=1,2,\dots, n$.

As always, the chiral ring $\mathcal{R}$ is defined by a set of relations which coincide with the classical vacuum equations
\begin{equation}\label{relrelchi}
d\mathcal{W}=0.
\end{equation}
Our first task is to solve these equations for $\mathcal{W}$ as in
\eqref{wwquivermmod}. The second task is to classify the models in which the solution is unique up to the action of the group \eqref{premprod}, which are the `Abelian' models we look
for. Then their exact $tt^*$ geometry is given by the formulae of section \ref{ss:abelian}.

\subsection{The Gaudin model, the Mukhin--Varchenko conjectures, and $tt^*$}

To solve the vacuum equations \eqref{relrelchi} one notices that, with the restriction discussed after eqn.\eqref{ooomn23}, they have an alternative interpretation as the Bethe ansatz equations for an integrable model, the Gaudin model with Lie algebra $\mathfrak{g}$ \cite{gaudin1,gaudin2,gaudin3,gaudin4,MVhessian,gaudin6,gaudin7} arising from the hypergeometric solutions of the  Knizhnik--Zamolodchikov equations \cite{KZ1,KZ2}. The hypergeometric solutions corresponding to the quiver LG model
specified by the dimension vector $\boldsymbol{N}=(N_1,N_2,\dots,N_r)$ and the highest weight representations
$\{L_{\boldsymbol{\Lambda}_a}\}_{a=1}^n$ live in the subspace
\begin{equation}
\mathsf{Sing}\!\left(\bigotimes\nolimits_{a=1}^n L_{\boldsymbol{\Lambda}_a};\boldsymbol{N}\right)\subset L_{\boldsymbol{\Lambda}_1}\otimes L_{\boldsymbol{\Lambda}_2}\otimes \cdots\otimes L_{\boldsymbol{\Lambda}_n},
\end{equation} 
of singular vectors (vectors annihilated by the Chevalley generators $e_\ell$) of weight
\begin{equation}
\sum_{a=1}^n \boldsymbol{\Lambda}_a-\sum_{\ell=1}^rN_\ell\,\alpha_\ell.
\end{equation}

For $\mathfrak{g}=A_r$ it is known \cite{lamegaudin} that the vacuum/Bethe ansatz equations \eqref{relrelchi} may be recast in the form of a linear ODE of order $(r+1)$ generalizing the Heine--Stieltjes second order equation for the $A_1$ case (cfr.\! \S\S.\,\ref{chiringva}, \ref{HSandvanvleck}). In particular, the model \eqref{wwquivermmod} with $\mathfrak{g}=A_r$ and differentials
\eqref{ooomn23} has Witten index\footnote{ Vacuum configurations in the same $\mathfrak{S}_{N_1}\times\cdots\times\mathfrak{S}_{N_r}$ orbit are identified.} $m$ not greater than \cite{BMV,lamegaudin}
\begin{equation}
\dim \mathsf{Sing}\!\left(\bigotimes\nolimits_{a=1}^n L_{\boldsymbol{\Lambda}_a};\boldsymbol{N}\right)\overset{\text{gen.}}{=}\dim \mathcal{U}(\mathfrak{n}_-)^{\otimes (n-1)}[\boldsymbol{N}],
\end{equation}
where $\mathfrak{n_-}$ is the nilpotent Lie subalgebra of
$\mathfrak{sl}_{r+1}$  of lower triangular matrices, 
$\mathcal{U}(\mathfrak{n}_-)$ its universal enveloping algebra,
$\mathcal{U}(\mathfrak{n}_-)^{\otimes (n-1)}$ its $(n-1)$--fold tensor power, which is endowed with a natural  
$\mathbb{Z}^r$--grading given by the $\mathfrak{sl}_{r+1}$ weight, 
$(\cdots)[\boldsymbol{N}]$ stands for the degree $\boldsymbol{N}$ subspace, and $\overset{\text{gen.}}{=}$ means that the equality holds for \emph{generic} weights
$\boldsymbol{\Lambda}_a$.

For general $\mathfrak{g}$, it is easy to see that $n>2$ implies $m>1$.
Hence for our purposes (that is, to find interesting $m=1$ models)
we may limit ourselves to $n=2$. In \cite{gaudin3} Mukhin and Varchenko state three deep and surprising conjectures for this case that we quote verbatim:
\medskip

\noindent\textbf{Conjecture 2.} \textit{If the space
$\mathsf{Sing}(L_{\boldsymbol{\Lambda}_1}\otimes L_{\boldsymbol{\Lambda}_2};\boldsymbol{N})$ is one--dimensional, then the corresponding superpotential $\mathcal{W}$ has exactly one critical point modulo 
$\mathfrak{S}_{N_1}\times \cdots\times \mathfrak{S}_{N_r}$.}
\vglue 6pt

\noindent\textbf{Conjecture 1.} \textit{If the space
$\mathsf{Sing}(L_{\boldsymbol{\Lambda}_1}\otimes L_{\boldsymbol{\Lambda}_2};\boldsymbol{N})$ is one--dimensional, then there exist a $N$--chain $\Delta$ such that the period integral $\int_\Delta e^\mathcal{W}\, d^NX$ can be computed explicitly and it is equal to an alternating product of Euler $\Gamma$--functions up to a rational number independent of $\boldsymbol{\Lambda}_{1,2}$ and $\beta$.}
\vglue 6pt

In particular if $\dim \mathsf{Sing}(L_{\boldsymbol{\Lambda}_1}\otimes L_{\boldsymbol{\Lambda}_2};\boldsymbol{N})=1$ the Bethe equations for the Gaudin model have a unique solution with
Bethe vector $X$.
\vglue 6pt

\noindent\textbf{Conjecture 3.} \textit{If $\dim \mathsf{Sing}(L_{\boldsymbol{\Lambda}_1}\otimes L_{\boldsymbol{\Lambda}_2};\boldsymbol{N})=1$, the length of the unique Bethe vector $X$ is given by the Hessian of the superpotential at its (unique) critical point.}
\medskip

The conjectures reduce for $\mathfrak{g}=A_1$ to the standard Selberg integral. They are proven for $\mathfrak{g}=A_r$ \cite{warnaar1,warnaar2,selberg2} and in some special cases for other Lie algebras \cite{selberg2}. 

In their paper \cite{gaudin3} Mukhin and Varchenko do not give any motivation for their conjectures, except for presenting a few explicit examples of their validity. From the $tt^*$ viewpoint, however, the reason why they should be true is pretty clear: if
$$\dim \mathsf{Sing}(L_{\boldsymbol{\Lambda}_1}\otimes L_{\boldsymbol{\Lambda}_2};\boldsymbol{N})$$ is (typically) equal to the Witten index of the associated quiver $(2,2)$ LG model, when it is equal $1$ the corresponding $tt^*$ geometry is Abelian, hence
encoded in a pluri--harmonic function 
$H=\log G_{tt^*}$ of the periods $\lambda_{\ell,a}$, $\lambda_0\equiv\beta$ and their corresponding vacuum angles $\phi_{\ell,a}$, and $\phi_0\equiv \theta$
\begin{equation}
\left(\frac{\partial^2}{\partial \overline{\lambda}_I\partial\lambda_J}+4\pi^2\frac{\partial^2}{\partial \phi_I\partial\phi_J}\right)H=0,\qquad \begin{aligned}&I, J= 0\ \text{or }(\ell,a)\\ 
&\ell=1,2,... ,r, \quad a=1,2, ... ,n.
\end{aligned}
\end{equation}
As $\lambda_I\to\infty$ the function $H$ goes to zero in almost all directions. $H$ is singular only at loci in the coupling space where a 2d BPS soliton becomes massless. Then the function $H$ is obtained by the techniques of \S.\ref{ss:abelian}\begin{equation}
H(\lambda_I,\phi_I)=\sum_s q_s \;h\!\left(\ell_{s,I}\,\lambda_I,\ell_{s,I}\,\bar\lambda_I, \ell_{s,I}\,\phi_I\right),
\end{equation}
where $h(\mu,\bar\mu,\theta)$ is the logarithm of the  Abelian $tt^*$ metric for the basic charge distribution $\delta_\mathbb{Z}(x)-1$, that is,
\begin{equation}
h(\mu,\bar\mu,2\pi x)=\frac{1}{\pi}\,\mathrm{Im}\!\int_0^\infty \frac{ds}{s}\, \log\big(1-e^{-2\pi (\mu/s-i x+\bar\mu\,s)}\big),\qquad \mathrm{Re}\,\mu>0,
\end{equation} and 
$\ell_{s,I}$, $q_s$  are integers corresponding to the charges and multiplicities of the 2d BPS particles;
the corresponding magnetic function (cfr. \S.\ref{ss:abelian}) is
\begin{equation}
F(z_I)=\sum_s q_s\,\log\!\big(1-e^{-2\pi \ell_{s,I}\,z_I}\big).
\end{equation}  
The integers $\ell_{s,I}$, $q_s$ may be computed from
$\mathcal{W}$ by any one of the four methods described in \S.\ref{ss:abelian}. By the `thumb rule' of section
\ref{ss:abelian}, the corresponding twistorial brane amplitude reads
\begin{equation}\label{whatXi}
\Psi(\lambda_I,\theta_I,\zeta)=\prod_s \Gamma\!\left(i\,\ell_{s,I}\,\lambda_I/\zeta,-i\ell_{s,I}\,\bar\lambda_I\,\zeta, \ell_{s,I}\,\phi_I,\zeta\right)^{\!q_s},
\end{equation} 
where $\Gamma(\mu,\bar\mu,\theta)$ is the twistorial Gamma function.
Taking the asymmetric limit, the \textsc{lhs} reduces to the period integral $\int_\Delta e^\mathcal{W}\,d^NX$ (up trivial factors) while the \textsc{rhs} becomes a product of 
$\Gamma$--functions (again up to trivial factors).
This is the statement of \textbf{Conjecture 1} which is essentially proven by Abelian $tt^*$ geometry.

In fact, mathematicians look for a refinement of \textbf{Conjecture 1} in which both the chain of integration $\Delta$ and the specific form of the product of $\Gamma$--functions (that is, the integers $q_s$, $\ell_{s,I}$) are given.
In particular, the chain $\Delta$ is typically rather involved, see refs.\cite{warnaar1,warnaar2,selberg2},
and one is interested in its \emph{a priori} characterization.
$tt^*$ geometry yields the required refinement:
\medskip

\noindent\textbf{Conjecture 1*.} \textit{If  $\dim \mathsf{Sing}(L_{\boldsymbol{\Lambda}_1}\otimes L_{\boldsymbol{\Lambda}_2};\boldsymbol{N})=1$ one has
\begin{equation}
\kappa\int_\Delta e^{\mathcal{W}(\lambda_I)}\;d^N\!X=\prod_s
\Gamma\!\Big( \ell_{s,I}\big(\lambda_I+J_I\big)\Big)^{\!\!q_s}
\end{equation}
where \begin{itemize}
\item the chain $\Delta$ is the support of the unique $D$--brane
(or, in the other Stokes sector {\rm\cite{Cecotti:2013mba}}, of the unique Neumann brane);
\item the integers $q_s$ and $\ell_{s,I}$ may be computed by any one of the four methods described in \S.\ref{SS:solvingabelian};
\item the integral shifts $J_I$'s may be fixed by studying the discontinuities of the brane amplitude as a function of the vacuum angles $\theta_I$;
\item the rational number $\kappa$ is a mere statistical factor. 
\end{itemize}}

\textbf{Conjecture 2} may also be understood physically in terms of the class $\mathcal{S}[\mathfrak{g}]$ 4d $\mathcal{N}=2$ which is associated to the given model by the AGT correspondence \cite{AGT,wyllard}. As in the
$\mathfrak{g}=A_1$ case discussed in section \ref{ss:abelian}, the condition $m=1$ is expected to be equivalent to the statement that the corresponding 4d class $\mathcal{S}[\mathfrak{g}]$ theory is free (which, in particular, means that there are no non--trivial magnetic charges).

\textbf{Conjecture 3} also looks quite suggestive 
from the $tt^*$ perspective. For the general case, specified by a Lie algebra $\mathfrak{g}$, a dimension vector $\boldsymbol{N}$, and a collection of highest weight representations $L_{\boldsymbol{\Lambda}_1},\cdots, L_{\boldsymbol{\Lambda}_n}$, the solutions of the Gaudin model Bethe ansatz equations define the chiral ring $\mathcal{R}$. $\mathcal{R}$ is a commutative Frobenius $\mathbb{C}$--algebra with dimension equal to the number $m$ of solutions to the Bethe equation.
Each solution defines a Bethe vector $X^{(\alpha)}$ ($\alpha=1,\dots, m$) and on the space of such vector we have a natural symmetric bilinear form $B(\cdot,\cdot)$ induced by the Shapovalov form \cite{gaudin3}.
On the other hand, $\mathcal{R}$, being Frobenius, is also equipped with a natural non--degenerate symmetric bilinear form $\langle\cdot,\cdot\rangle$ (\textit{i.e.}\! the TFT 2--point function). $B(\cdot,\cdot)$ and $\langle\cdot,\cdot\rangle$ are symmetric bilinear forms on the same space, $\mathbb{C}^m$, and are determined by the same set of equations, namely the
Gaudin model Bethe ansatz ones. It is natural to guess that 
they are one and the same. To compare them explicitly we need to fix basis. For generic couplings, $\mathcal{R}$ is semisimple, and a natural basis is given by a complete system of orthogonal idempotents\footnote{ Since we identify configurations in the same orbit of $\mathfrak{S}_{\boldsymbol{N}}\equiv \prod_\ell \mathfrak{S}_{N_\ell}$, our chiral ring is actually $R^{\mathfrak{S}_{\boldsymbol{N}}}$, where $R$ is the usual chiral ring of the LG model, and the indecomposable idempotents $e_\alpha\in\mathcal{R}$ are averages over the orbits of $\mathfrak{S}_{\boldsymbol{N}}$ of the indecomposable
idempotents of $R$.}  $e_\alpha$ ($\alpha=1,2,\dots,m$)
\begin{equation}
e_\alpha\,e_\beta=\delta_{\alpha\beta}\,e_\alpha,
\end{equation} 
(this is called the `point basis' in \cite{Cecotti:1992rm,Cecotti:1991me}). Note that there is a natural one--to--one correspondence between classical vacua, \textit{i.e.}\! Bethe vectors $X_\alpha$ and idempotents $e_\alpha$.
Another canonical basis is the dual point basis, $\{e^\alpha\}$ defined by
\begin{equation}
\langle e^\alpha, e_\beta\rangle= {\delta^\alpha}_\beta.
\end{equation} 
It is well known that
\begin{equation}
\langle e^\alpha, e^\beta\rangle = \delta^{\alpha\beta}\,\det(\partial\partial\mathcal{W})\Big|_\alpha,
\end{equation} 
where in the \textsc{rhs} we have the Hessian of $\mathcal{W}$
evaluated at the $\alpha$--th critical point. For $m=1$ this 
gives back \textbf{Conjecture 3} provided we identify the
Shapovalov form on the Bethe vectors with the TFT $2$--point function in the dual point basis. This suggests the following
extension
\medskip

\noindent\textbf{Conjecture 3*.} \textit{In general,}
\begin{equation}\label{456qq}
B(X^{(\alpha)},X^{(\beta)})=\delta^{\alpha\beta}\;
\det(\partial\partial\mathcal{W})\Big|_\alpha.
\end{equation}
\medskip

In fact, for $\mathfrak{g}=A_r$ eqn.\eqref{456qq} is a proven theorem \cite{MVhessian}; in fact, in this case one also proves the stronger statement (also physically expected) that  the $(2,2)$ chiral ring $\mathcal{R}$ \emph{is isomorphic} 
to the Bethe algebra of the associated Gaudin model
\cite{MViso1,MViso2}.

\subsection{Explicit $tt^*$ amplitudes for the $A_r$ quiver matrix LG models}

For $\mathfrak{g}=A_r$ the 
Mukhin--Varchenko conjectures have been proven \cite{warnaar1,warnaar2,selberg2}. The detailed form of the result may be used to determine explicitly the magnetic
function $F(\boldsymbol{z})$ of the corresponding Abelian $tt^*$ geometry, from which we may read the exact twistorial amplitudes for the $SU(r+1)$ Toda field theory.
\smallskip

For $\mathfrak{g}=A_r$ \textbf{Conjecture 1} becomes the following 
\medskip

\noindent\textbf{Theorem} \cite{warnaar1,warnaar2,selberg2}.
\textit{Consider the quiver LG model with $\mathfrak{g}=A_r$ and
$n=2$ with\footnote{ The simple roots $\alpha_\ell$ of $A_r$ are numbered in the natural order.} 
\begin{equation}
\beta\,\boldsymbol{\Lambda}_1=(0,0,\cdots,0,\xi_r-1),\qquad 
\beta\,\boldsymbol{\Lambda}_2=(\eta_1-1,\eta_2-1,\cdots,\eta_r-1),
\end{equation} $\xi_r,\eta_1,\cdots,\eta_r\in\mathbb{C}$, and $0\leq N_1\leq N_2\leq \cdots\leq N_r$.
Then there is a chain $\Delta$ so that
\begin{multline}\label{selbegear}
\int_\Delta e^{\mathcal{W}(\xi_r,\eta_\ell,\beta)}\;d^NX=
\prod_{\ell=1}^r\prod_{i=1}^{N_\ell}\;\frac{\Gamma\!\big(\xi_\ell+(i-N_{\ell+1}-1)\beta\big)\,\Gamma(i\beta)}{\Gamma(\beta)} \times\\
\times\prod_{1\leq \ell \leq \ell^\prime\leq r}\prod_{i=1}^{N_\ell-N_{\ell-1}}\frac{\Gamma\!\big(\eta_\ell+\cdots+\eta_{\ell^\prime}+(i+\ell-\ell^\prime-1)\beta\big)}{
\Gamma\!\big(\xi_{\ell^\prime}+\eta_\ell+\cdots+\eta_{\ell^\prime}+(i+\ell-\ell^\prime+N_{\ell^\prime}-N_{\ell^\prime+1}-2)\beta\big)}
\end{multline}
where
\begin{equation}\label{rrvzn}
\xi_\ell=\begin{cases}1 & \ell\neq r\\
\xi_r & \ell=r,\end{cases}
\end{equation}
and the various parameters are chosen so that the integral is convergent (\emph{i.e.}\! we are in the `right' Stokes sector {\rm \cite{Cecotti:2013mba}}).}
\medskip

The chain $\Delta$ is explicitly known but rather involved, see 
\cite{warnaar1,warnaar2,selberg2}.

In view of eqn.\eqref{selbegear}, the the techniques of \S.\ref{SS:solvingabelian} yield}
 for the twistorial brane amplitude
for these models the expression obtained by replacing in the product in the \textsc{rhs} of \eqref{selbegear}
each $\Gamma$--function factor of the form
\begin{equation}
\Gamma\!\Big(\sum\nolimits_\ell a_\ell\,\xi_\ell+\sum\nolimits_\ell b_\ell\, \eta_\ell+c\beta\Big)
\end{equation}
with a twistorial Gamma function factor of the corresponding arguments\footnote{ For simplicity, we omit writing the barred variables in the argument.}
\begin{equation}
\Gamma\!\Big(\sum\nolimits_\ell a_\ell\,(\xi_\ell-1)+\sum\nolimits_\ell b_\ell\, (\eta_\ell-1)+c\beta,\; \sum\nolimits_\ell a_\ell\,\psi_\ell+\sum\nolimits_\ell b_\ell\, \phi_\ell+c\,\theta\Big),
\end{equation}
where $\psi_\ell$, $\phi_\ell$ and $\theta$ are
the angles associated, respectively, to the couplings $\xi_\ell$, $\eta_\ell$ and $\beta$. In view of \eqref{rrvzn}, one has
$\psi_\ell\equiv0$ for $\ell<r$.

Then, to get the twistorial Toda amplitudes one performs
the products of twistorial Gamma functions corresponding to the \textsc{rhs} of \eqref{selbegear} in terms of
ratios of twistorial \emph{double} Gamma functions
(compare with eqn.
\eqref{gaugaubrane}). The arguments of the 
 twistorial Gamma functions are the 't Hooft parameters to be kept fixed as $N\to\infty$. One analytically continues to arbitrary values of those parameters, and writes them in term of Toda quantities using the AGT correspondence as dictionary. The resulting objects are the explicit twistorial extensions of the corresponding Toda $3$--point holomorphic blocks.

\bibliography{references}

\end{document}